\documentclass[a4paper,11pt]{article}
\usepackage{jheppub} 
\usepackage[utf8]{inputenc}
\usepackage{physics}
\usepackage{slashed}
\usepackage{caption}
\usepackage{xcolor}
\usepackage{comment}
\usepackage{multirow}
\usepackage{graphics}
\usepackage{float}
\usepackage{cancel}
\usepackage{soul}
\usepackage{cases}
\usepackage{array}
\usepackage{mathtools}  
\usepackage{amsfonts}
\usepackage{hyperref}
\usepackage{amsmath}
\usepackage{amssymb}
\usepackage{tcolorbox}
\usepackage{tikz} 
\usepackage{tikz-cd}
\usepackage{tcolorbox}
\usepackage{booktabs}

\usetikzlibrary{arrows.meta, positioning}

\usepackage{mdframed}
\usepackage{tikz}
\usepackage{xcolor}
\usepackage{amsmath}
\usepackage{amssymb}

\definecolor{TwistorPurple}{RGB}{88,36,130} 
\definecolor{TwistorBlue}{RGB}{36,72,130} 
\definecolor{WittenGreen}{RGB}{0,100,100}
\definecolor{PenroseGray}{RGB}{80,80,80}
\definecolor{penroseblue}{RGB}{30, 60, 90}
\definecolor{penrosegrey}{RGB}{200, 200, 200}

\newenvironment{penrosebox}
{%
\mdfsetup{%
    skipabove=12pt,
    skipbelow=12pt,
    innertopmargin=1.2\baselineskip,
    innerbottommargin=1\baselineskip,
    innerleftmargin=1em,
    innerrightmargin=1em,
    linecolor=penroseblue,
    backgroundcolor=penroseblue!10,
    linewidth=1pt,
    roundcorner=10pt,
    frametitleaboveskip=0pt,
    frametitlealignment=\raggedright,
    frametitlefont=\bfseries\color{white},
    frametitlebackgroundcolor=penroseblue,
    frametitle={\strut The Penrose Transform}
}
\begin{mdframed}
}
{\end{mdframed}}

\newenvironment{wittenbox}
{%
\mdfsetup{%
    skipabove=12pt,
    skipbelow=12pt,
    innertopmargin=1.2\baselineskip,
    innerbottommargin=1\baselineskip,
    innerleftmargin=1em,
    innerrightmargin=1em,
    linecolor=WittenGreen,
    backgroundcolor=WittenGreen!5,
    linewidth=1pt,
    roundcorner=10pt,
    frametitleaboveskip=0pt,
    frametitlealignment=\raggedright,
    frametitlefont=\bfseries\color{white},
    frametitlebackgroundcolor=WittenGreen,
    frametitle={\strut The Witten Transform}
}
\begin{mdframed}
}
{\end{mdframed}}

\newenvironment{superpenrosebox}
{%
\mdfsetup{%
    skipabove=12pt,
    skipbelow=12pt,
    innertopmargin=1.2\baselineskip,
    innerbottommargin=1\baselineskip,
    innerleftmargin=1em,
    innerrightmargin=1em,
    linecolor=TwistorBlue,
    backgroundcolor=TwistorBlue!5,
    linewidth=1pt,
    roundcorner=10pt,
    frametitleaboveskip=0pt,
    frametitlealignment=\raggedright,
    frametitlefont=\bfseries\color{white},
    frametitlebackgroundcolor=TwistorBlue,
    frametitle={\strut Super-Penrose Transform}
}
\begin{mdframed}
}
{\end{mdframed}}

\newenvironment{superpenroseboxscalar}
{%
\mdfsetup{%
    skipabove=12pt,
    skipbelow=12pt,
    innertopmargin=1.2\baselineskip,
    innerbottommargin=1\baselineskip,
    innerleftmargin=1em,
    innerrightmargin=1em,
    linecolor=TwistorBlue,
    backgroundcolor=TwistorBlue!5,
    linewidth=1pt,
    roundcorner=10pt,
    frametitleaboveskip=0pt,
    frametitlealignment=\raggedright,
    frametitlefont=\bfseries\color{white},
    frametitlebackgroundcolor=TwistorBlue,
    frametitle={\strut Super-Penrose Transform for scalars}
}
\begin{mdframed}
}
{\end{mdframed}}

\definecolor{TwistorPurple}{RGB}{102,51,153}

\newenvironment{superwittenbox}
{%
\mdfsetup{%
    skipabove=12pt,
    skipbelow=12pt,
    innertopmargin=1.2\baselineskip,
    innerbottommargin=1\baselineskip,
    innerleftmargin=1em,
    innerrightmargin=1em,
    linecolor=WittenGreen,
    backgroundcolor=WittenGreen!5,
    linewidth=1pt,
    roundcorner=10pt,
    frametitleaboveskip=0pt,
    frametitlealignment=\raggedright,
    frametitlefont=\bfseries\color{white},
    frametitlebackgroundcolor=WittenGreen,
    frametitle={\strut The Super Witten Transform}
}
\begin{mdframed}
}
{\end{mdframed}}

\newenvironment{parityoddsuperwittenbox}
{%
\mdfsetup{%
    skipabove=12pt,
    skipbelow=12pt,
    innertopmargin=1.2\baselineskip,
    innerbottommargin=1\baselineskip,
    innerleftmargin=1em,
    innerrightmargin=1em,
    linecolor=WittenGreen,
    backgroundcolor=WittenGreen!5,
    linewidth=1pt,
    roundcorner=10pt,
    frametitleaboveskip=0pt,
    frametitlealignment=\raggedright,
    frametitlefont=\bfseries\color{white},
    frametitlebackgroundcolor=WittenGreen,
    frametitle={\strut The parity odd Super Witten Transform}
}
\begin{mdframed}
}
{\end{mdframed}}

\newenvironment{superwittenboxscalar}
{%
\mdfsetup{%
    skipabove=12pt,
    skipbelow=12pt,
    innertopmargin=1.2\baselineskip,
    innerbottommargin=1\baselineskip,
    innerleftmargin=1em,
    innerrightmargin=1em,
    linecolor=WittenGreen,
    backgroundcolor=WittenGreen!5,
    linewidth=1pt,
    roundcorner=10pt,
    frametitleaboveskip=0pt,
    frametitlealignment=\raggedright,
    frametitlefont=\bfseries\color{white},
    frametitlebackgroundcolor=WittenGreen,
    frametitle={\strut The Super Witten Transform for scalars}
}
\begin{mdframed}
}
{\end{mdframed}}

\newenvironment{paritybox}
{%
\mdfsetup{%
    skipabove=12pt,
    skipbelow=12pt,
    innertopmargin=1.2\baselineskip,
    innerbottommargin=1\baselineskip,
    innerleftmargin=1em,
    innerrightmargin=1em,
    linecolor=TwistorPurple,
    backgroundcolor=TwistorPurple!10,
    linewidth=1pt,
    roundcorner=10pt,
    frametitleaboveskip=0pt,
    frametitlealignment=\raggedright,
    frametitlefont=\bfseries\color{white},
    frametitlebackgroundcolor=TwistorPurple,
    frametitle={\strut Parity Even Current Three-Point Functions in Twistor Space}
}
\begin{mdframed}
}
{\end{mdframed}}

\newtcolorbox{simpleboxenv}[1][]{
  colframe=black!30,
  colback=black!5,
  coltitle=black!85!blue,
  fonttitle=\bfseries,
  boxrule=1pt,
  arc=6pt,
  top=6pt,
  bottom=6pt,
  left=6pt,
  right=6pt,
  title={#1}
}

\definecolor{LegendreBlue}{RGB}{38, 94, 156}

\newenvironment{Legendrebox}
{%
\mdfsetup{%
    skipabove=12pt,
    skipbelow=12pt,
    innertopmargin=1.2\baselineskip,
    innerbottommargin=1\baselineskip,
    innerleftmargin=1em,
    innerrightmargin=1em,
    linecolor=LegendreBlue,
    backgroundcolor=LegendreBlue!5,
    linewidth=1pt,
    roundcorner=10pt,
    frametitleaboveskip=0pt,
    frametitlealignment=\raggedright,
    frametitlefont=\bfseries\color{white},
    frametitlebackgroundcolor=LegendreBlue,
    frametitle={\strut The Legendre Transform}
}
\begin{mdframed}
}
{\end{mdframed}}

\definecolor{LegendreBlue}{RGB}{38, 94, 156}

\newenvironment{D for Odelta scalar}
{%
\mdfsetup{%
    skipabove=12pt,
    skipbelow=12pt,
    innertopmargin=1.2\baselineskip,
    innerbottommargin=1\baselineskip,
    innerleftmargin=1em,
    innerrightmargin=1em,
    linecolor=LegendreBlue,
    backgroundcolor=LegendreBlue!5,
    linewidth=1pt,
    roundcorner=10pt,
    frametitleaboveskip=0pt,
    frametitlealignment=\raggedright,
    frametitlefont=\bfseries\color{white},
    frametitlebackgroundcolor=LegendreBlue,
    frametitle={\strut Action of D on $\hat{O}_{\Delta}$ scalar}
}
\begin{mdframed}
}
{\end{mdframed}}

\title{\boldmath\boldmath An Ode to the Penrose and Witten transforms in Twistor space for 3D CFT}

\author{Aswini Bala and Dhruva K.S.}

\affiliation{Indian Institute of Science Education and Research,\\ Dr Homi Bhabha Road, Pashan, Pune, India}

\emailAdd{aswini.bala@students.iiserpune.ac.in}

\emailAdd{k.s.dhruva@students.iiserpune.ac.in}

\abstract{Here we discuss the construction of Sp$(4;\mathbb{R})$ invariant objects in the twistor space for three dimensional conformal field theories. The Sp$(4;\mathbb{R})$ invariant projective delta function, alongside the Twistor symplectic dot product invariants form the basis for conformal Wightman functions involving conserved currents and $\Delta=1$ scalars. For correlators involving scalars with $\Delta\ne 1$, generic spinning primaries and parity odd correlators we show that the infinity twistor of $\mathbb{R}^{2,1}$ must be incorporated into the analysis. We show that this feature can be traced to the Penrose and Witten transforms of these operators that we derive.  We then discuss the super-twistor space construction and derive the supersymmetric Penrose transform for $\mathcal{N}=1$ theories using the Fourier transform and the supersymmetric Witten transform. We construct OSp$(\mathcal{N}|4;\mathbb{R})$ invariants and its application to several super-Wightman functions. Similar to the non supersymmetric case, we find an important role played by the (super) infinity twistor which we exemplify through parity odd super-correlators and a supersymmetric contact term.}

\begin{document}
\maketitle

\section{Introduction}
 Twistor theory, since it was first conceived by Penrose \cite{Penrose:1967wn}, has led to many developments such as in the context of four dimensional flat-space scattering amplitudes \cite{Nair:1988bq,Witten:2003nn,Mason:2009sa,Arkani-Hamed:2009hub}. More recently, there have been developments to study three dimensional conformal field theories (CFTs) in twistor space \cite{Baumann:2024ttn,Bala:2025gmz} which through the lens of holography, also describes scattering in four dimensional (anti)de-Sitter spacetime. The study of CFT$_3$ in twistor space is complementary to the traditional position space approaches \cite{Belavin:1984vu,Rattazzi:2008pe,Poland:2022qrs,Hartman:2022zik} and the more recent momentum space and spinor helicity point of view \cite{Maldacena:2011nz,McFadden:2011kk,Ghosh:2014kba,Coriano:2013jba,Bzowski:2013sza,Bzowski:2015pba,Bzowski:2017poo,Farrow:2018yni,Bzowski:2018fql,Bautista:2019qxj,Lipstein:2019mpu,Baumann:2020dch,Jain:2020rmw,Jain:2020puw,Jain:2021wyn,Baumann:2021fxj,Jain:2021qcl,Jain:2021vrv,Jain:2021gwa,Jain:2021whr,Isono:2019ihz,Gillioz:2019lgs,Baumann:2019oyu}. (Real) Twistor space in some sense sits in between position space and momentum space where it is connected to the former via a Penrose transform \cite{Penrose:1967wn} and to the latter through Witten's half-Fourier transform \cite{Witten:2003nn}. Thus, this presents an opportunity to import the best tools and techniques from either side and use them to set up the twistor space bootstrap. In particular, the simplicity of two and three point conserved current correlators in twistor space \cite{Baumann:2024ttn,Bala:2025gmz} suggests that it might provide a nice stage to bootstrap spinning correlators.

In this paper, we build on the constructed twistor space for 3D CFTs and derive several new results. One point to note is that previous works \cite{Baumann:2024ttn,Bala:2025gmz} used an ambidextrous approach employing both twistors and their Fourier duals to represent two and three point Wightman functions involving conserved currents. On the other hand, we shall work purely with twistor variables by employing additional invariants of the conformal group which also allows us to extend the results to include twistor space Wightman functions involving $\Delta=1$ scalar operators. An interesting observation that we make is that in order to describe correlators involving scalars with $\Delta\ne 1$, one needs to use the infinity twistor of $\mathbb{R}^{2,1}$ that breaks the natural conformal invariance in twistor space. However, we shall see that this is compensated by that fact that scalars with $\Delta\ne 1$ transform in a representation of Sp$(4)$ that involves non-local terms and with these generators, their correlators are indeed conformally invariant. The infinity twistor also features in parity odd correlators. However, the dependence is mild in the sense that it appears only in sign factors and in such a way that on the support of conformally invariant delta functions that it multiplies, the whole expression is conformally invariant as a distribution. We also derive the Penrose transform for generic non-conserved spinning operators. In all cases, the infinity twistor features. One of the main messages of this paper is that in order to accommodate general representations of the conformal group, one must extend the space of Sp$(4)$ invariants to allow for those that involve the infinity twistor.

We then switch gears and turn towards superconformal field theories. We derive the supersymmetric Penrose transform for $\mathcal{N}=1$ theories and discuss its relation to the supersymmetric Witten transform. An interesting feature is that the infinity twistor of $\mathbb{R}^{2,1}$ is naturally incorporated by the super-incidence relations. We then discuss the different super-conformal OSp$(\mathcal{N}|4)$ invariants such as the orthosymplectic dot products and the projective super-delta functions that serve as building blocks for correlators. We explicitly determine various examples in $\mathcal{N}=1$ theories including super-correlators involving scalar super-fields. 

 A more detailed outline of the paper is as follows:
\subsection*{Outline of the main-text}
In section \ref{sec:Geometry}, we discuss the basics of the geometry of twistor space associated to $\mathbb{R}^{2,1}$, its connection to position space via the Penrose transform and its connection to momentum space via Witten's half Fourier transform. We then derive the Penrose transform for conserved currents from the Fourier transform coupled with the Witten transform. We apply these ideas in Section \ref{sec:WightmanConserved3pt} to solve for two and three point Wightman functions of conserved currents and $\Delta=1$ scalars. In section \ref{sec:InfinityTwistorInvariants}, we show that the infinity twistor of $\mathbb{R}^{2,1}$ plays an essential role in defining the Penrose transform for generic primary operators. It is divided into three parts: We first obtain the Penrose transform and Wightman functions involving $\Delta=2$ scalars in subsection \ref{subsec:genscalar} via the Legendre transform. We then generalize the Penrose transform to accommodate arbitrary $\Delta$ scalars and show that the twistor space action of special conformal transformations is non-local. The second part, subsection \ref{subsec:parityoddsubsec} deals with the epsilon transform in twistor space and parity odd Wightman functions. The third part viz subsection \ref{subsec:nonconsTwistor} deals with the Penrose transform for generic non-conserved spinning operators. We switch gears in section \ref{sec:SuperTwistorGeo} and discuss super-twistor space. We construct the super-Penrose transform in two ways: 1)From the super-field expansion and the component level Penrose transform and 2) Via the Fourier and super-Witten transform employing the Faddeev Popov method. We then discuss the super-twistor space generators and construct the super-conformal invariants that are annihilated by them. Finally, we construct several super-twistor space Wightman functions in section \ref{sec:SUSYWightmanTwistor}. We also extend the super-Penrose transform to accomodate super-scalar operators and derive correlators involving them including supersymmetric contact terms. Discussion and some future directions are presented in section \ref{sec:Discussion}.

\subsection*{Outline of the appendices}
We have several appendices to expand upon and complement the main text. Appendix \ref{app:Notation} sets the notation of the paper. In appendix \ref{app:4dto3d} we perform a dimensional reduction of twistor space associated to $\mathbb{R}^{2,2}$ to obtain the one associated to $\mathbb{R}^{2,1}$. Appendix \ref{app:ProjectiveVSnonproj} deals with some very useful identities involving projective integrals. Appendices \ref{app:DerivationofDelta4} and \ref{app:TwistortoDual} present two derivations of the Sp$(4)$ invariant projective delta function involving three twistors. Appendix \ref{app:TwistorSpaceResults} tabulates all helicity configuration results for three point functions of currents. Appendix \ref{app:FixingWightman} serves as a test for our expressions for Wightman functions involving $\Delta=1$ scalars.  Appendix \ref{app:EPTdetails} presents details of the epsilon transform used in the construction of parity odd twistor space correlators. In appendix \ref{app:TwistorGenerators}, we sketch the derivation of the representation of the conformal algebra acting on arbitrary scalars. We then prove in appendix \ref{app:ODeltaConf}, the conformal invariance of two point functions of the $O_{\Delta}$. Finally, we compute some explicit two point Penrose transforms in appendix \ref{app:ExplicitPenrose}.  

\section{The Geometry of Twistor Space}\label{sec:Geometry}
In this section, we shall discuss the essential features of real twistor space which we shall use to study $2+1$ dimensional CFTs. We first introduce coordinates that we use to chart the space and its connection to $\mathbb{R}^{2,1}$ through the incidence relations. The discussion here is quite reminiscent of the four dimensional case \cite{Adamo:2017qyl} since one can interpret the twistor space for $\mathbb{R}^{2,1}$ as being derived from a dimensional reduction of the twistor space of $\mathbb{R}^{2,2}$. For more discussion, please see appendix \ref{app:4dto3d}.

The projective twistor space $\mathbb{RP}^3$ is spanned by projective coordinates $Z^A$ which are in the fundamental representation of Sp$(4)$, the double cover of the $2+1$ dimensional conformal group $SO(3,2)$. It can be written as a direct sum of fundamental representations of $SL(2,\mathbb{R})$  which is itself the double cover of the $2+1$ dimensional Lorentz group $SO(2,1)$:
\begin{align}\label{defoftwistorZA}
    Z^A=(\lambda^a,\Bar{\mu}_{a'}).
\end{align}
 The connection of this twistor space to the spacetime $\mathbb{R}^{2,1}$ is through the \textit{incidence relations},
\begin{align}\label{IncidenceRelation}
    \Bar{\mu}_a=-x_{ab} \lambda^b.
\end{align}
Here, $x_{ab}=(\sigma_\mu)_{ab}x^\mu$ is the usual contraction of the position vector $x^\mu$ with the Pauli matrices $\sigma_\mu$. Please see appendix \ref{app:Notation} for our conventions.
Note that the incidence relations \eqref{IncidenceRelation} after modding out by projective rescalings defines a $\mathbb{RP}^{1}\subset \mathbb{RP}^3$. Thus, given any point in $x_{ab}\in \mathbb{R}^{2,1}$, we can associate a $\mathbb{RP}^{1}$ in twistor space which is topologically equivalent to a circle. The non-locality of this correspondence works both ways. Let us define a point in twistor space as the intersection of two lines \eqref{IncidenceRelation} associated to points $x,y\in\mathbb{R}^{2,1}$. We have,
\begin{align}\label{xablambdabeq0}
x_{ab}\lambda^b=y_{ab}\lambda^b\implies (x-y)_{ab}\lambda^b=0.
\end{align}
This implies\footnote{\eqref{xablambdabeq0} tells us that $(x-y)^{ab}=\alpha^a\lambda^b$ for some $\alpha^a\in\mathbb{R}^2$. However since the Pauli matrices and thus the position bispinors are symmetric in their indices, one should rather say that $(x-y)^{ab}=\alpha^{(a}\lambda^{b)}$ with the constraint $\alpha\cdot \lambda=0$ to ensure \eqref{xablambdabeq0}. However, in the two dimensional space spanned by these spinors this implies that $\alpha\propto \lambda$ thus yielding \eqref{nullpoints1}.},
\begin{align}\label{nullpoints1}
    (x-y)_{ab}=\lambda^a \lambda^b,
\end{align}
 \eqref{nullpoints1} corresponds to null separated points $x,y$ as can easily be seen by squaring the equation. Therefore, we see that a point in twistor space corresponds to null line in $\mathbb{R}^{2,1}$. This leads us naturally to the concept of the infinity twistor. 

 \subsection{The Infinity Twistor}
 Since null rays are invariant under conformal transformations, we see that twistor space as it stands is insensitive to the overall conformal factor of  the metric in $\mathbb{R}^{2,1}$. This is where the \textit{infinity} twistor enters the fray. It breaks the conformal invariance and encodes the structure of spacetime at infinity. First, let us see how we can construct the $\mathbb{R}^{2,1}$ metric up to an overall scale from two twistors $Z_1^A,Z_2^A$ associated to a single point $x$ using the incidence relations \eqref{IncidenceRelation}. We construct the skew combination,
\begin{align}\label{XAB}
        &X^{AB}=Z_1^{[A}Z_2^{B]}=\frac{1}{2}\begin{pmatrix}
            \lambda_1^a\lambda_2^b-\lambda_1^b\lambda_2^a &&\lambda_1^a\Bar{\mu}_{2b'}-\Bar{\mu}_{1b'}\lambda_2^a\\
            \Bar{\mu}_{1a'}\lambda_2^b-\lambda_1^b\Bar{\mu}_{2a'}&&\Bar{\mu}_{1a'}\Bar{\mu}_{2b'}-\Bar{\mu}_{1b'}\Bar{\mu}_{2a'}.
        \end{pmatrix}=\frac{\langle 1 2\rangle}{2}\begin{pmatrix}
            \epsilon^{ab}&&-x^{a}_{b'}\\
            x^{b}_{a'}&& -\epsilon_{a'b'}x^2
        \end{pmatrix}.
\end{align}
Given this quantity $X^{AB}$, we can construct the following natural line element
\begin{align}\label{conformalmetric}
    d\tilde{s}^2=\frac{1}{2}\epsilon_{ABCD}dX^{AB}dX^{CD}=\frac{1}{2}\text{Pf}\big(dX\big).
\end{align}
The four index Levi-Civita symbol in $\mathbb{RP}^{3}$ is constructed out of the Sp$(4)$ conformally invariant tensor $\Omega$ via,
\begin{align}
    \epsilon_{ABCD}=-\bigg(\Omega_{AB}\Omega_{CD}-\Omega_{AC}\Omega_{BD}+\Omega_{AD}\Omega_{BC}\bigg),
\end{align}
where,
\begin{align}\label{Omega}
    \Omega_{AB}=\begin{pmatrix}
        0&&\delta_{a}^{b'}\\
        -\delta_{a'}^b &&0
    \end{pmatrix}.
\end{align}
$\text{Pf}(dX)$ is the Pfaffian of the matrix $dX^{AB}$ formed using \eqref{XAB}.
Opening up \eqref{conformalmetric} using the variation of \eqref{XAB} and \eqref{Omega} we obtain,
\begin{align}\label{conformalmetric1}
    d\tilde{s}^2=\langle 1 2\rangle^2(-dt^2+dx^2+dz^2)=\langle 1 2\rangle^2\eta_{\mu\nu}dx^\mu dx^\nu.
\end{align}
Note that \eqref{conformalmetric1} depends on $\langle 12\rangle^2$ and thus reproduces the Minkowski metric up to an overall conformal factor. To obtain the flat metric we must introduce another bi-twistor $I_{AB}$ to cancel out this factor. Define instead of \eqref{conformalmetric},
\begin{align}\label{conformalmetric2}
    ds^2=\frac{\epsilon_{ABCD}dX^{AB}dX^{CD}}{2(I_{AB}X^{AB})^2}=\frac{\text{Pf}(dX)}{2(I\cdot X)^2}.
\end{align}
This also ensures that this metric is invariant under projective rescalings $Z\to r Z, r\in\mathbb{R}$ as is appropriate in $\mathbb{RP}^3$.
If we choose,
\begin{align}\label{infinitytwistor}
    I_{AB}=\begin{pmatrix}
        \epsilon_{ab}&&0\\
        0&&0,
    \end{pmatrix}
\end{align}
we obtain,
\begin{align}
    (I_{AB}X^{AB})^2=\langle 1 2\rangle^2.
\end{align}
Substituting this in \eqref{conformalmetric2} and using \eqref{conformalmetric1} results in the flat Minkowski metric of $\mathbb{R}^{2,1}$ viz,
\begin{align}
    ds^2=-dt^2+dx^2+dz^2.
\end{align}
Thus, we see that \eqref{infinitytwistor} is the infinity twistor that breaks the Sp$(4)$ conformal invariance and picks out the  $\mathbb{R}^{2,1}$ flat Minkowski metric that has only the usual Poincare invariance. We shall return to the infinity twistor in sections \ref{sec:InfinityTwistorInvariants} where we shall see that it is required to describe scalars with arbitrary scaling dimensions, parity odd Wightman functions as well as non-conserved currents.

We shall now proceed to discuss the construction of conserved currents and $\Delta=1$ scalars. First through the Penrose transform that relates twistor space to position space and second, through Witten's half Fourier transform that relates twistor space to momentum space (spinor helicity variables). To show the equivalence of these constructions, we then derive the Penrose transform from the Witten transform.

\subsection{The Penrose transform}
Let us first introduce the Penrose transform. Just like how the spinor formalism provides an unconstrained way of describing null vectors (like in \eqref{nullpoints1})\footnote{Given any spinor $\lambda$, the moment we form a vector using $x_{ab}=\lambda_a\lambda_b$, the vector is automatically null. Contrast this to the standard definition of a null vector $x^\mu$. It is a vector $x^\mu$ with the constraint that $-t^2+x^2+z^2=0$. However, when we write $x_{ab}=\lambda_a \lambda_b$, there is no further constraint on $\lambda$ and hence the terminology unconstrained.}, the Penrose transform provides an unconstrained way to describe conserved currents.

In \cite{Baumann:2024ttn}, the authors introduce the Penrose transform from the embedding space perspective. Here, we stick to the $2+1$ dimensional viewpoint\footnote{Of course, it is easy to check that their Penrose transform when restricted to the Poincare section of the embedding space null cone gives rise to \eqref{PenroseTransform}.}. Given a symmetric traceless conserved current $J_s^{a_1\cdots a_{2s}}(x)$ we can represent it as follows to make manifest its conservation:

\begin{penrosebox}
\begin{align}\label{PenroseTransform}
    J_s^{a_1\cdots a_{2s}}(x) = \int \langle \lambda d\lambda \rangle \lambda^{a_1} \cdots \lambda^{a_{2s}} \hat{J}_s^{+}(\lambda, \bar{\mu}) \big|_X,
\end{align}
where $X$ denotes imposing the incidence relation \eqref{IncidenceRelation} and the measure is the natural one on $\mathbb{RP}^1$.
\end{penrosebox}

Let us check that \eqref{PenroseTransform} leads to a conserved current. Taking a divergence we obtain,
\begin{align}\label{PenroseTransformConservation}
    \frac{\partial}{\partial x^{a_1 a_2}}J_s^{a_1\cdots a_{2s}}(x)=\int \langle \lambda d\lambda\rangle \lambda^{a_1}\cdots\lambda^{a_{2s}}\frac{\partial \Bar{\mu}^a}{\partial x^{a_1a_2}}\frac{\partial}{\partial \Bar{\mu}^a}\hat{J}_s^{+}(\lambda,\Bar{\mu})|_{X}
\end{align}
Using the incidence relation \eqref{IncidenceRelation} we find that,
\begin{align}\label{mubarxderivative}
   \frac{\partial \Bar{\mu}^a}{\partial x^{a_1a_2}}=\lambda_b\big(-2\delta^b_{a_2}\delta^a_{a_1}+\epsilon_{a_1a_2}\epsilon^{ab}\big)=-2\lambda_{a_2}\delta^{a}_{a_1}+\epsilon_{a_1a_2}\lambda^a.
\end{align}
Substituting \eqref{mubarxderivative} in \eqref{PenroseTransformConservation} yields,
\begin{align}
     \frac{\partial}{\partial x^{a_1 a_2}}J_s^{a_1\cdots a_{2s}}(x)=\int \langle \lambda d\lambda\rangle \lambda^{a_1}\cdots\lambda^{a_{2s}}\big(-2\lambda_{a_2}\delta^a_{a_1}+\epsilon_{a_1a_2}\lambda^a)\frac{\partial}{\partial \Bar{\mu}^a}\hat{J}_s^{+}(\lambda,\Bar{\mu})|_{X}=0,
\end{align}
since $\lambda^{a_2}\lambda_{a_1}=0$ and $\epsilon_{a_1a_2}\lambda^{a_1}\lambda^{a_2}=0$. Note that this is true for any $J_s^{+}(\lambda,\Bar{\mu})$, thus showing that the Penrose transform \eqref{PenroseTransform} allows us to represent conserved currents in an unconstrained way. An important property of \eqref{PenroseTransform} is that the integrand is invariant under projective rescalings $Z^A=(\lambda^a,\Bar{\mu}_{a'})\to r Z^A=(r\lambda^a,r \Bar{\mu}_{a'})$ which also requires that the twistor space current scales as,
\begin{align}\label{rescalingformulaJsone}
    \hat{J}_s^{+}(r Z)=\frac{1}{r^{2s+2}}\hat{J}_s^{+}(Z).
\end{align}

There also exists another type of Penrose transform \cite{Baumann:2024ttn} which is called the derivative based Penrose transform in contrast to its product based counterpart \eqref{PenroseTransform}.
\begin{align}\label{PenroseTransformDerivativeType}
    J_s^{a_1\cdots a_{2s}}(x)=\int \langle \Bar{\lambda} d\Bar{\lambda}\rangle\frac{\partial}{\partial \mu_{a_1}}\cdots \frac{\partial}{\partial \mu_{a_{2s}}}\hat{J}_s^{+}(\mu,\Bar{\lambda})|_{X'},
\end{align}
where $X'$ denotes the dual incidence relation,
\begin{align}
    \mu_a=x_{ab}\Bar{\lambda}^b.
\end{align}
It is easy to check that this Penrose transform is also conserved just like \eqref{PenroseTransformConservation}. The current in dual-twistor variables $W_A=(\mu_a,\Bar{\lambda}^{a'})$ also satisfies,
\begin{align}
    \hat{J}_s^{+}(r W)=\frac{1}{r^{-2s+2}}\hat{J}_s^{+}(W).
\end{align}
 In this work, we stick to the product based Penrose transform \eqref{PenroseTransform}. Before we proceed, one important point to note is that both in \eqref{PenroseTransform} and \eqref{PenroseTransformDerivativeType}, the twistor space current that appears inside the $\mathbb{RP}^1$ integral is the \textit{positive} helicity component of the current which we shall define in \eqref{Jspm}. Recall that symmetric traceless conserved currents in three dimensions have only two independent components which we call positive and negative helicity\footnote{One easy way to understand this is in the context of holography. Each spin$-s$ symmetric traceless conserved current is dual to a spin$-s$ massless gauge field in AdS$_4$. The latter have only two physical degrees of freedom which translates to the former having only two independent components.}. There also exist similar Penrose transforms that make use of the negative rather than the positive helicity currents but we do not require the same for this paper and do not present it.
 
 Let us now discuss another aspect of twistor space: Its connection to momentum space through Witten's half Fourier transform.

\subsection{Witten's half Fourier transform}\label{subsec:WittenTransform}
Witten's half Fourier transform is a map from spinor helicity variables to twistor space. For a review of three dimensional spinor helicity variables in this context please refer to section $2$ of \cite{Bala:2025gmz}. The essential feature is that the momentum vector, the polarization spinors are all expressed in terms of $SL(2,\mathbb{R})$ spinors $\lambda$ and $\Bar{\lambda}$ which are assigned negative and positive helicity respectively. The explicit formula are given by,
\begin{align}\label{SHvariables}
    p_{ab}=\lambda_{(a}\Bar{\lambda}_{b)}~~,~~
\zeta_{-}^a=\frac{\lambda^a}{|\frac{\lambda\cdot\Bar{\lambda}}{2}|^{\frac{1}{2}}},\zeta_{+}^a=\frac{\Bar{\lambda}^a}{|\frac{\lambda\cdot\Bar{\lambda}}{2}|^{\frac{1}{2}}}.
\end{align}
Given a symmetric traceless conserved current in momentum space, one can convert it into spinor helicity variables via,
\begin{align}\label{Jspm}
    J_s^{\pm}(\lambda,\Bar{\lambda})=\zeta_{\pm}^{a_1}\cdots \zeta_{\pm}^{a_{2s}}J_{s~a_1\cdots a_{2s}}(p),
\end{align}
and thus $J_s^{\pm}$ denote the two independent components of the current. The remaining mixed helicity components are zero due to conservation.
The operation of a half-Fourier transform to twistor space due to Witten \cite{Witten:2003nn} is as follows:
\begin{wittenbox}
\begin{align}\label{WittenTransform}
   \hat{J}_s^{+}(Z)=\hat{J}_s^{+}(\lambda,\Bar{\mu})=\int \frac{d^2 \Bar{\lambda}}{(2\pi)^2}e^{i\Bar{\lambda}\cdot \Bar{\mu}}\frac{J_s^{+}(\lambda,\Bar{\lambda})}{|p|^{s-1}},
\end{align}
where $p=-\frac{1}{2}\langle \lambda \Bar{\lambda}\rangle$. The rescaling with the momentum magnitude ensures a natural and simple action of special conformal transformations; see \cite{Bala:2025gmz} for more details.
\end{wittenbox}

There is another half-Fourier transform that takes us from spinor helicity variables to dual-twistor space\footnote{\label{footnote:spacelike}Both Witten transforms \eqref{WittenTransform} and \eqref{WittendualTransform} are defined for space-like momenta due to the Lorentzian reality conditions that enforce $\lambda=\lambda^*,\Bar{\lambda}=\Bar{\lambda}^*$ for space-like momenta but $\lambda^*=\Bar{\lambda}$ for time-like momenta, see \cite{Bala:2025gmz}. Therefore, to obtain spinor-helicity results valid for general momenta, one must perform an analytic continuation at the end. Since, in Lorentzian signature $p=-\frac{1}{2}\lambda\cdot \Bar{\lambda}$ can be real or imaginary corresponding to timelike or spacelike momenta respectively. Since we are working with the spacelike momenta, $|p| =\sqrt{p^2}$ is always positive.},
\begin{align}\label{WittendualTransform}
    \hat{J}_s^{+}(W)=\hat{J}_s^{+}(\mu,\Bar{\lambda})=\int \frac{d^2 \lambda}{(2\pi)^2}e^{-i\lambda\cdot\mu}\frac{J_s^{+}(\lambda,\Bar{\lambda})}{|p|^{s-1}},
\end{align}
It is easy to see that the relation between this transform and \eqref{WittenTransform} is as follows:
\begin{align}\label{ZtoWfullFourier}
    \hat{J}_s^{+}(Z)=\int \frac{d^4 W}{(2\pi)^2}e^{i W\cdot Z}\hat{J}_s^{+}(W).
\end{align}
Analogous formulae exist for negative helicity currents and can be found in \cite{Bala:2025gmz}.

The next important thing to check is that the twistor space obtained via a half-Fourier transform \eqref{WittenTransform} and the twistor space discussed earlier in the context of the Penrose transform \eqref{PenroseTransform} are equivalent. As we shall see, the Witten transform \eqref{WittenTransform} when used in tandem with the ordinary Fourier transform can be used to derive the Penrose transform \eqref{PenroseTransform}. This will also make it clear that the spinor helicity currents \eqref{Jspm} are indeed what appears in the Penrose transform \eqref{PenroseTransform}.

\subsection{Derivation of Penrose transform }\label{subsec:PenroseFromWitten}
The Fourier transform for a spin-s symmetric traceless conserved current is given by,
\begin{align}\label{FourierTrans1}
    J_s^{a_1\cdots a_{2s}}(x^\mu)=\int \frac{d^3 p}{(2\pi)^3}e^{-2ip\cdot x} J_s^{a_1\cdots a_{2s}}(p^\mu).
\end{align}
Let us now express the momentum in spinor helicity variables \eqref{SHvariables} and re-write the integral \eqref{FourierTrans1} in spinor helicity variables using\footnote{As remarked in footnote \ref{footnote:spacelike}, we work with space-like momenta for which $\lambda$ and $\Bar{\lambda}$ are real and thus all the components of the integral in \eqref{projvsnonproj1} are over the real line.},
\begin{align}\label{projvsnonproj1}
    \int d^3 p=\frac{1}{4\text{Vol}(GL(1,\mathbb{R}))}\int d^2\lambda d^2\Bar{\lambda}|\lambda\cdot\Bar{\lambda}|.
\end{align}
We derive this formula in appendix \ref{app:ProjectiveVSnonproj}. \eqref{FourierTrans1} becomes,
\begin{align}\label{FouriertoPenrosestep1}
    J_s^{a_1\cdots a_{2s}}(x^\mu)&=\frac{1}{4\text{Vol}(GL(1,\mathbb{R}))}\int \frac{d^2\lambda d^2\Bar{\lambda}}{(2\pi)^3}|\lambda\cdot \Bar{\lambda}|e^{i\Bar{\lambda}_a\lambda_b x^{ab}}J_s^{a_1\cdots a_{2s}}(\lambda,\Bar{\lambda})\notag\\
    &=\frac{1}{4\text{Vol}(GL(1,\mathbb{R}))}\int \frac{d^2\lambda d^2\Bar{\lambda}}{(2\pi)^3}|\lambda\cdot \Bar{\lambda}|e^{i\Bar{\lambda}_a\lambda_b x^{ab}}\epsilon^{a_1b_1}\cdots \epsilon^{a_{2s}b_{2s}}J_{sb_1\cdots b_{2s}}(\lambda,\Bar{\lambda}).
\end{align}
Let us now use the identity,
\begin{align}\label{epsilonabid}
    \epsilon^{ab}=\frac{\lambda^a\Bar{\lambda}^b-\Bar{\lambda}^a\lambda^b}{\lambda\cdot \Bar{\lambda}}.
\end{align}
Substituting this in \eqref{FouriertoPenrosestep1} yields,
\small
\begin{align}\label{FouriertoPenroseStep1next}
    J_s^{a_1\cdots a_{2s}}(x)=\frac{1}{4\text{Vol}(GL(1,\mathbb{R}))}\int \frac{d^2\lambda d^2\Bar{\lambda}}{(2\pi)^3}|\lambda\cdot \Bar{\lambda}|e^{i\Bar{\lambda}_a\lambda_b x^{ab}}\frac{(\lambda^{a_1}\Bar{\lambda}^{b_1}-\Bar{\lambda}^{a_1}\lambda^{b_1})\cdots (\lambda^{a_{2s}}\Bar{\lambda}^{b_{2s}}-\Bar{\lambda}^{a_{2s}}\lambda^{b_{2s}})}{((\lambda\cdot\Bar{\lambda})^2)^s}J_{sb_1\cdots b_{2s}}(\lambda,\Bar{\lambda}).
\end{align}
\normalsize
Using the fact that a conserved current has only two helicity components viz positive and negative helicity \eqref{Jspm}, we can re-write the above as,
\begin{align}\label{FouriertoPenroseinterstep}
    J_s^{a_1\cdots a_{2s}}(x)=\frac{1}{4\text{Vol}(GL(1,\mathbb{R}))}\int \frac{d^2\lambda d^2\Bar{\lambda}}{(2\pi)^3}\frac{|\lambda\cdot \Bar{\lambda}|}{((\lambda\cdot\Bar{\lambda}))^s}&e^{i\Bar{\lambda}_a\lambda_b x^{ab}}\bigg(\lambda^{a_1}\cdots \lambda^{a_{2s}}\Bar{\lambda}^{b_1}\cdots \Bar{\lambda}^{b_{2s}}J_{sb_1\cdots b_{2s}}(\lambda,\Bar{\lambda})\notag\\
    &+(-1)^{2s}\Bar{\lambda}^{a_1}\cdots \Bar{\lambda}^{a_{2s}}\lambda^{b_1}\cdots \lambda^{b_{2s}}J_{sb_1\cdots b_{2s}}(\lambda,\Bar{\lambda})\bigg).
\end{align}
We can now re-label $\lambda\leftrightarrow \Bar{\lambda}$ in the second term which yields\footnote{We also use the fact that $J_{s b_1\cdots b_{2s}}(\lambda,\Bar{\lambda})$ is symmetric under $\lambda\leftrightarrow \Bar{\lambda}$ as it is a function of momentum \eqref{SHvariables}. The first (second) term in \eqref{FouriertoPenroseinterstep} is a positive (negative) helicity component which are the two independent components of the current. However, since the spinors $\lambda,\Bar{\lambda}$ are integrated over, these two components are equal inside a Penrose transform. Also, quite importantly, as is easy to check, the other terms we have thrown away are  proportional to the divergence of the current. If one attempts to construct time-ordered correlators, the divergence of the current leads to contact terms in the form of Ward-Takahashi identities. However, our focus is on Wightman functions which always have zero Ward-Takahashi identity, see \cite{Bala:2025gmz}. Therefore, we are free to set the divergence of the current to zero.},
\begin{align}\label{FouriertoPenrosestep2}
    J_s^{a_1\cdots a_{2s}}(x)&=\frac{1}{2\text{Vol}(GL(1,\mathbb{R}))}\int \frac{d^2\lambda d^2\Bar{\lambda}}{(2\pi)^3}\lambda^{a_1}\cdots \lambda^{a_{2s}}e^{i\Bar{\lambda}_a\lambda_b x^{ab}}\frac{\Bar{\lambda}^{b_1}\cdots \Bar{\lambda}^{b_{2s}}}{|\lambda\cdot \Bar{\lambda}|^{s}}\frac{J_{sb_1\cdots b_{2s}}(\lambda,\Bar{\lambda})}{|\lambda\cdot \Bar{\lambda}|^{s-1}}\notag\\
    &=\frac{1}{2^{2s}\text{Vol}(GL(1,\mathbb{R}))}\int \frac{d^2\lambda d^2\Bar{\lambda}}{(2\pi)^3}\lambda^{a_1}\cdots \lambda^{a_{2s}}e^{i\Bar{\lambda}_a\lambda_b x^{ab}}\hat{J}_s^{+}(\lambda,\Bar{\lambda}),
\end{align}
where we have used the definition of the polarizations \eqref{SHvariables} and identified the rescaled positive helicity current,
\begin{align}
    \hat{J}_s^{+}(\lambda,\Bar{\lambda})=\zeta_{+}^{b_1}\cdots \zeta_{+}^{b_{2s}}\frac{J_{sb_1\cdots b_{2s}}}{|\frac{\lambda\cdot \Bar{\lambda}}{2}|^{s-1}},
\end{align}
which is what features in the integrand of Witten's transform \eqref{WittenTransform}. Note that this quantity satisifes,
\begin{align}\label{rescaledJsprojective}
    \hat{J}_s^{+}(r \lambda,\frac{\Bar{\lambda}}{r})=\frac{1}{r^{2s}}\hat{J}_s^{+}( \lambda,\Bar{\lambda})
\end{align}

Therefore we can use the inverse of the Witten transform \eqref{WittenTransform},
\begin{align}\label{JsinverseWitten1}
    \hat{J}_s^{+}(\lambda,\Bar{\lambda})=\int d^2\Bar{\mu}~e^{-i\Bar{\lambda}\cdot \Bar{\mu}}\hat{J}_s^{+}(\lambda,\Bar{\mu}). 
\end{align}
Using \eqref{rescaledJsprojective}, this leads to the projective property,
\begin{align}\label{Jshatprojective1}
    \hat{J}_s^{+}(r\lambda,r\Bar{\mu})=\frac{1}{r^{2s+2}}\hat{J}_s^{+}(\lambda,\Bar{\mu}).
\end{align}
Using \eqref{JsinverseWitten1} in \eqref{FouriertoPenrosestep2} results in,
\begin{align}\label{FouriertoPenrosestep3}
    J_s^{a_1\cdots a_{2s}}(x)&=\frac{1}{2^{2s}\text{Vol}(GL(1,\mathbb{R})}\int \frac{d^2\lambda d^2\Bar{\mu} d^2\Bar{\lambda}}{(2\pi)^3}\lambda^{a_1}\cdots \lambda^{a_{2s}}e^{-i\Bar{\lambda}_a(\Bar{\mu}^a-\lambda_b x^{ab})}\hat{J}_s^{+}(\lambda,\Bar{\mu})\notag\\
    &=\frac{1}{2^{2s}2\pi\text{Vol}(GL(1,\mathbb{R})}\int d^2\lambda d^2\Bar{\mu}\lambda^{a_1}\cdots \lambda^{a_{2s}}\delta^2(\Bar{\mu}^a-x^{ab}\lambda_b)\hat{J}_s^{+}(\lambda,\Bar{\mu})\notag\\
    &=\frac{1}{2^{2s}2\pi\text{Vol}(GL(1,\mathbb{R})}\int d^2\lambda \lambda^{a_1}\cdots \lambda^{a_{2s}} \hat{J}_s^{+}(\lambda,\Bar{\mu})|_X,
\end{align}
Thus we see that the incidence relation $X$ we discussed earlier \eqref{IncidenceRelation} appears naturally in \eqref{FouriertoPenrosestep2}. To bring this to the form of the Penrose transform \eqref{PenroseTransform}, we require one final projective integral identity. As we show in appendix \ref{app:ProjectiveVSnonproj},
\begin{align}\label{projID1}
    f(r \lambda,r\Bar{\mu})=\frac{1}{r^2}f(\lambda,\Bar{\mu})\implies \frac{1}{\text{Vol}(GL(1,\mathbb{R}))}\int d^2\lambda f(\lambda,\Bar{\mu})=\int \langle \lambda d\lambda\rangle f( \lambda,\Bar{\mu}).
\end{align}
 We see that the integrand in \eqref{FouriertoPenrosestep3} satisifes this property by virtue of \eqref{Jshatprojective1} resulting in,
\begin{align}
    J_s^{a_1\cdots a_{2s}}(x)=\frac{1}{2^{2s}2\pi}\int \langle \lambda d\lambda\rangle \lambda^{a_1}\cdots \lambda^{a_{2s}}\hat{J}_s^{+}(\lambda,\Bar{\mu})|_X,
\end{align}
which is exactly the Penrose transform \eqref{PenroseTransform}! This concludes our proof of the equivalence of the Penrose and Witten transforms.
\subsection{The Conformal Generators}
Having obtained formulae that connect twistor space currents to their position space \eqref{PenroseTransform} and momentum space \eqref{WittenTransform} counterparts with their equivalence proved in subsection \ref{subsec:PenroseFromWitten}, let us now see how the conformal generators act on twistor space currents. The space $\mathbb{RP}^3$ carries a natural action of the conformal group $Sp(4)$ \cite{Baumann:2024ttn,Bala:2025gmz}. Acting on conserved currents we have,
\begin{align}\label{TABZ}
    [T^{AB},\hat{J}_s^{\pm}(Z)]=Z^{(A}\frac{\partial}{\partial Z_{B)}}\hat{J}_s^{\pm}(Z).
\end{align}
In component language, we find \cite{Bala:2025gmz},
\begin{align}\label{TABcomponents}
    T_{AB} \equiv \begin{pmatrix}
    -\bar{\mu}_{(a'} \frac{\partial}{\partial \lambda^{b)}} &  -\bar{\mu}_{a'} \frac{\partial}{\partial \bar{\mu}_{b}} + \lambda^{b'} \frac{\partial}{\partial \lambda^{a}}  \\[1em]
    -\bar{\mu}_{a'} \frac{\partial}{\partial \bar{\mu}_{b}}+ \lambda^{b} \frac{\partial}{\partial \lambda^{a'}} & +\lambda^{(a} \frac{\partial}{\partial\bar{\mu}_{b')}} 
    \end{pmatrix}
    =
    \begin{pmatrix}
    \textit{i} K_{a'b} &  -\textit{i} M^b_{a'} + \frac{2}{\textit{i}}\delta_{a'}^{b} D \\
    -\textit{i} M^b_{a'}+ \frac{2}{\textit{i}}\delta^{b}_{a'} D & -\textit{i} P^{ab'}
    \end{pmatrix},
\end{align}
with,
\begin{align}\label{action1twistora}
    P_{ab}=i\lambda_{(a}\frac{\partial}{\partial\Bar{\mu}^{b)}},&\qquad K_{ab}=i\Bar{\mu}_{(a}\frac{\partial}{\partial \lambda^{b)}},\notag\\
\Tilde{M}_{ab}=i \bigg(\lambda_{(a}\frac{\partial}{\partial\lambda^{b)}}+\Bar{\mu}_{(a}\frac{\partial}{\partial\Bar{\mu}^{b)}}\bigg),&\qquad  D=\frac{i}{2}\bigg(\lambda^a\frac{\partial}{\partial\lambda^a}-\Bar{\mu}^a\frac{\partial}{\partial\Bar{\mu}^{a}}\bigg).
\end{align}
For the dual twistor currents $\hat{J}_s^{\pm}(W)$ we find using \eqref{ZtoWfullFourier} that the generators \eqref{TABZ} translate into,
\begin{align}\label{TABW}
    [T^{AB},\hat{J}_s^{\pm}(W)]=W^{(A}\frac{\partial}{\partial W_{B)}}\hat{J}_s^{\pm}(W),
\end{align}
with component expansions analogous to \eqref{TABcomponents}. Thus the conformal $n-$ point Wightman functions satisfy,
\begin{align}\label{conformalardId}
    \sum_{i=1}^{n}\langle 0| \cdots [T_{AB},\hat{J}^{\pm}_{s_i}]\cdots|0\rangle=0
\end{align}
Another important operator that we must consider is the helicity operator. It reads the helicity of a symmetric traceless conserved current which can be $\pm s$. Explicitly we have,
\begin{align}\label{helicityZandW}
    &h_i\langle 0|\cdots \hat{J}_{s_i}^{\pm}(Z_i)\cdots|0\rangle=-\frac{1}{2}\big(Z_i^A\frac{\partial}{\partial Z_i^A}+2)\big)\langle 0|\cdots \hat{J}_{s_i}^{\pm}(Z_i)\cdots|0\rangle=\pm s_i \langle 0|\cdots \hat{J}_{s_i}^{\pm}(Z_i)\cdots|0\rangle,\notag\\
    &h_i\langle 0|\cdots \hat{J}_{s_i}^{\pm}(W_i)\cdots|0\rangle=\frac{1}{2}\big(W_i^A\frac{\partial}{\partial W_i^A}+2)\big)\langle 0|\cdots \hat{J}_{s_i}^{\pm}(W_i)\cdots|0\rangle=\pm s_i \langle 0|\cdots \hat{J}_{s_i}^{\pm}(W_i)\cdots|0\rangle.
\end{align}
Imposing \eqref{helicityZandW} is equivalent to imposing the projective properties such as \eqref{rescalingformulaJsone}.
Now, with the stage being set, we move towards the construction of conformally invariant objects in twistor space that shall feature in correlators of conserved currents and scalars with $\Delta=1$. More general classes of invariants involving the infinity twistor will be considered in section \ref{sec:InfinityTwistorInvariants}.

\subsection{Conformal invariants in twistor space}\label{Conformal delta3 in twistor space}
Given the conformal generators \eqref{TABZ},\eqref{TABW}, one can construct invariants out of different twistors $Z_i$ and dual twistors $W_j$ that are annihilated by them. In \cite{Baumann:2024ttn,Bala:2025gmz}, the authors construct conformally invariant solutions by taking them to depend only on twistor dot products which belong to the set,
\begin{align}\label{twistordotprods}
    \big\{Z_i\cdot Z_j=-Z_{i}^{A}\Omega_{AB}Z_j^B~,~W_i\cdot W_j=W_{iA}\Omega^{AB}W_{jB}~,~W_i\cdot Z_j=W_{iA}Z_{j}^{A}\big\}.
\end{align}
However, there also exist another class of natural invariants: Projective delta functions.
The object,
\begin{align}\label{firstdefdelta4}
    \delta^4(c_1 Z_1+\cdots+c_{n-1}Z_{n-1}+ Z_n),
\end{align}
is an invariant of Sp$(4)$. To show this, consider an Sp$(4)$ transformation $Z_i\to M Z_i, M\in \text{Sp}(4)$. We have,
\begin{align}
    &\delta^4(c_1 Z_1+\cdots+c_{n-1}Z_{n-1}+Z_n)\to \delta^4(c_1 M Z_1+\cdots+c_{n-1}M Z_{n-1}+M Z_n)\notag\\&=\frac{1}{|\text{Det}(M)|} \delta^4(c_1 Z_1+\cdots+c_{n-1}Z_{n-1}+Z_n)=\delta^4(c_1 Z_1+\cdots+c_{n-1}Z_{n-1}+Z_n),
\end{align}
where we used the fact that symplectic transformations preserve the volume element and hence $\text{Det}(M)=1$.
To erase the arbitrariness in the parameters $c_i$, we integrate them on the support of a function $f(c_1,c_2,\cdots c_{n-1})$,
\begin{align}\label{nptansatz}
    \mathcal{F}(Z_1,\cdots Z_n)=\int dc_1\cdots dc_{n-1}~ f(c_1,\cdots,c_{n-1})\delta^4(c_1 Z_1+\cdots  Z_n),
\end{align}
where the function $f$ can also depend on symplectic dot products of twistors \eqref{twistordotprods}. Since we are working in the real projective space $\mathbb{RP}^3$, we need to ensure that this quantity has good projective properties. More precisely we demand,
\begin{align}\label{nptprojective}
    \mathcal{F}(Z_1,\cdots,r Z_k,\cdots Z_n)=\frac{1}{r^{2\alpha_k+2}}\mathcal{F}(Z_1,\cdots,Z_k,\cdots Z_n),
\end{align}
for some $\alpha_k\in \mathbb{R}$. When constructing Wightman functions of conserved currents this amounts to imposing the helicity counting identity \eqref{helicityZandW} where $\alpha_k$ will be identified with the helicity of the current with argument $Z_k$. 

With three twistors, the unique invariant using the projective delta function that we can form is,
\small
\begin{align}\label{delta4def}
&\delta^3(Z_1,Z_2,Z_3;\alpha_{12},\alpha_{23},\alpha_{31})=(-i)^{-\alpha_{12}-\alpha_{23}-\alpha_{31}}\delta^{[-\alpha_{12}-\alpha_{23}-\alpha_{31}]}(Z_1\cdot Z_2)\int dc_{23}dc_{31}c_{23}^{\alpha_{23}}c_{31}^{\alpha_{31}}\delta^4(c_{23}Z_1+c_{31}Z_2+Z_3)\notag\\&=\int dc_{12}dc_{23}dc_{31}c_{12}^{\alpha_{12}}c_{23}^{\alpha_{23}}c_{31}^{\alpha_{31}}\delta^4(c_{23} Z_1+c_{31} Z_2+c_{12} Z_3)e^{\frac{i}{3}\big(\frac{Z_1\cdot Z_2}{c_{12}}+\frac{Z_2\cdot Z_3}{c_{23}}+\frac{Z_3\cdot Z_1}{c_{31}}\big)}.
\end{align}
\normalsize
The first line is assymmetric with respect to the three twistors and in going to the second line, we restored it by introducing an integral over $c_{12}$ on support of the four dimensional delta function. Note that the final result is very reminscent of the $SL(4,\mathbb{C})$ invariant delta function that occurs in four dimensional scattering amplitudes \cite{Mason:2009sa}.
\eqref{delta4def} defines the three argument $Sp(4)$ invariant projective delta function which is also three dimensional as is appropriate for a delta function in $\mathbb{RP}^3$. The important point to note about \eqref{delta4def} is the projective property,
\begin{align}
    &\delta^3(r_1Z_1,r_2 Z_2,r_3 Z_3;\alpha_{12},\alpha_{23},\alpha_{31})=\frac{1}{r_1^{-(\alpha_{12}+\alpha_{31})+2}r_2^{-(\alpha_{12}+\alpha_{23})+2}r_3^{-(\alpha_{23}+\alpha_{31})+2}}\delta^3(Z_1,Z_2,Z_3;\alpha_{12},\alpha_{23},\alpha_{31}).
\end{align}
For the derivation of \eqref{delta4def}, please see appendix \ref{app:DerivationofDelta4}.
This projective delta function will allow us to construct correlation functions involving operators with definite helicity using \eqref{helicityZandW}.

At the level of four points, there exist infinitely many solutions as it should be since conformal invariance does not suffice to fix the functional form of the correlator. One example of a four point correlator involving a possible $f(c_1,c_2,c_3,c_4)$ is\footnote{We introduced an additional auxiliary integral over $c_4$ compared to \eqref{nptansatz} in order to represent the expression in a more symmetric manner.},
\small
\begin{align}\label{4ptdelta4}
    \frac{\delta^{[\alpha_{12}]}(Z_1\cdot Z_2)\delta^{[\alpha_{34}]}(Z_3\cdot Z_4)}{\text{Vol}(GL(1,\mathbb{R}))}(\prod_{i=1}^{4}\int dc_i) c_1^{\alpha_1-\alpha_{12}} c_2^{\alpha_2-\alpha_{12}} c_3^{\alpha_3-\alpha_{34}}c_4^{\alpha_4-\alpha_{34}}\delta^4(c_1 Z_1+c_2 Z_2+c_3 Z_3+ c_4Z_4),
\end{align}
\normalsize
with the constraint $\alpha_1+\alpha_2+\alpha_3+\alpha_4=2(\alpha_{12}+\alpha_{34})$.
Note that \eqref{4ptdelta4} satisfies the projective property \eqref{nptprojective} for each twistor. Moreover, we see that it is satisfied for any value of the free parameter $\alpha_{12}$ in \eqref{4ptdelta4}. We have also divided by the volume of $GL(1,\mathbb{R})$ (please see \eqref{volumeGL1R} for the definition) since one of the integrals in \eqref{4ptdelta4} just provides an overall infinite factor that is canceled out by $\text{Vol}(GL(1,\mathbb{R}))$. It would be very interesting to study the class of CFT correlators that enjoy more symmetry such as dual conformal and Yangian invariance. In such cases, one might be able to solve for the function $f$ in \eqref{nptansatz} such that it is Yangian invariant. However, we shall not attempt the same in this work. We now turn to the study of two and three point Wightman functions and how they are built using the invariants \eqref{twistordotprods} and \eqref{delta4def}.

\section{Wightman functions of conserved currents and $\Delta=1$ scalars}\label{sec:WightmanConserved3pt}
In this section, we shall present the general forms of parity even two and three point Wightman functions of conserved currents. In contrast to previous works \cite{Baumann:2024ttn, Bala:2025gmz} which employ an ambidextrous approach, we shall only use twistors and not dual-twistors in this work. As we shall see, this will necessitate the introduction of the projective delta function \eqref{delta4def} in addition to the dot product invariants \eqref{twistordotprods}. As a bonus, this description will also allow us to accommodate $\Delta=1$ scalar operators in twistor space.

\subsection{Two point functions}
At the level of two point functions of currents, only the twistor dot products \eqref{twistordotprods} feature in the expressions \cite{Baumann:2024ttn,Bala:2025gmz}. Moreover, it is only non-zero when both currents are identical and have the same helicity \cite{Bala:2025gmz}. We have,
\begin{align}\label{JsJstwopoint}
    \langle 0\hat{J}_s^{\pm}(Z_1)\hat{J}_s^{\pm}(Z_2)|0\rangle=\frac{c_s}{(Z_1\cdot Z_2)^{2(\pm s+1)}}.
\end{align}
This is the unique result.

\subsection{Three point functions}
Moving onto three points, we shall see that the projective delta function \eqref{delta4def} plays an important role in contrast to previous ambidextrous works \cite{Baumann:2024ttn,Bala:2025gmz}. Let us first review the same. At the level of three points with all three operators having non-zero spin, there exist two distinct parity even solutions: \textit{homogeneous} and \textit{non-homogeneous}. More explanation on the reasoning behind the terminology can be found in \cite{Bala:2025gmz}. The expressions in the eight possible helicity configurations take the form,
\small
\begin{align}\label{twistorspace3points}
  &\langle 0| \hat{J}_{s_1}^{h_1}(T_1)\hat{J}_{s_2}^{h_2}(T_2)\hat{J}_{s_3}^{h_3}(T_3)|0\rangle_{h}=c_{s_1s_2s_3}^{(h)}i^{s_1+s_2+s_3}\delta^{[s_1+s_2-s_3]}(T_1\cdot T_2)\delta^{[s_2+s_3-s_1]}(T_2\cdot T_3)\delta^{[s_3+s_1-s_2]}(T_3\cdot T_1),\notag\\
      &\langle 0| \hat{J}_{s_1}^{h_1}(U_1)\hat{J}_{s_2}^{h_2}(U_2)\hat{J}_{s_3}^{h_3}(U_3)|0\rangle_{nh}=i^{-s_1-s_2-s_3}c_{s_1s_2s_3}^{(nh)} \delta^{[-s_1-s_2+s_3]}(U_1\cdot U_2)\delta^{[-s_2-s_3+s_1]}(U_2\cdot U_3)\delta^{[-s_3-s_1+s_2]}(U_3\cdot U_1),
\end{align}
\normalsize
where,
\begin{align}\label{TUnotation}
    &T_i=Z_i~\text{if}~h_i=+s_i ~\text{and}~ T_i=W_i ~\text{if}~ h_i=-s_i,\notag\\
    &U_i=W_i~\text{if}~h_i=+s_i ~\text{and}~ U_i=Z_i ~\text{if}~ h_i=-s_i.
\end{align}
Note in particular that \eqref{twistorspace3points} makes use of both twistors $Z$ and dual twistors $W$ to capture all eight helicity configurations. Again, we emphasize the point that \eqref{twistorspace3points} is purely made up of elements of the set \eqref{twistordotprods}. For example, consider the $(+++)$ helicity configuration. Using \eqref{twistorspace3points}, we see that the homogeneous correlator is a function of twistor variables whereas the non-homogeneous correlator depends on dual-twistor variables. However, what we are going to show now is that by using the projective delta function \eqref{delta4def}, we can bring both solutions on equal footing and represented in just twistor variables. Solving the helicity identities \eqref{helicityZandW} for $(+++)$ helicity for \eqref{delta4def} fixes,
\begin{align}
    \alpha_{ij}=-s_i-s_j+s_k~,i\ne j\ne k,
\end{align}
and yields the result,
\small
\begin{align}\label{pppZZZnh}
    &\langle 0|\hat{J}^{+}_{s_1}(Z_1)\hat{J}^{+}_{s_2}(Z_2)\hat{J}^{+}_{s_3}(Z_3)|0\rangle_{nh}\notag\\&=c_{s_1s_2s_3}^{(nh)}\int dc_{12}dc_{23}dc_{31}c_{12}^{-s_1-s_2+s_3}c_{23}^{-s_2-s_3+s_1}c_{31}^{-s_3-s_1+s_2}\delta^4(c_{23} Z_1+c_{31} Z_2+c_{12} Z_3)e^{\frac{i}{3}\big(c_{12}Z_1\cdot Z_2+c_{23}Z_2\cdot Z_3+c_{31}Z_3\cdot Z_1\big)}.
\end{align}
\normalsize
The reason for the subscript $nh$ is because this solution is the Fourier transform of the non-homogeneous $(WWW),(+++)$ helicity solution \eqref{twistorspace3points} as we show in appendix \ref{app:TwistortoDual}:
\small
\begin{align}
    &\langle 0|\hat{J}^{+}_{s_1}(Z_1)\hat{J}^{+}_{s_2}(Z_2)\hat{J}^{+}_{s_3}(Z_3)|0\rangle_{nh}=\int d^4 W_1 d^4 W_2 d^4 W_3 e^{iZ_1\cdot W_1+iZ_2\cdot W_2+i Z_3\cdot W_3}\langle 0|\hat{J}^{+}_{s_1}(W_1)\hat{J}^{+}_{s_2}(W_2)\hat{J}^{+}_{s_3}(W_3)|0\rangle_{nh}\notag\\
    &\propto\int d^4 W_1 d^4 W_2 d^4 W_3 e^{iZ_1\cdot W_1+iZ_2\cdot W_2+i Z_3\cdot W_3}\delta^{[-s_1-s_2+s_3]}(W_1\cdot W_2)\delta^{[-s_2-s_3+s_1]}(W_2\cdot W_1)\delta^{[-s_3-s_1+s_2]}(W_3\cdot W_1).
\end{align}
\normalsize
Both the delta function product dual twistor space result \eqref{twistorspace3points} and the projective delta function twistor space result \eqref{pppZZZnh} go over to the same spinor helicity variables result after Witten's half-Fourier transform \eqref{WittenTransform} as can be understood from the below diagram as well as the discussion in subsection \ref{subsec:WittenTransform}: 
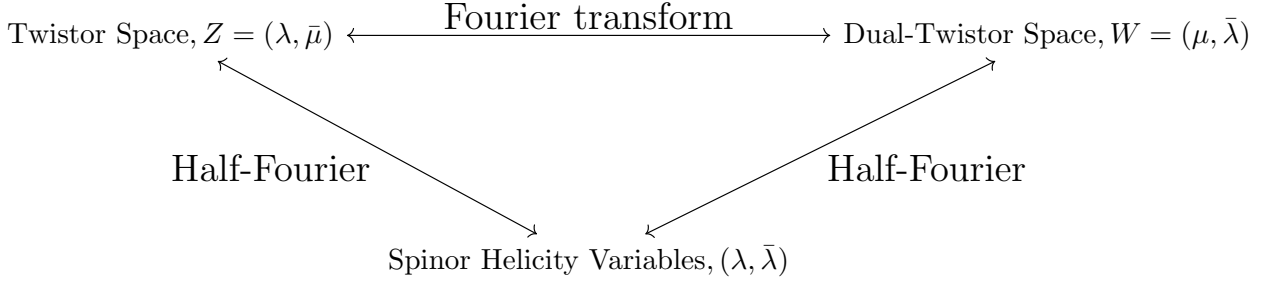
\begin{figure}[h]
    \centering
    \[
    \begin{tikzcd}[column sep=1 em, row sep=6em]
        \text{Twistor Space},  Z = (\lambda, \Bar{\mu}) 
        \arrow[dr, swap, "\text{\Large{Half-Fourier}}", leftrightarrow] 
        \arrow[rr, "\text{\Large{Fourier transform}}", leftrightarrow] 
        & & 
        \text{Dual-Twistor Space},  W = (\mu, \Bar{\lambda}) 
        \arrow[dl, "\text{\Large{Half-Fourier}}", leftrightarrow] \\
        & 
        \text{Spinor Helicity Variables}, (\lambda, \Bar{\lambda}) 
        &
    \end{tikzcd}
    \]
    \caption{The trio of Twistors, Dual-Twistors and Spinor-Helicity variables}
    \label{fig:twistor_diagram}
\end{figure}

Thus, we can represent the most general parity even solution for the $(+++)$ helicity correlator purely in the $(ZZZ)$ variables as a sum of the homogeneous solution in \eqref{twistorspace3points} (which is already given in $Z$ variables) and the non-homogeneous solution \eqref{pppZZZnh} which we can now represent in $Z$ variables thanks to the projective delta function \eqref{delta4def}.

\begin{align}
    \langle 0|\hat{J}^{+}_{s_1}(Z_1)\hat{J}^{+}_{s_2}(Z_2)\hat{J}^{+}_{s_3}(Z_3)|0\rangle&=c_{s_1s_2s_3}^{(h)}\langle 0|\hat{J}^{+}_{s_1}(Z_1)\hat{J}^{+}_{s_2}(Z_2)\hat{J}^{+}_{s_3}(Z_3)|0\rangle_{h}+c_{s_1s_2s_3}^{(nh)}\langle 0|\hat{J}^{+}_{s_1}(Z_1)\hat{J}^{+}_{s_2}(Z_2)\hat{J}^{+}_{s_3}(Z_3)|0\rangle_{nh}.
\end{align}
Putting in the explicit expressions for these quantities we get,
\begin{paritybox}
\small
\begin{align}\label{pppZZZallsols}
    &\langle 0|\hat{J}^{+}_{s_1}(Z_1)\hat{J}^{+}_{s_2}(Z_2)\hat{J}^{+}_{s_3}(Z_3)|0\rangle \notag\\
    & = c_{s_1s_2s_3}^{(h)} i^{s_1 + s_2 + s_3} \delta^{[\alpha_{12}]}(Z_1 \cdot Z_2) \delta^{[\alpha_{23}]}(Z_2 \cdot Z_3) \delta^{[\alpha_{31}]}(Z_3 \cdot Z_1) \notag\\
    & \quad + c_{s_1s_2s_3}^{(nh)} \int dc_{12} \, dc_{23} \, dc_{31} \, c_{12}^{-\alpha_{12}} c_{23}^{-\alpha_{23}} c_{31}^{-\alpha_{31}} \delta^4(c_{23} Z_1 + c_{31} Z_2 + c_{12} Z_3) \, e^{\frac{iZ_1\cdot Z_2}{c_{12}}},
\end{align}
where $\alpha_{ij}=s_i+s_j-s_k,i\ne j\ne k$.
\normalsize
\end{paritybox}
Let us once again emphasize that this expression is purely in terms of twistor variables without using dual twistors in contrast to the earlier results \eqref{twistorspace3points}.
The remaining helicity expressions purely in the $Z$ twistor variables without employing dual twistors $W$ and vice versa can be found similarly. Their explicit forms are provided in appendix \ref{app:TwistorSpaceResults}. Before moving on to correlators involving $\Delta=1$ scalars, let us comment about the CPT properties of these solutions. For a review of CPT in twistor space please see \cite{Bala:2025gmz}.
\subsection{CPT properties}
As we have discussed, one can choose to work purely in twistor space with $CPT$ taking us to dual twistor-space as shown in appendix \ref{app:TwistorSpaceResults}. On the other hand the earlier results \eqref{twistorspace3points} use a representation of correlators that employ both twistors and dual twistors and $CPT$ invariance keeps us within that class. Thus, depending on the application and convenience, one may choose to work with just twistor variables, just dual twistor variables or a mix of both. For Wightman functions involving scalars, we shall shortly see that both classes of distributional solutions in twistor space such as in \eqref{pppZZZallsols} are necessary. The relation between Wightman functions with different choices of variables is encapsulated in figure \ref{fig:cpt_fourier}.

\begin{figure}[h]
    \centering
    \[
    \begin{tikzcd}[row sep=large, column sep=16em] 
    \langle 0 | \hat{J}_{s_1}^{+}(W_1) \hat{J}_{s_2}^{+}(W_2) \hat{J}_{s_3}^{+}(W_3) | 0 \rangle
        \arrow[r, "\text{\large Fourier transform}" above, leftrightarrow]
        \arrow[d, "CPT" left, leftrightarrow]
    & \langle 0 | \hat{J}_{s_1}^{+}(Z_1) \hat{J}_{s_2}^{+}(Z_2) \hat{J}_{s_3}^{+}(Z_3) | 0 \rangle
        \arrow[d, "CPT" right, leftrightarrow] \\
    \langle 0 | \hat{J}_{s_1}^{-}(Z_1) \hat{J}_{s_2}^{-}(Z_2) \hat{J}_{s_3}^{-}(Z_3) | 0 \rangle
        \arrow[r, "\text{\large Fourier transform}" above, leftrightarrow]
    & \langle 0 | \hat{J}_{s_1}^{-}(W_1) \hat{J}_{s_2}^{-}(W_2) \hat{J}_{s_3}^{-}(W_3) | 0 \rangle
    \end{tikzcd}
    \]
    \caption{CPT and the Twistor Fourier transform. CPT exchanges positive and negative helicities and $W$ and $Z$ while keeping the functional form fixed. The Fourier transform does not change the helicity but converts a function of $W$ into a different function of $Z$.}
    \label{fig:cpt_fourier}
\end{figure}
For example, we start with the $(+++)$ helicity non-homogeneous Wightman function \eqref{twistorspace3points} in the top left of the figure. A CPT transformation takes it to the $(---)$ non-homogeneous correlator in the $Z$ variables \eqref{twistorspace3points} which is in the bottom left of the figure. A twistor Fourier transform \eqref{ZtoWfullFourier} on the other hand takes us from the top left to the $\delta^3$ type solution \eqref{pppZZZnh} which is in the top right of figure \ref{fig:cpt_fourier}. The CPT transform of the latter takes us to non-homogeneous $(---)$ helicity $\delta^3$ solution (that is provided in appendix \ref{app:TwistorSpaceResults}) which is in the bottom right. This result is related to the $(---)$ Z representation non-homogeneous correlator \eqref{twistorspace3points} by a Fourier transform, thus completing the rectangle.

Having discussed in detail parity even Wightman functions of conserved currents, we shall now extend the results to include cases involving $\Delta=1$ scalar operators.

\subsection{Extension to $\Delta=1$ Scalars}
Let us reconsider the Penrose transform \eqref{PenroseTransform}. Setting $s=0$ it reads,
\begin{align}\label{O1PenroseTrans}
    O_1(x)=\int \langle \lambda d\lambda\rangle \hat{O}_1(Z)|_X.
\end{align}
The reason for the subscript $1$ is due to the fact that this procedure results in a $\Delta=1$ scalar operator. Given the representation of the conformal algebra \eqref{TABZ}, it is clear that the twistor space currents (which for us now includes the $s=0$ scalar case) are dimensionless. Therefore, in \eqref{O1PenroseTrans}, we see by dimensional analysis that the integrand has dimension $1$ due to the $\mathbb{RP}^1$ measure and thus it outputs a scalar operator with $\Delta=1$. Let us now analyze two and three point twistor space Wightman functions involving these operators.

\subsubsection{Two point functions}
The two point result is simply obtained by setting $s=0$ in \eqref{JsJstwopoint} thus yielding,
\begin{align}\label{O1O1twopoint}
    \langle 0|\hat{O}_1(Z_1)\hat{O}_1(Z_2)|0\rangle=\frac{c_0}{(Z_1\cdot Z_2)^2}.
\end{align}
It is easy to check that \eqref{O1O1twopoint} goes to the correct position space two point function as we show in appendix \ref{app:ExplicitPenrose}.

\subsubsection{Three point functions}
Moving on to three points, let us consider the case with two currents and one scalar first. 
\subsection*{$\underline{\langle 0|\hat{J}_{s_1}\hat{J}_{s_2}\hat{O}_1|0\rangle}$}
Let us write down the most general ansatz involving the two classes of solutions such as those in \eqref{pppZZZallsols}.
\begin{align}
    \langle 0|\hat{J}_{s_1}(Z_1)\hat{J}_{s_2}(Z_2)\hat{O}_1(Z_3)|0\rangle&=\alpha~i^{s_1+s_2}\delta^{[\alpha_{12}]}(Z_1\cdot Z_2)\delta^{[\alpha_{23}]}(Z_2\cdot Z_3)\delta^{[\alpha_{31}]}(Z_3\cdot Z_1)\notag\\
    &+\beta \delta^{3}(Z_1,Z_2,Z_3;\beta_{12},\beta_{23},\beta_{31}).
\end{align}
The helicity identity \eqref{helicityZandW} with $s=0$ constrains the coefficients to take the following values:
\begin{align}
&\alpha_{12}=s_1+s_2,\alpha_{23}=s_2-s_1,\alpha_{31}=s_1-s_2,\notag\\
    &\beta_{12}=-s_1-s_2,\beta_{23}=-s_2+s_1,\beta_{31}=-s_2+s_1.
\end{align}
The result is thus,
\begin{align}\label{Js1Js2O1ansatz}
    \langle 0|\hat{J}_{s_1}(Z_1)\hat{J}_{s_2}(Z_2)\hat{O}_1(Z_3)|0\rangle&=\alpha~i^{s_1+s_2}\delta^{[s_1+s_2]}(Z_1\cdot Z_2)\delta^{[s_2-s_1]}(Z_2\cdot Z_3)\delta^{[s_1-s_2]}(Z_3\cdot Z_1)\notag\\
    &+\beta \delta^{3}(Z_1,Z_2,Z_3;-s_1-s_2,-s_2+s_1,-s_2+s_1).
\end{align}
The coefficients $\alpha$ and $\beta$ are independent at this point and not fixed by conformal invariance and the helicity counting identity. However, after converting \eqref{Js1Js2O1ansatz} to spinor helicity variables, we see that the value of $\beta=\alpha$ reproduces the correct Wightman functions as we show with a characteristic example in appendix \ref{app:FixingWightman}. Therefore, the functional form of a three point function involving two spinning operators is given by,
\begin{simpleboxenv}[Parity Even Current Three-Point Functions in Twistor Space]
\begin{align}\label{Js1Js2O1result}
    \langle 0|\hat{J}_{s_1}(Z_1)\hat{J}_{s_2}(Z_2)\hat{O}_1(Z_3)|0\rangle &\propto \big(~i^{s_1+s_2}\delta^{[s_1+s_2]}(Z_1\cdot Z_2)\delta^{[s_2-s_1]}(Z_2\cdot Z_3)\delta^{[s_1-s_2]}(Z_3\cdot Z_1) \notag \\
    &\quad + \delta^{3}(Z_1,Z_2,Z_3;-s_1-s_2,-s_2+s_1,-s_2+s_1)\big).
\end{align}
\end{simpleboxenv}
At this point, it is important to note that the authors of \cite{Baumann:2024ttn} did not consider the $\delta^3$ type solution in \eqref{Js1Js2O1result} and considered only the first term. However, they showed that just that term suffices to reproduce the correct position space result. The resolution to this apparent paradox is that both terms in \eqref{Js1Js2O1result} are required to reproduce the correct spinor helicity expression for the Wightman function as we have taken. However, as we detail in appendix \ref{app:FixingWightman}, both solutions in \eqref{Js1Js2O1result} are equal inside the Penrose transform \eqref{PenroseTransform} and therefore lead to the same position space result.

The following diagram encapsulates this interplay between these transformations and also their corresponding Wick rotations to Euclidean signature.

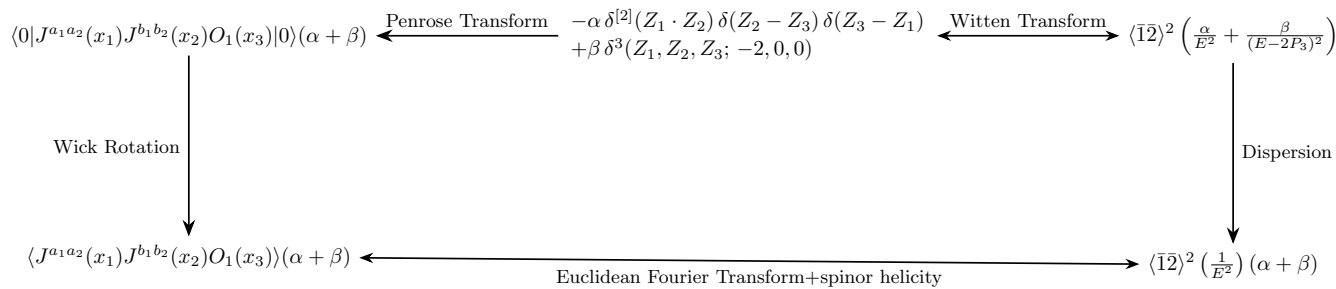
\begin{figure}[htbp]
\centering
\scalebox{0.80}{ 
\begin{tikzpicture}[
    node distance=2.5cm and 3cm,
    every node/.style={align=center, font=\small},
    arrow/.style={thick, -{Stealth[scale=1.2]}},
    doublearrow/.style={thick, <->, >=Stealth},
    font=\itshape
]

\node (A) at (0, 0) {$\langle 0|J^{a_1a_2}(x_1)J^{b_1b_2}(x_2)O_1(x_3)|0\rangle(\alpha + \beta)$};

\node (B) [right=of A] {%
$\begin{array}{l}
  -\alpha\, \delta^{[2]}(Z_1 \cdot Z_2)\, \delta(Z_2 - Z_3)\, \delta(Z_3 - Z_1) \\
  +\beta\, \delta^3(Z_1, Z_2, Z_3;\,-2,0,0)
\end{array}$};

\node (C) [right=of B] {$\langle \Bar{1}\Bar{2} \rangle^{2} \left( \frac{\alpha}{E^2} + \frac{\beta}{(E - 2P_3)^2} \right)$};

\node (A2) [below=3cm of A] {$\langle J^{a_1a_2}(x_1)J^{b_1b_2}(x_2)O_1(x_3)\rangle(\alpha + \beta)$};

\node (C2) [below=3cm of C] {$\langle \Bar{1}\Bar{2} \rangle^{2} \left( \frac{1}{E^2} \right)(\alpha+\beta)$};

\draw[arrow] (B) -- (A) node[midway, above] {\footnotesize Penrose Transform};
\draw[doublearrow] (B) -- (C) node[midway, above] {\footnotesize Witten Transform};

\draw[arrow] (A) -- (A2) node[midway, left] {\footnotesize Wick Rotation};
\draw[arrow] (C) -- (C2) node[midway, right] {\footnotesize Dispersion};

\draw[doublearrow] (A2) -- (C2) node[midway, below] {\footnotesize Euclidean Fourier Transform+spinor helicity};

\end{tikzpicture}
}
\caption{The interplay between the Penrose transform and Witten transform}
\label{fig:twistor-transform-flow}
\end{figure}
Starting from the linear combination of the two solutions in twistor space one can perform a Penrose transform to obtain the position space Wightman function\footnote{The explicit expressions for the correlator $\langle JJO_1\rangle$ can be found for instance in \cite{Giombi:2011kc}. Also recall that the Wightman function and Euclidean correlator in position space differ only by an $i\epsilon$ prescription.} (top-left in figure \ref{fig:twistor-transform-flow}) which is insensitive to the twistor representative used. However, Witten's half-Fourier distinguishes these solutions and only for $\alpha=\beta$ does it produce the correct Wightman function viz \eqref{JJO1correctWightman}. However, analytically continuing to Euclidean momentum space using the methods of \cite{Baumann:2024ttn,Bala:2025gmz} yields a result that is yet again insensitive to the two solutions. This is also consistent with the Fourier transform of the Wick rotated postition space Euclidean correlator (bottom left of figure \ref{fig:twistor-transform-flow}).
\begin{simpleboxenv}[Two scalars and one conserved current]
A similar analysis can be performed for Wightman functions involving two $\Delta=1$ scalar operators and one conserved current. The result is given by,
\begin{align}\label{JsO1O1ansatz}
    \langle 0|\hat{J}_{s}(Z_1)\hat{O}_1(Z_2)\hat{O}_1(Z_3)|0\rangle&\propto \big(~i^{s}\delta^{[s]}(Z_1\cdot Z_2)\delta^{[-s]}(Z_2\cdot Z_3)\delta^{[s]}(Z_3\cdot Z_1)\notag\\&+\delta^{3}(Z_1,Z_2,Z_3;-s,s,s)\big).
\end{align}
Yet again, we verified this formula by converting to spinor helicity variables and checking that it reproduces the correct Wightman function.

Finally, the case where all three operators are $\Delta=1$ is given by,
\begin{align}\label{O1O1O1Wightman1}
    &\langle 0|\hat{O}_1(Z_1)\hat{O}_1(Z_2)\hat{O}_1(Z_3)|0\rangle\propto \big(~\delta(Z_1\cdot Z_2)\delta(Z_2\cdot Z_3)\delta(Z_3\cdot Z_1)+ \delta^{3}(Z_1,Z_2,Z_3;0,0,0)\big).
\end{align}
Both of these after a half-Fourier transform give rise to the same result and thus we may choose to work with either of these solutions.
\end{simpleboxenv}

This concludes our discussion about parity even Wightman functions involving conserved currents and also $\Delta=1$ scalars. In the next section, we shall enlarge our space of invariants to include the infinity twistor \eqref{infinitytwistor} and as we shall see, this is required for correlators involving scalars with arbitrary scaling dimension, parity odd Wightman functions as well as general non-conserved spinning operators.

\section{Conformal invariants involving the Infinity Twistor}\label{sec:InfinityTwistorInvariants}
In section \ref{sec:Geometry}, we introduced the infinity twistor as the bi-twistor that breaks the natural conformal invariance of twistor space down to its Poincare' subgroup that preserves the Minkowski metric. So far, in our discussion of parity even Wightman functions of conserved currents and $\Delta=1$ scalars, the infinity twistor did not appear. However, we shall see in this section that whenever there are scalars with $\Delta\ne 1$ present in the correlator or for parity odd Wightman functions, the infinity twistor is a required ingredient. Although it breaks conformal invariance by itself, we shall see that it occurs in correlators in such a way that the whole expression is rendered conformally invariant. The first part of this section focuses on general scalars whereas the second part deals with parity odd Wightman functions. In the third part of this section, we derive the Penrose transform for generic primary operators with arbitrary scaling dimension and spin. We also briefly discuss some preliminary results regarding their correlation functions in twistor space.

\subsection{Wightman functions of general scalar operators}\label{subsec:genscalar}
As we saw in the previous section, $\Delta=1$ scalars are naturally described in twistor space. Before discussing the construction of correlators of generic scalar operators, we start with the $\Delta=2$ case. This case is special since a $\Delta=2$ scalar is related to a $\Delta=1$ scalar via a \textit{Legendre} transform, which is a conformally covariant transformation that preserves the eigenvalues of the quadratic and quartic conformal Casimirs.
\subsubsection{The Legendre transform}
Given a scalar operator $O_{\Delta}$ in three dimensions, its Legendre or shadow transform is defined as \cite{Simmons-Duffin:2012juh},
\begin{align}
    O_{3-\Delta}(x)=\int\frac{d^3 y}{|x-y|^{2(3-\Delta)}}O_{\Delta}(y).
\end{align}
Taking $\Delta=1$ results in,
\begin{align}
    O_2(x)=\int\frac{d^3 y}{|x-y|^{4}}O_{1}(y).
\end{align}
Using the Penrose transform \eqref{O1PenroseTrans} for $\Delta=1$ scalars, let us now implement this transformation in twistor space. We have,
\begin{align}
    O_2(x)=\int \frac{d^3 y}{|x-y|^{4}}\int \langle \lambda d\lambda\rangle d^2\Bar{\mu}~\delta^2(\Bar{\mu}^a-y^{ab}\lambda_b)\hat{O}_1(\lambda,\Bar{\mu}),
\end{align}
where we have implemented the incidence relations \eqref{IncidenceRelation} using a delta function. Exponentiating the two dimensional delta function and switching the order of integrals results in,
\begin{align}
    &O_2(x)=\int\langle\lambda d\lambda\rangle d^2\Bar{\mu} \int \frac{d^2\rho}{(2\pi)^2} \int \frac{d^3 y}{|x-y|^{4}}e^{-i\rho_a(\Bar{\mu}^a-y^{ab}\lambda_b)}\hat{O}_1(\lambda,\Bar{\mu})\notag\\
    &=\int \langle\lambda d\lambda\rangle d^2\Bar{\mu}\int\frac{d^2\rho}{(2\pi)^2}e^{-i\rho_a(\Bar{\mu}^a-x^{ab}\lambda_b)}\int d^3 y'\frac{e^{i\rho_a y'^{ab}\lambda_b}}{|y'|^{4}}\hat{O}_1(\lambda,\Bar{\mu})\notag\\
    &=\int \langle\lambda d\lambda\rangle d^2\Bar{\mu}\int\frac{d^2\rho}{(2\pi)^2}e^{-i\rho_a(\Bar{\mu}^a-x^{ab}\lambda_b)}\frac{|\lambda\cdot \rho|}{2}\hat{O}_1(\lambda,\Bar{\mu})\notag\\
    &=-\int \langle\lambda d\lambda\rangle d^2\Bar{\mu}\int \frac{dc}{4\pi~c^2}\int \frac{d^2\rho}{(2\pi)^2}e^{-i\rho_a(\Bar{\mu}^a-x^{ab}\lambda_b-c\lambda^a)}\hat{O}_1(\lambda,\Bar{\mu})\notag\\
    &=\int \langle\lambda d\lambda\rangle d^2\Bar{\mu}\delta^2(\Bar{\mu}^a-x^{ab}\lambda_b)\int \frac{-dc}{4\pi c^2}\hat{O}_1(\lambda,\Bar{\mu}+c\lambda).
\end{align}
\begin{Legendrebox}
Therefore, up to a normalization factor, we find that the twistor space Legendre transform is given by,
\begin{align}\label{TwistorLegendre}
    \hat{O}_2(\lambda,\Bar{\mu})=\hat{O}_2(Z)=\int_{-\infty}^{\infty}\frac{dc}{c^2}\hat{O}_1(\lambda,\Bar{\mu}+c\lambda),
\end{align}
with the Penrose transform for $O_2$ given by,
\begin{align}\label{O2PenroseTrans}
    O_2(x)=\int \langle\lambda d\lambda\rangle \hat{O}_2(\lambda,\Bar{\mu})|_X.
\end{align}
\end{Legendrebox}

Note the important difference between this Penrose transform and the one for $\Delta=1$ scalars and conserved currents \eqref{PenroseTransform}. There, the twistor space operators were dimensionless as can be seen by dimensional analysis or the action of the dilatation generator in \eqref{TABcomponents}. On the other hand, we see that in \eqref{O2PenroseTrans} that the twistor space current $O_2(\lambda,\Bar{\mu})$ has dimension $1$ and therefore carries a representation of the conformal algebra different than \eqref{TABcomponents}. We will discuss this more when we generalize to arbitrary scaling dimensions. Now, we shall discuss two and three point correlators involving the $O_2$ operator obtained via the Legendre transform.

\subsubsection{A contact term and the two point function}
Consider the two point function of the $O_2$ operator with an $O_1$ operator. Generically, the two point function of an operator and its shadow is non-zero and is in fact equal to the $d-$dimensional spacetime delta function in position space \cite{Nakayama:2019mpz}. Our aim now is to reproduce this in twistor space using the twistor space Legendre transform \eqref{TwistorLegendre}. The $\langle 0|\hat{O}_1(Z_1)\hat{O}_1(Z_2)|0\rangle$ two point function is given by \eqref{O1O1twopoint}. Performing a Legendre transform of the first operator results in,
\begin{align}\label{O2O1Twistor}
    &\langle 0|\hat{O}_2(Z_1)\hat{O}_1(Z_2)|0\rangle=\int\frac{dc}{c^2}\frac{1}{(\lambda_1\cdot \Bar{\mu}_2-\lambda_2\cdot \Bar{\mu}_1+c \lambda_1\cdot \lambda_2)^2}\notag\\
    &=\int \frac{dc}{c^2}\int dc_{12} |c_{12}|e^{-ic_{12}(\lambda_1\cdot \Bar{\mu}_2-\lambda_2\cdot \Bar{\mu}_1+c\lambda_1\cdot \lambda_2)}=\int dc_{12} |c_{12}|e^{-ic_{12}Z_1\cdot Z_2}\int \frac{dc}{c^2}e^{-ic_{12} c\langle 1 2\rangle}\notag\\&
    = \int dc_{12}|c_{12}||c_{12}\langle 12\rangle|e^{-ic_{12}Z_1\cdot Z_2}=|\langle 1 2\rangle|\int dc_{12}c_{12}^2 e^{-ic_{12}Z_1\cdot Z_2}=|\langle 1 2\rangle| \delta^{[2]}(Z_1\cdot Z_2).
\end{align}
Recasting the result in terms of the infinity twistor \eqref{infinitytwistor} and using the fact that $\langle 12\rangle=\langle Z_1 I Z_2\rangle=Z_1^A I_{AB} Z_2^B$ with $I_{AB}$ the infinity twistor \eqref{infinitytwistor} and $Z^A$ given by \eqref{defoftwistorZA} yields,
\begin{simpleboxenv}[Two point conformal contact term]
\begin{align}\label{O2O1Twistor}
    \langle 0|\hat{O}_2(Z_1)\hat{O}_1(Z_2)|0\rangle=|\langle Z_1 I Z_2\rangle|\delta^{[2]}(Z_1\cdot Z_2)
\end{align}
\end{simpleboxenv}
Notice that the infinity twistor has entered the fray for this correlator. Note also that (derivatives) of delta functions enforcing $Z_1\cdot Z_2=0$ were discarded in earlier analysis for correlators of conserved currents \cite{Baumann:2024ttn,Bala:2025gmz}. However, we see here that it naturally occurs in this two point contact term. One can also verify as we do in appendix \ref{app:ExplicitPenrose} that upon a Penrose transform, \eqref{O2O1Twistor} results in the correct position space contact term. 

Let us now consider the two point function of the $O_2$ operator. Performing the Legendre transform \eqref{TwistorLegendre} for both operators on the $\langle 0|\hat{O}_1(Z_1)\hat{O}_1(Z_2)|0\rangle$ correlator \eqref{O1O1twopoint} results in,
\begin{simpleboxenv}[Two point function of $\Delta=2$ scalar]
\begin{align}\label{O2O2Twistor}
    \langle 0|\hat{O}_2(Z_1)\hat{O}_2(Z_2)|0\rangle=\frac{\langle Z_1 I Z_2\rangle^2}{(Z_1\cdot Z_2)^4}.
\end{align}
\end{simpleboxenv}
Yet again, the infinity twistor is present in the result \eqref{O2O2Twistor}.

\subsubsection{Three point functions}
The Legendre transform formula \eqref{TwistorLegendre} makes it very easy to obtain three point functions with an arbitrary number of $O_2$ insertions given the corresponding correlator with $O_1$. For example, the three point function of this operator using \eqref{O1O1O1Wightman1} is given by,
\begin{simpleboxenv}[Three point function of $\Delta=2$ scalar]
\small
\begin{align}\label{O2O2O2twistor}
    &\langle 0|\hat{O}_2(Z_1)\hat{O}_2(Z_2)\hat{O}_2(Z_3)|0\rangle=\notag\\&\prod_{i=1}^{3}(\int\frac{dk_i}{k_i^2})\delta(Z_1\cdot Z_2+(k_1+k_2)Z_1 I Z_2)\delta(Z_2\cdot Z_3+(k_2+k_3)Z_2 IZ_3)\delta(Z_3\cdot Z_1+(k_3+k_1)Z_3 I Z_1).
\end{align}
\normalsize
\end{simpleboxenv}
This result can also be written as,
\begin{align}
    \frac{|\langle Z_1IZ_2\rangle\langle Z_2 I Z_3\rangle\langle Z_3 I Z_1\rangle|}{\bigg(\frac{Z_1\cdot Z_2}{\langle Z_1 I Z_2\rangle}-\frac{Z_2\cdot Z_3}{\langle Z_2 I Z_3\rangle}-\frac{Z_3\cdot Z_1}{\langle Z_3 I Z_1\rangle}\bigg)^2\bigg(-\frac{Z_1\cdot Z_2}{\langle Z_1 I Z_2\rangle}+\frac{Z_2\cdot Z_3}{\langle Z_2 I Z_3\rangle}-\frac{Z_3\cdot Z_1}{\langle Z_3 I Z_1\rangle}\bigg)^2\bigg(-\frac{Z_1\cdot Z_2}{\langle Z_1 I Z_2\rangle}-\frac{Z_2\cdot Z_3}{\langle Z_2 I Z_3\rangle}+\frac{Z_3\cdot Z_1}{\langle Z_3 I Z_1\rangle}\bigg)^2}.
\end{align}
Similarly, one can obtain the expressions for cases where the $O_2$ operator appears once or twice with conserved currents. We note that the below answers are valid only inside a Penrose transform as we ignore the second contribution in equations \eqref{Js1Js2O1result} and \eqref{JsO1O1ansatz} although its easy enough to include them if required.
\begin{simpleboxenv}[Correlators with two or one $\Delta=2$ scalars] 
For two $\Delta=2$ scalars and one conserved current we have,
\small
\begin{align}\label{JsO2O2twistor}
    &\langle 0|\hat{J}_s^+(Z_1)\hat{O}_2(Z_2)\hat{O}_2(Z_3)|0\rangle\notag\\&=\int\frac{dk_1~dk_2}{(k_1k_2)^2}\delta^{[s]}(Z_1\cdot Z_2+k_1 Z_1 I Z_2)\delta^{[-s]}(Z_2\cdot Z_3+(k_1+k_2)Z_2 I Z_3)\delta^{[s]}(Z_3\cdot Z_1+k_2 Z_3 I Z_1),
\end{align}
\normalsize
whereas the one $\Delta=2$ insertion yields,
\begin{align}\label{Js1Js2O2twistor}
    &\langle 0|\hat{J}_{s_1}^{+}(Z_1)\hat{J}_{s_2}^{+}(Z_2)\hat{O}_2(Z_3)|0\rangle\notag\\&=\int\frac{dk}{k^2}\delta^{[s_1+s_2]}(Z_1\cdot Z_2)\delta^{[s_2-s_1]}(Z_2\cdot Z_3+k Z_2 I Z_3)\delta^{[s_1-s_2]}(Z_3\cdot Z_1+k Z_3 I Z_1).
\end{align}
\end{simpleboxenv}
Let us now move on to the generalization of the Penrose transform to scalars with arbitrary scaling dimension.

\subsubsection{The Penrose transform for arbitrary $\Delta$}

Consider a scalar operator $O_{\Delta}(x)$. Our aim is to derive a Penrose transform for this operator. We look for an expression of the form,
\begin{tcolorbox}[
    colback=blue!5!white,
    colframe=blue!60!black,
    title=Penrose transform for arbitrary scalar operators,
    sharp corners=south,
    boxrule=0.8pt,
    arc=4pt,
    fonttitle=\bfseries
]
\begin{align}\label{ODeltaPenrose}
    O_{\Delta}(x)=\int \langle\lambda d\lambda\rangle~\hat{O}_{\Delta}(Z)|_X,
\end{align}
where $X$ denotes the usual incidence relations \eqref{IncidenceRelation}.
\end{tcolorbox}
By dimensional analysis, we see that $\hat{O}_\Delta(Z)$ must have scaling dimension $\Delta-1$. The projectiveness of the integrand \eqref{ODeltaPenrose} also implies that $\hat{O}_{\Delta}(Z)$ has homogeneity $-2$. Given the fact that twistor space is connected to spinor helicity variables by a Witten transform \eqref{WittenTransform}, let us obtain $\hat{O}_{\Delta}(Z)$ via the same. Consider,
\begin{tcolorbox}[
    colback=blue!5!white,
    colframe=green!60!black,
    title=Witten transform for arbitrary scalar operators,
    sharp corners=south,
    boxrule=0.8pt,
    arc=4pt,
    fonttitle=\bfseries
]
\begin{align}\label{ODeltaWitten}
    \hat{O}_{\Delta}(Z)=\hat{O}_{\Delta}(\lambda,\Bar{\mu})=\int \frac{d^2\Bar{\lambda}}{(2\pi)^2}e^{i\Bar{\lambda}\cdot\Bar{\mu}}O_{\Delta}(\lambda,\Bar{\lambda})\bigg|\frac{\lambda\cdot\Bar{\lambda}}{2}\bigg|.
\end{align}
\end{tcolorbox}
Note that we have rescaled the operator by $|\frac{\lambda\cdot\Bar{\lambda}}{2}|$ since only then does the twistor space operator $\hat{O}_{\Delta}(\lambda,\Bar{\mu})$ have dimension $\Delta-1$ (The operator $O_{\Delta}(\lambda,\Bar{\lambda})$ has scaling dimension $\Delta-3$ due to the Fourier transform and the measure contributes $+1$ which along with the rescaling factor makes the total $\Delta-1$). One can now follow steps very similar to our Penrose transform derivation for the conserved currents and $\Delta=1$ scalars in subsection \ref{subsec:PenroseFromWitten} to see that it leads to the Penrose transform \eqref{ODeltaPenrose}. The Witten transform \eqref{ODeltaWitten} enables us to derive the action of the conformal generators on the twistor space operator $\hat{O}(\lambda,\Bar{\mu})$. The details of the computation are sketched in appendix \ref{app:TwistorGenerators}. Translations and Lorentz transformations act exactly like in \eqref{TABcomponents} whereas dilatations are modified to,
\begin{simpleboxenv}
  \begin{align}\label{DODelta}
    [D,\hat{O}_{\Delta}(\lambda,\Bar{\mu})]=\frac{i}{2}\bigg(\lambda^a\frac{\partial}{\partial\lambda^a}-\Bar{\mu}^a\frac{\partial}{\partial\Bar{\mu}^a}+2(1-\Delta)\bigg)\hat{O}_{\Delta}(\lambda,\Bar{\mu}),
  \end{align}
  \end{simpleboxenv}
which is expected since the dimensionality of the twistor space current is $\Delta-1$ as we saw after establishing the Penrose transform \eqref{ODeltaPenrose}. The special conformal transformations, on the other hand, act in a very interesting \textit{non-local} manner. We find
\begin{tcolorbox}[
    colback=blue!5!white,
    colframe=yellow!60!black,
    title=Special conformal transformations for arbitrary scalars,
    sharp corners=south,
    boxrule=0.8pt,
    arc=4pt,
    fonttitle=\bfseries
]
\small
\begin{align}\label{KODelta}
    &[K_{ab},\hat{O}_{\Delta}(\lambda,\Bar{\mu})]\notag\\&=2i\bigg[\Bar{\mu}_{(a}\frac{\partial}{\partial\lambda^{b)}}+(\Delta-1)\bigg((\lambda\cdot\frac{\partial}{\partial\Bar{\mu}})^{-1}\big(\lambda_{(a}\frac{\partial}{\partial \lambda^{b)}}-\Bar{\mu}_{(a}\frac{\partial}{\partial\Bar{\mu}^{b)}}\big)-2(\lambda\cdot\frac{\partial}{\partial\Bar{\mu}})^{-2}\lambda_{(a}\frac{\partial}{\partial\Bar{\mu}^{b)}}\bigg)\bigg]\hat{O}_{\Delta}(\lambda,\Bar{\mu}).
\end{align}
\normalsize
 We define the inverse derivatives appearing in \eqref{KODelta} viz,
 \begin{align}\label{inversederivative}
     (\lambda\cdot\frac{\partial}{\partial\Bar{\mu}})^{-1}f(\lambda,\Bar{\mu})=-\int_0^\infty ds f(\lambda,\Bar{\mu}+s\lambda).
 \end{align}
 \end{tcolorbox}
For the inverse derivatives to make sense we require that $f(\lambda,\Bar{\mu})\not\in \text{Ker}(\lambda\cdot\frac{\partial}{\partial\Bar{\mu}})$. For the twistor space operators $\hat{O}_{\Delta}(\lambda,\Bar{\mu})$ which satisfy the unitary bound $(\Delta\ge \frac{\Delta-2}{2})$, this requirement is indeed satisfied and therefore this operator is invertible on the subspace spanned by the twistor space operators. One can easily check using the definition \eqref{inversederivative} that it satisfies,
\begin{align}
    (\lambda\cdot\frac{\partial}{\partial\Bar{\mu}})(\lambda\cdot\frac{\partial}{\partial\Bar{\mu}})^{-1}f(\lambda,\Bar{\mu})=f(\lambda,\Bar{\mu}),
\end{align}
provided $f(\lambda,\Bar{\mu})$ vanishes when $\Bar{\mu}\to \infty$ which is true of the examples we consider in this paper such as the two point function to be derived soon. We also note that the quantity $\lambda\cdot\frac{\partial}{\partial\Bar{\mu}}$ is the same as $Z^A I_{AB}\frac{\partial}{\partial Z_B}$ as can be seen from \eqref{infinitytwistor} and \eqref{defoftwistorZA} thus showing that the infinity twistor explicitly appears in the special conformal generator \eqref{KODelta}. Let us remark at this point that such a non-local representation has been considered in the context of \textit{conformal quantum mechanics} \cite{Khodaee:2017tbk} where the authors show that the conformal $\mathfrak{sl}(2,\mathbb{R})$ algebra admits a representation where the special conformal transformation contains a non-local term quantified by a parameter $\rho$. We see that in our case \eqref{KODelta}, the analogous parameter is $\Delta-1$. When $\Delta=1$, both the dilatation and SCT generator collapse down to the local generators \eqref{TABcomponents} and for $\Delta\ne 1$, we need to deal with the non-local transformations in \eqref{KODelta}\footnote{In \cite{Khodaee:2017tbk}, the dilatation operator is unaffected by the non-local term in the SCT operator. However, in our construction one can see that the second term in the RHS of \eqref{KODelta} precisely produces the additional term in \eqref{DODelta} by taking the commutator of $P$ with $K$ which shows that the conformal algebra is obeyed.}. The $O_2$ operator discussed in the first part of this section is a special case of the general $\Delta$ case and one can check that the results \eqref{O2O1Twistor}, \eqref{O2O2Twistor},\eqref{O2O2O2twistor},\eqref{JsO2O2twistor} and \eqref{Js1Js2O2twistor} satisfy the conformal Ward identities dictated by \eqref{DODelta}, \eqref{KODelta} for $\Delta=2$.

Let us now investigate two and some three point functions involving $\hat{O}_{\Delta}$.
\subsection*{Two point functions}
Consider a two point function of $\hat{O}_{\Delta}(Z_1)$. A general ansatz allowing for the infinity twistor (due to its presence in the generator \eqref{KODelta}) takes the form,
\begin{align}
    \langle 0|\hat{O}_{\Delta}(Z_1)\hat{O}_{\Delta}(Z_2)|0\rangle=(Z_1\cdot Z_2)^{\alpha}|\langle Z_1 I Z_2\rangle |^{\beta}.
\end{align}
Covariance under dilatations \eqref{DODelta} fixes $\beta=2(\Delta-1)$ and invariance under projective rescaling in \eqref{ODeltaPenrose} fixes $\alpha=-2\Delta$. Thus our candidate two point function reads,
\begin{simpleboxenv}
\begin{align}\label{ODelta2point}
    \langle 0|\hat{O}_{\Delta}(Z_1)\hat{O}_{\Delta}(Z_2)|0\rangle=\frac{|\langle Z_1 I Z_2\rangle |^{2(\Delta-1)}}{(Z_1\cdot Z_2)^{2\Delta}}.
\end{align}
\end{simpleboxenv}
We show in appendix \ref{app:ODeltaConf} that \eqref{ODelta2point} is invariant under special conformal transformations \eqref{KODelta}. Further, in appendix \ref{app:ExplicitPenrose}, we show that it goes over to the correct position space result. We do not pursue the construction of three point functions of arbitrary scalars in this paper. One way to proceed is to solve the conformal Ward identity \eqref{KODelta} which in this case is an integro-differential equation and thus is quite involved. Another inviting possibility is to use weight-shifting operators \cite{Karateev:2017jgd,Baumann:2019oyu} and the results for the correlators involving $O_1$ and $O_2$ to bootstrap higher $\Delta$ correlators. We leave such an exercise to the future.

\subsection{Parity odd Wightman functions}\label{subsec:parityoddsubsec}
In three dimensional CFT, conformal invariance allows for the existence of parity odd contributions to two and three point functions of conserved currents.
These were obtained in twistor space in \cite{Bala:2025gmz}. For instance, at the level of three points they obtained,
\begin{align}\label{twistorspaceodd3points}
   &\langle 0|\hat{J}_{s_1}^{h_1}(T_1)\hat{J}_{s_2}^{h_2}(T_2)\hat{J}_{s_3}^{h_3}(T_3)|0\rangle_{\text{odd}}\notag\\&=+i~\text{Sgn}\big(\sum_{j=1}^{3} \frac{h_j}{s_j}\big) c_{s_1s_2s_3}^{(\text{odd})}i^{s_1+s_2+s_3}\delta^{[s_1+s_2-s_3]}(T_1\cdot T_2)\delta^{[s_2+s_3-s_1]}(T_2\cdot T_3)\delta^{[s_3+s_1-s_2]}(T_3\cdot T_1),
\end{align}
\normalsize
where,
\begin{align}\label{TUnotation}
    &T_i=Z_i~\text{if}~h_i=+s_i ~\text{and}~ T_i=W_i ~\text{if}~ h_i=-s_i.
\end{align}
The result \eqref{twistorspaceodd3points} is consistent with the correct spinor helicity results after a half-Fourier transform \cite{Bala:2025gmz}. However, one must take caution before using it in a Penrose transform \eqref{PenroseTransform}, the reason for which we shall discuss now. The point that we want to make here is that starting with \eqref{twistorspaceodd3points} (lets say choosing the $(+++)$ helicity) and setting up the traditional Penrose transform \eqref{PenroseTransform}, we get the same expression as the parity even homogeneous correlator in \eqref{twistorspace3points} up to an overall coefficient! As it stands \eqref{twistorspaceodd3points} is a suitable twistor space representation for the parity odd correlator when viewed in the guise of the half-Fourier transform to spinor helicity variables for all space-like momenta. However, it is also desirable to have an expression whose Penrose transform to position space yields the parity odd correlator. That is the aim of this subsection.

\subsubsection{The Epsilon transform}
Let us 
start with a discussion of the epsilon transform. It is a conformally invariant (preserves eigenvalues of quadratic and quartic Casimirs) transformation in three dimensional CFTs \cite{Jain:2021gwa,Caron-Huot:2021kjy}. However, its novelty is that it flips the discrete parity label of a representation. Given a current $J^{\mu_1\cdots \mu_s}(x)$, its epsilon transform is defined as,
\begin{align}\label{EPT}
    (\epsilon\cdot J)^{\mu_1\cdots\mu_s}(x)=\epsilon^{\nu\sigma (\mu_1 }\int \frac{d^3 y}{|y-x|^2}\frac{\partial}{\partial y^{\nu}}J^{\mu_2\cdots \mu_s)}_{\nu}(y).
\end{align}
Similar to the Legendre transform, we desire to cast this transformation in twistor space. To illustrate the point, we consider a spin-1 current. We shall also employ spinor notation by contracting any vector indices with the Pauli matrices, since it is more natural in this framework. After some simple sigma matrix manipulation we get for the spin-1 case,
\begin{align}\label{EPTspinornotation}
    (\epsilon\cdot J)^{a_1 a_2}(x)=-i\int \frac{d^3 y}{|y-x|^2}\frac{\partial}{\partial y^c_{(b_1}}J^{a_1) c}(y)
\end{align}

Using the Penrose transform \eqref{PenroseTransform} for the spin-1 current in the RHS of \eqref{EPTspinornotation} we obtain,

\begin{align}\label{EPTPenrosetransform1}
    (\epsilon.J)_{ab}(x)=\int \langle\lambda d\lambda\rangle \lambda_{(a}\lambda_{b)} \int d^2 \bar{\mu}~\delta^2(\bar{\mu}^a-x^{ad}\lambda_d) (-i)\int \frac{d^3 y}{y^2} Z^AI_{AB} \frac{\partial}{\partial Z_B} \hat{J}^{+}(\lambda,\bar{\mu})\bigg|_{\bar{\mu}^a\to \bar{\mu}^a + y^{ad}\lambda_d}.
\end{align}
This identifies the twistor space epsilon transform as,
\begin{align}\label{twistorEPTspin1}
    (\epsilon\cdot \hat{J})^+(Z)=-i\int \frac{d^3 y}{y^2}Z^A I_{AB}\frac{\partial}{\partial Z_B}\hat{J}^{+}(\lambda,\Bar{\mu})|_{\Bar{\mu}^a\to \Bar{\mu}^a+y^{ab}\lambda_b}.
\end{align}
We note that the infinity twistor \eqref{infinitytwistor} is part of the very definition of the epsilon transform \eqref{twistorEPTspin1}! As we will see now, it will thus feature in parity odd Wightman functions. Before we proceed, we present the generalization to arbitrary spin which is a simple generalization of \eqref{twistorEPTspin1}.
\begin{tcolorbox}[
    colback=blue!5!white,
    colframe=blue!75!black,
    title=Epsilon Transform in Twistor Space,
    sharp corners=south,
    boxrule=0.8pt,
    arc=4pt,
    fonttitle=\bfseries
]
The epsilon transform in twistor space for a spin-s current is given by,
\begin{align}\label{twistorEPTspins}
    (\epsilon \cdot \hat{J}_s)^+(Z) = 
    -i \int \frac{d^3 y}{y^2} \,
    Z^A I_{AB} \frac{\partial}{\partial Z_B}
    \hat{J}_s^+(\lambda, \bar{\mu}) \bigg|_{\bar{\mu}^a \to \bar{\mu}^a + y^{ab} \lambda_b}.
\end{align}
The ``parity odd" Penrose transform is thus given by,
\begin{align}\label{parityoddPenrosetransform1}
    (\epsilon\cdot J_s)^{a_1\cdots a_{2s}}(x)=\int \langle\lambda d\lambda\rangle\lambda^{a_1}\cdots \lambda^{a_{2s}} (\epsilon \cdot \hat{J}_s)^+(Z)|_{X},
\end{align}
where $ (\epsilon \cdot \hat{J}_s)^+(Z)$ is given in \eqref{twistorEPTspins} and $X$ are the usual incidence relations \eqref{IncidenceRelation}.
\end{tcolorbox}
\begin{tcolorbox}[
    colback=blue!5!white,
    colframe=green!60!black,
    title=Witten transform for the ``parity" odd case,
    sharp corners=south,
    boxrule=0.8pt,
    arc=4pt,
    fonttitle=\bfseries
]
For convenience, we also present the epsilon transformed version of the Witten transform \eqref{WittenTransform} that can be obtained using the Fourier transform of \eqref{EPT}. The result is,
\begin{align}\label{wittentransparityodd}
    (\epsilon\cdot \hat{J}_s)(\lambda,\Bar{\mu})=+i\int \frac{d^2\Bar{\lambda}}{(2\pi)^2}e^{i\Bar{\lambda}\cdot\Bar{\mu}}\text{Sign}(\sqrt{p^2})\frac{J_s^{+}(\lambda,\Bar{\lambda})}{|p|^{s-1}},
\end{align}
where as usual $p=-\frac{1}{2}\lambda\cdot\Bar{\lambda}$.
\end{tcolorbox}
\subsubsection{Two point functions}
In three dimensions, conserved currents can have a parity odd contribution to the two point function that is a contact term. In twistor space, one can perform using \eqref{twistorEPTspins} an epsilon transform with respect to the first operator of the parity even positive helicity two point function \eqref{JsJstwopoint}. This results in
\begin{simpleboxenv}
\begin{align}\label{twopointodd}
   \langle 0|\hat{J}_s^{+}(Z_1)\hat{J}_s^{+}(Z_2)|0\rangle_{\text{odd}}=-i c_{\text{s,odd}}\text{Sgn}(\langle Z_1IZ_2\rangle)\delta^{[2s+1]}(Z_1\cdot Z_2).
\end{align}
\end{simpleboxenv}
Note that delta functions of twistor dot products were discarded in earlier works \cite{Baumann:2024ttn,Bala:2025gmz} since they do not have the correct rescaling properties under rescaling by negative numbers and do not give rise to the correct spinor helicity correlators. Here, on the other hand, the Sign factor involving the infinity twistor in \eqref{twopointodd} takes care of both of these issues simultaneously. Under rescalings by an amount $r$ we have,
\begin{align}
    \text{Sgn}(Z_1IZ_2)\delta^{[3]}(Z_1\cdot Z_2)\to \text{Sgn}(r)\frac{1}{r^{2s+2}\text{Sgn}(r)}\text{Sgn}(Z_1IZ_2)\delta^{[3]}(Z_1\cdot Z_2)=\text{Sgn}(Z_1IZ_2)\delta^{[3]}(Z_1\cdot Z_2).
\end{align}
One can verify that \eqref{twopointodd} is conformally invariant with respect to the generators \eqref{TABZ} as a \textit{distribution} (that is, on the support of the delta function that it multiplies) despite the presence of the sign factor whose argument breaks conformal invariance. We give details of the same in appendix \ref{app:EPTdetails}.

Similarly, when converted to spinor helicity variables, $\text{Sgn}(Z_1IZ_2)$ cancels out the $\text{Sgn}(\langle 12\rangle)$ that appears in the Witten transform \eqref{WittenTransform} as was discussed in \cite{Bala:2025gmz}. Moreover, as we show in appendix \ref{app:ExplicitPenrose} it goes over to the correct conformally invariant parity odd contact term after performing the Penrose transform in \eqref{twopointodd} thereby validating the result.

\subsubsection{Three point functions}
Performing an epsilon transform of the parity even homogeneous Wightman function with respect to the first current, we obtain
\begin{simpleboxenv}
\begin{align}\label{Js1s2s3TwistorNEW}
   &\langle 0|\hat{J}_{s_1}^{+}(Z_1)\hat{J}_{s_2}^{+}(Z_2)\hat{J}_{s_3}^{+}(Z_3)|0\rangle_{\text{odd}}\notag\\&=\frac{i}{2}\int dc_{12}dc_{23}dc_{31}c_{12}^{s_1+s_2-s_3}c_{23}^{s_2+s_3-s_1}c_{31}^{s_3+s_1-s_2}\text{Sgn}\big(c_{12}\langle Z_1 IZ_2\rangle+c_{31}\langle Z_3 IZ_1\rangle\big)\notag\\&\qquad\qquad \times\big(e^{-ic_{12}Z_1\cdot Z_2-ic_{23}Z_2\cdot Z_3-ic_{31}Z_3\cdot Z_1}\big).
\end{align}
\end{simpleboxenv}
The infinity twistor similar to in the two point function \eqref{twopointodd} appears via sign factors. One can show similar to the two point case that \eqref{Js1s2s3TwistorNEW} satisfies the conformal Ward identities \eqref{TABZ} as a distribution.

Let us now compare our new result \eqref{Js1s2s3TwistorNEW} with the parity odd three point function of \cite{Bala:2025gmz} given in \eqref{twistorspaceodd3points}, in Schwinger parametrization:
\begin{align}\label{Js1s2s3TwistorOLD}
    &\langle 0|\hat{J}_{s_1}^{+}(Z_1)\hat{J}_{s_2}^{+}(Z_2)\hat{J}_{s_3}^{+}(Z_3)|0\rangle_{\text{odd from \cite{Bala:2025gmz}}}\notag\\&=\frac{i}{2}\int dc_{12}dc_{23}dc_{31}c_{12}^{s_1+s_2-s_3}c_{23}^{s_2+s_3-s_1}c_{31}^{s_3+s_1-s_2}e^{-ic_{12}Z_1\cdot Z_2-ic_{23}Z_2\cdot Z_3-ic_{31}Z_3\cdot Z_1}.
\end{align}
Note in particular that \eqref{Js1s2s3TwistorOLD} differs from \eqref{Js1s2s3TwistorNEW} by a factor of $\text{Sgn}\big(c_{12}\langle 12\rangle+c_{31}\langle 31\rangle\big)$ in the Schwinger parameter integrals. When converted to spinor-helicity variables via Witten's half-Fourier transform \eqref{WittenTransform}, \eqref{Js1s2s3TwistorOLD} and \eqref{Js1s2s3TwistorNEW} respectively yield,
\begin{align}
    &\langle 0|\hat{J}_{s_1}^{+}(p_1)\hat{J}_{s_2}^{+}(p_2)\hat{J}_{s_3}^{+}(p_3)|0\rangle_{\text{odd from} \cite{Bala:2025gmz}}=i\frac{\langle \Bar{1}\Bar{2}\rangle^{s_1+s_2-s_3}\langle \Bar{2}\Bar{3}\rangle^{s_2+s_3-s_1}\langle \Bar{3}\Bar{1}\rangle^{s_3+s_1-s_2}}{E^{s_1+s_2+s_3}}\notag\\
    &\langle 0|\hat{J}_{s_1}^{+}(p_1)\hat{J}_{s_2}^{+}(p_2)\hat{J}_{s_3}^{+}(p_3)|0\rangle_{\text{odd} }=i\text{Sgn}\big(\sqrt{p_1^2}\big)\frac{\langle \Bar{1}\Bar{2}\rangle^{s_1+s_2-s_3}\langle \Bar{2}\Bar{3}\rangle^{s_2+s_3-s_1}\langle \Bar{3}\Bar{1}\rangle^{s_3+s_1-s_2}}{E^{s_1+s_2+s_3}}.
\end{align}
Since the half-Fourier transform is performed for space-like momenta the sign factor is equal to $1$ in such cases and thus both results agree. However, it is \eqref{Js1s2s3TwistorNEW} that also gives rise to the correct position space result after a Penrose transform \eqref{PenroseTransform}. Therefore, as the authors themselves commented, the results of \cite{Bala:2025gmz} for parity-odd correlators are in the context of the half-Fourier transform for space-like momenta. Our new result \eqref{Js1s2s3TwistorNEW} is valid in the context of both the Witten and Penrose transforms. The sign factors appearing in our formulae \eqref{twopointodd} and \eqref{Js1s2s3TwistorNEW} are responsible for resulting in a parity odd expression which can be seen by using these expressions in the Penrose transform \eqref{PenroseTransform}\footnote{Recall that the parity operator in $2+1$ dimensions takes the coordinate $z\to -z$. One can check by plugging in \eqref{twopointodd} and \eqref{Js1s2s3TwistorNEW} into the Penrose transform and using the incidence relations that they pick up a sign under a parity operation, affirming that the expression is parity odd. Similarly, one can check that they are also odd under time-reversal $t\to -t,i\to -i$. This also ensures that they are invariant under $PT$.}. The main result of this subsection was the twistor space epsilon transform \eqref{twistorEPTspins}. It allows us to define a new Penrose transform like \eqref{EPTPenrosetransform1} that involves the infinity twistor in contrast to \eqref{PenroseTransform}. The generelization to other helicities is straightforward so we do not present the details here.

\subsection{Non conserved currents in twistor space}\label{subsec:nonconsTwistor}
In the third and final part of this section, we shall generalize the Penrose transform \eqref{PenroseTransform} to accommodate general symmetric traceless integer spin primary representations of the conformal group. We then investigate two point Wightman functions involving such operators. Let us concentrate on the spin-1 case for technical simplicity and then generalize to arbitrary spin. Consider a spin-1 operator with dimension $\Delta$. We propose the following Penrose transform for this operator:
\begin{tcolorbox}[
    colback=blue!5!white,
    colframe=blue!80!black,
    title=Penrose Transform for spin-1 non conserved current,
    sharp corners=south,
    boxrule=0.8pt,
    arc=4pt,
    fonttitle=\bfseries
]
\begin{align}\label{nonconsPenrose}
    \mathcal{O}^{ab}_{\Delta}(x)=\int \langle \lambda d\lambda\rangle\bigg(\lambda^{a_1}\lambda^{a_2}\hat{\mathcal{O}}_{\Delta}^{+1}(\lambda,\Bar{\mu})-i\lambda_{(a}\frac{\partial}{\partial\Bar{\mu}_{b)}}\hat{\mathcal{O}}_{\Delta}^{0}(\lambda,\Bar{\mu})\bigg)|_{X},
\end{align}
where $X$ denotes the usual incidence relations \eqref{IncidenceRelation}.
\end{tcolorbox}
Some words about \eqref{nonconsPenrose} are in order. First of all, unlike the Penrose transforms for currents \eqref{PenroseTransform} or $O_{\Delta}$ scalars \eqref{ODeltaPenrose} which output the position space operator given a single twistor space operator, the formula \eqref{nonconsPenrose} requires two twistor space operators for the same. Both these operators $\hat{\mathcal{O}}^{+1}(\lambda,\Bar{\mu})$ as well as $\hat{\mathcal{O}}^{0}(\lambda,\Bar{\mu})$ have the scaling dimension $\Delta-2$ as we can read off from \eqref{nonconsPenrose}. However, they have different projective properties. For the integrand in \eqref{nonconsPenrose} to be invariant under projective rescalings $(\lambda,\Bar{\mu})\to (r\lambda,r\Bar{\mu})$ we require,
\begin{align}
    \hat{\mathcal{O}}_{\Delta}^{+1}(r\lambda,r\Bar{\mu})=\frac{1}{r^4}\hat{\mathcal{O}}_{\Delta}^{+1}(\lambda,\Bar{\mu}),~~\hat{\mathcal{O}}_{\Delta}^{0}(r\lambda,r\Bar{\mu})=\frac{1}{r^2}\hat{\mathcal{O}}_{\Delta}^{0}(\lambda,\Bar{\mu}).
\end{align}
These operators are obtained via the Witten transform \eqref{WittenTransform} as follows:
\begin{tcolorbox}[
    colback=blue!5!white,
    colframe=green!60!black,
    title=Witten Transform for spin-1 non conserved current,
    sharp corners=south,
    boxrule=0.8pt,
    arc=4pt,
    fonttitle=\bfseries
]
\begin{align}\label{nonconsWittenTransform}
    \hat{\mathcal{O}}_{\Delta}^{+1}(\lambda,\Bar{\mu})=\int \frac{d^2\Bar{\lambda}}{(2\pi)^2}e^{i\Bar{\lambda}\cdot\Bar{\mu}}\mathcal{O}_{\Delta}^{+1}(\lambda,\Bar{\lambda}),~~ \hat{\mathcal{O}}_{\Delta}^{0}(\lambda,\Bar{\mu})=\int \frac{d^2\Bar{\lambda}}{(2\pi)^2}e^{i\Bar{\lambda}\cdot\Bar{\mu}}\mathcal{O}_{\Delta}^{0}(\lambda,\Bar{\lambda}).
\end{align}
\end{tcolorbox}
The spinor helicity operators appearing in the integrals in the above equation are given by (see \eqref{SHvariables} for the definitions of the polarizations),
\begin{align}\label{nonconsspin1SH}
    \mathcal{O}_{\Delta}^{+1}(\lambda,\Bar{\lambda})=\zeta_{+}^{a_1}\zeta_{+}^{a_2}\mathcal{O}_{\Delta a_1 a_2}(\lambda,\Bar{\lambda}),~~\mathcal{O}_{\Delta}^{0}(\lambda,\Bar{\lambda})=\zeta_{+}^{a_1}\zeta_{-}^{a_2}\mathcal{O}_{\Delta a_1 a_2}(\lambda,\Bar{\lambda})
\end{align}
Note that the second term in \eqref{nonconsspin1SH} is identically zero for conserved currents as it is simply equal to $p^{a_1 a_2}\mathcal{O}_{\Delta a_1 a_2}(\lambda,\Bar{\lambda})$ which is the divergence of the current. Let us now derive the formula \eqref{nonconsPenrose} following similar steps to subsection \ref{subsec:PenroseFromWitten}. Till \eqref{FouriertoPenroseStep1next}, the steps remain the same. The difference appears in the next step \eqref{FouriertoPenroseinterstep} where we threw away terms proportional to the divergence of the current. For a spin-1 non conserved current the analog of \eqref{FouriertoPenroseStep1next} after some simple variable-relabelling in some of the terms reads,
\begin{align}
    \mathcal{O}_{\Delta}^{a_1 a_2}(x)=\frac{1}{\text{Vol}(GL(1,\mathbb{R}))}\int \langle\lambda d\lambda\rangle|\frac{\lambda\cdot\Bar{\lambda}}{2}|&\bigg(\lambda^{a_1}\lambda^{a_2}\Bar{\lambda}^{b_1}\Bar{\lambda}^{b_2}\mathcal{O}_{\Delta b_1 b_2}(\lambda,\Bar{\lambda})-\lambda^{(a_1}\Bar{\lambda}^{a_2)}\lambda^{b_1}\Bar{\lambda}^{b_2}\mathcal{O}_{\Delta b_1 b_2}(\lambda,\Bar{\lambda})\bigg).
\end{align}
The second term in the above equation is the zero helicity component of the current which we saw in \eqref{nonconsspin1SH}. Combining this formula with the Witten transform \eqref{nonconsWittenTransform} then results in the Penrose transform \eqref{nonconsPenrose}. The generalization to higher spin is similar. For a spin s symmetric traceless operator with scaling dimension $\Delta$, we find a Penrose transform involving $s+1$ terms which include the `helicities" $+s,s-1,s-2,\cdots 1,0$. This is due to the fact that such an operator has $2s+1$ different independent components (which is the also the same as the number of degrees of freedom of a massive spin-s gauge boson in four dimensions) and inside the Penrose transform, the component with equal and opposite ``helicities" are equivalent similar to what we saw in going from \eqref{FouriertoPenroseinterstep} to \eqref{FouriertoPenrosestep2} . It is much simpler to write this expression this after contracting with arbitrary polarization spinors $\zeta_{a}$. The result is,
\begin{tcolorbox}[
    colback=blue!5!white,
    colframe=blue!70!black,
    title=Penrose Transform for spin-s non conserved current,
    sharp corners=south,
    boxrule=0.8pt,
    arc=4pt,
    fonttitle=\bfseries
]
\begin{align}\label{PenroseTransformNonConservedSPins}
    \mathcal{O}_{\Delta,s}(x,\zeta)=\int \langle \lambda d\lambda\rangle \sum_{k=0}^{s}c_k(\zeta\cdot \lambda)^{2s-k}(\zeta\cdot \frac{\partial}{\partial\Bar{\mu}})^{k}\hat{\mathcal{O}}^{s-k}_{\Delta}(\lambda,\Bar{\mu})|_X,
\end{align}
where $X$ denotes the incidence relation \eqref{IncidenceRelation} and $c_k$ are coefficients that can easily be calculated like in \eqref{nonconsPenrose} and performing the Witten transform \eqref{nonconsWittenTransformspins} for each component.
\end{tcolorbox}
The Witten transform for each component is given by,
\begin{tcolorbox}[
    colback=blue!5!white,
    colframe=green!60!black,
    title=Witten Transform for spin-s non conserved current,
    sharp corners=south,
    boxrule=0.8pt,
    arc=4pt,
    fonttitle=\bfseries
]
\begin{align}\label{nonconsWittenTransformspins}
    \hat{\mathcal{O}}_{\Delta}^{+k}(\lambda,\Bar{\mu})=\int \frac{d^2\Bar{\lambda}}{(2\pi)^2}e^{i\Bar{\lambda}\cdot\Bar{\mu}}\frac{\mathcal{O}_{\Delta}^{+k}(\lambda,\Bar{\lambda})}{|\frac{\lambda\cdot\Bar{\lambda}}{2}|^{s-1}},
\end{align}
where $\mathcal{O}^{+k}(\lambda,\Bar{\lambda})$ contains $s+k$ positive helicity polarization spinors and $s-k$ negative  helicity polarization spinors \eqref{SHvariables} contracted with the momentum space operator $\mathcal{O}_{\Delta a_1\cdots a_{2s}}$.
\end{tcolorbox}
 We can obtain the two point functions for each component of the spin-s operator using dimensional analysis and invariance under projective rescalings. For example, the component $\hat{\mathcal{O}}$ has the scaling dimension $\Delta-(s+1)$ and scales with a factor of $\frac{1}{r^{2(s-k)+2}}$.
 \begin{simpleboxenv}
 \begin{align}
     \langle 0|\hat{\mathcal{O}}_{\Delta}^{+k}(Z_1)\hat{\mathcal{O}}_{\Delta}^{+l}(Z_2)|0\rangle=c_{kl}\delta^{k l}\frac{\langle Z_1 I Z_2\rangle^{2(\Delta-(s+1))}}{(Z_1\cdot Z_2)^{2(\Delta-k)}}.
 \end{align}
  \end{simpleboxenv}
 $c_{kl}$ are coefficients that are all related and are multiples of the two point function coefficient $c_{\mathcal{O}_{\Delta,s}}$. One can check that this formula leads to the correct non-conserved current two point function by performing the Penrose transform \eqref{nonconsPenrose}\footnote{For instance, we explicitly matched the spin-1 case with the result in \cite{Marotta:2022jrp} by converting the result of the Penrose transform \eqref{nonconsPenrose} to momentum space. We hope to return to a systematic analysis in the future.}. Similarly, it would be interesting to solve for three point functions involving at least one non-conserved current which will greatly help in setting up the conformal bootstrap in twistor space. One way is to solve the integro-differential conformal Ward identities (which is equal to \eqref{KODelta}+ a contribution that depends on the spin which can be derived from the extra term in footnote \eqref{footnoteODeltas}). Alternatively using weight shifting and spin raising operators \cite{Karateev:2017jgd,Baumann:2019oyu} also seems like an interesting and viable way to proceed. We hope to return to this in the future.
\section{The Geometry of Super-Twistor Space}\label{sec:SuperTwistorGeo}

In this section, we first introduce the basics of the super-twistor space developed in \cite{Bala:2025jbh}. We then derive the corresponding super-incidence relations and the super-Penrose transform by using the super-field expansion and the Penrose transform \eqref{PenroseTransform} for each component current. We compare this with the Super-Witten transform and show that when combined with the ordinary Fourier transform and using the Faddeev-Popov method, the super-Penrose transform can also be derived. We then discuss the form of the super-conformal generators and derive the OSp$(\mathcal{N},4)$ invariants that serve as building blocks for super-twistor space Wightman functions.

\subsection{A Brief review of super-twistor space}
Let us begin with a brief review of the super-twistor space constructed in \cite{Bala:2025jbh} for theories $\mathcal{N}-$extended supersymmetry in $2+1$ dimensions. Super twistor space is an open subset of the projective space $\mathbb{RP}^{3|\mathcal{N}}$. It is spanned by bosonic coordinates $Z^A$ and fermionic coordinates $\psi^N$ where $A=1,2,3,4$ is the $Sp(4)$ fundamental index and $N=1,\cdots,\mathcal{N}$ is the $O(\mathcal{N})$ fundamental index. Together they form the super twistor space coordinates $\mathcal{Z}^{\mathcal{A}}=(Z^A,\psi^N)=(\lambda^a,\Bar{\mu}_{a'},\psi^N)$ which is in the fundamental representation of the the super-conformal group $OSp(\mathcal{N}|4;\mathbb{R})$.
The coordinates $Z^A$ are real and $\psi^N$ satisfy $(\psi^N)^*=i \psi^N$. Similar to the non-supersymmetric case, there exist dual twistor space coordinates $\mathcal{W}_{\mathcal{A}}=(W_A,\bar{\psi}_N)=(\mu_{a},\bar{\lambda}^{a'}, \bar{\psi}_N)$ with $W^{A}$ being real and $(\bar{\psi}^N)*= i \Bar{\psi}^N$. Indices of super-twistors are raised and lowered using the OSp($\mathcal{N}|4;\mathbb{R})$ invariant graded symplectic form $\Omega_{\mathcal{AB}}$ as in the appendix \ref{app:Notation}. With these ingredients and the Penrose transform \eqref{PenroseTransform}, we are now in a position to make a connection to the position super-space. We focus on the $\mathcal{N}=1$ case, but the methods and results should be easily generalizable to extended supersymmetry.

\subsection{The Super-Penrose Transform}\label{subsec:SuperPenrose1}

A conserved super-current in position super-space takes the following form \cite{Nizami:2013tpa}:
\begin{align}\label{SuperFieldExpansion}
\mathbf{J}_s^{a_1\cdots a_{2s}}(x,\theta)=J_s^{a_1\cdots a_{2s}}(x)+\theta_a J_{s+\frac{1}{2}}^{aa_1\cdots a_{2s}}(x)-\frac{i\theta^2}{4}\slashed{\partial}_{a}^{a_1}J_s^{a_2\cdots a_{2s}a}(x).
\end{align}
The statement of conservation reads,
\begin{align}\label{SuperCons}
    D_{a_1}J_s^{a_1\cdots a_{2s}}=0,
\end{align}
where the super-covariant derivative is given by,
\begin{align}
    D_{a}=\frac{\partial}{\partial\theta^a}-\frac{i}{2}\theta_b\slashed{\partial}_a^b.
\end{align}
Given the super-field expansion \eqref{SuperFieldExpansion} and the fact that we know the Penrose transform of each component current appearing there \eqref{PenroseTransform}, we can re-write \eqref{SuperFieldExpansion} as,
\small
\begin{align}\label{SuperCurrentcomponent1}
    &\mathbf{J}_s^{a_1\cdots a_{2s}}(x,\theta)=\int \langle \lambda d\lambda\rangle \lambda^{a_1}\cdots \lambda^{a_{2s}}\bigg(\hat{J}_s^{+}(\lambda,\Bar{\mu})+\theta_a\lambda^a \hat{J}_{s+\frac{1}{2}}^{+}(\lambda,\Bar{\mu})\bigg)\bigg|_{X}-\frac{i\theta^2}{4}\slashed{\partial}_a^{a_1}\int \langle \lambda d\lambda \rangle \lambda^{a_2}\cdots \lambda^{a_{2s}}\lambda^a \hat{J}_s^{+}(\lambda,\Bar{\mu})|_{X}\notag\\
    &=\int \langle \lambda d\lambda\rangle \lambda^{a_1}\cdots \lambda^{a_{2s}}\bigg(\hat{J}_s^{+}(\lambda,\Bar{\mu})+\theta_a\lambda^a \hat{J}_{s+\frac{1}{2}}^{+}(\lambda,\Bar{\mu})\bigg)\bigg|_{X}-\frac{i\theta^2}{4}\int \langle\lambda d\lambda\rangle \lambda^{a_2}\cdots \lambda^{a_{2s}}\lambda^a\frac{\partial\Bar{\mu}^b}{\partial x^{a}_{a_1}}\frac{\partial}{\partial \Bar{\mu}^b}\hat{J}_s^{+}(\lambda,\Bar{\mu})|X\notag\\&=\int \langle \lambda d\lambda\rangle \lambda^{a_1}\cdots \lambda^{a_{2s}}\bigg(\hat{J}_s^{+}(\lambda,\Bar{\mu})+\theta_a\lambda^a \hat{J}_{s+\frac{1}{2}}^{+}(\lambda,\Bar{\mu})\bigg)\bigg|_{X}-\frac{i\theta^2}{4}\int \langle\lambda d\lambda\rangle \lambda^{a_2}\cdots \lambda^{a_{2s}}\lambda^a(-2\lambda^{a_1}\delta^{b}_a+\lambda^b \delta^{a_1}_{a})\frac{\partial}{\partial\Bar{\mu}^b}\hat{J}_s^{+}(\lambda,\Bar{\mu})|_X\notag\\
    &=\int \langle \lambda d\lambda\rangle \lambda^{a_1}\cdots \lambda^{a_{2s}}\bigg(\hat{J}_s^{+}(\lambda,\Bar{\mu})+\theta_a\lambda^a \hat{J}_{s+\frac{1}{2}}^{+}(\lambda,\Bar{\mu})+\frac{i\theta^2}{4}\lambda^a\frac{\partial}{\partial\Bar{\mu}^a}\hat{J}_s^{+}(\lambda,\Bar{\mu})\bigg)\bigg|_{X}.
\end{align}
\normalsize
where $X$ as usual denote that the incidence relation \eqref{IncidenceRelation} should be imposed. We now want to rewrite this covariantly as an integral over the super-twistor space currents $\mathbf{\hat{J}}_s^{+}(\mathcal{Z})=\mathbf{\hat{J}}_s^{+}(\lambda,\Bar{\mu},\psi)$ first introduced in \cite{Bala:2025jbh}. First of all $\mathbf{J}_s^{+}(\mathcal{Z})$ similar to its non-supersymmetric counterpart \eqref{rescalingformulaJsone} satisfies \cite{Bala:2025jbh},
\begin{align}
    \mathbf{\hat{J}}_s^{+}(r \mathcal{Z})=\frac{1}{r^{2s+2}}\mathbf{\hat{J}}_s^{+}(\mathcal{Z}).
\end{align}
This leads us to the following well defined projective integral involving this current:
\begin{align}\label{ansatzSUSYPenrose}
    \mathbf{J}_s^{a_1\cdots a_{2s}}(x,\theta)=\int \langle \lambda d\lambda\rangle \lambda^{a_1}\cdots \lambda^{a_{2s}} \mathbf{J}_s^{+}(\lambda,\Bar{\mu},\psi)|_{\mathcal{X}}=\int \langle \lambda d\lambda\rangle \lambda^{a_1}\cdots \lambda^{a_{2s}}\big(\hat{J}_s^{+}(\lambda,\Bar{\mu})+\frac{e^{\frac{i\pi}{4}}\psi}{\sqrt{2}}\hat{J}_{s+\frac{1}{2}}^{+}(\lambda,\Bar{\mu})\big)|_{\mathcal{X}}.
\end{align}
We have also expanded the super-twistor space super-current into its two components in \eqref{ansatzSUSYPenrose} using the formula in \cite{Bala:2025jbh}.
The subscript $\mathcal{X}$ indicates the superincidence relations that we need to derive. Let us write down an ansatz for the same using dimension analysis and the projectiveness:
\begin{align}\label{Superincidenceansatz}
    \mathcal{X}=\big\{\Bar{\mu}_a=-x_{ab}\lambda^b+\alpha \theta^2 \lambda_a~,~\psi=\beta\theta^a\lambda_a\big\}.
\end{align}
Substituting \eqref{Superincidenceansatz} in \eqref{ansatzSUSYPenrose} and performing the Grassmann expansion yields,
\begin{align}\label{SUSYPenroseansatz2}
    \mathbf{J}_s^{a_1\cdots a_{2s}}(x,\theta)=\int \langle \lambda d\lambda\rangle\lambda^{a_1}\cdots \lambda^{a_{2s}}\bigg(\hat{J}_s^{+}(\lambda,\Bar{\mu})-\frac{e^{\frac{i\pi}{4}}}{\sqrt{2}}\beta \theta_a\lambda^a~\hat{J}_{s+\frac{1}{2}}^{+}(\lambda,\Bar{\mu})+\alpha\theta^2\lambda^a\frac{\partial}{\partial\Bar{\mu}^a}\hat{J}_s^{+}(\lambda,\Bar{\mu})\bigg)\bigg|_{X},
\end{align}
where $X$ now denotes the usual component level usual incidence relations \eqref{IncidenceRelation}.
Comparing the component expansion \eqref{SuperCurrentcomponent1} to our ansatz \eqref{SUSYPenroseansatz2} yields the values,
\begin{align}
    \alpha=\frac{i}{4},\beta=-\sqrt{2}e^{-\frac{i\pi}{4}}.
\end{align}
\begin{superpenrosebox}
Putting everything together, we obtain the Super-Penrose transformation:
\begin{align}\label{SuperPenroseTransform}
    \mathbf{J}_s^{a_1\cdots a_{2s}}(x,\theta)=\int \langle \lambda d\lambda\rangle\lambda^{a_1}\cdots \lambda^{a_{2s}}\hat{\mathbf{J}}_s^{+}(\lambda,\Bar{\mu},\psi)\bigg|_{\mathcal{X}},
\end{align}
where the super-incidence relations are given by,
\begin{align}\label{Superincidence}
    \mathcal{X}=\big\{\Bar{\mu}_a=-x_{ab}\lambda^b+\frac{i}{4} \theta^2 \lambda_a~,~\psi=-\sqrt{2}e^{-\frac{i\pi}{4}}\theta^a\lambda_a\big\}.
\end{align}
\end{superpenrosebox}

As a consistency check, it is easy to see that \eqref{SuperPenroseTransform} along with \eqref{Superincidence} satisfies the super-conservation \eqref{SuperCons}.

Let us reflect on \eqref{Superincidence} briefly. Given a point in position superspace $(x_{ab},\theta_c)$ we see that \eqref{Superincidence} defines a projective line $\mathbb{RP}^{1|1}\in \mathbb{RP}^{3|1}$. The generalization to extended supersymmetry should be straightforward. The only difference is that the Grassmann coordinate $\psi$ becomes a vector of $SO(\mathcal{N})$ which is the R-symmetry group and thus one needs to impose incidence relations for each component. Further, for $\mathcal{N}\ge 2$, one can form contractions in different ways using the invariants of the $R-$symmetry group. We leave such an exercise for the future.

\subsection{ Fourier$+$ Super-Witten transform$\implies$ Super-Penrose transform}
In subsection \ref{subsec:SuperPenrose1}, we have derived the Super-Penrose transform using the component level Penrose tranform \eqref{PenroseTransform} and the super-twistor currents of \cite{Bala:2025jbh}. Similar to the non-supersymmetric exercise of subsection \ref{subsec:PenroseFromWitten}, we shall now derive the Super-Penrose transform starting from the Fourier transform and using the super-Witten transform developed in \cite{Bala:2025jbh} thus proving their equivalence. A word on notation before we proceed: Our conventions for the Fourier transform \eqref{FourierTrans1} differ from those in \cite{Bala:2025jbh} by a factor of $-2$ in the exponent of the plane wave. This is done for convenience but will result in intermediate formulae including those involving Grassmann coordinates to differ compared to \cite{Bala:2025jbh}.

Our starting point is the ordinary Fourier transform which is simply obtained by replacing the current by the super-current in \eqref{FourierTrans1}:
\begin{align}\label{superFourierTrans1}
    \mathbf{J}_s^{a_1\cdots a_{2s}}(x,\theta)=\int \frac{d^3 p}{(2\pi)^3}e^{-2ip\cdot x}\mathbf{J}_s^{a_1\cdots a_{2s}}(p^\mu).
\end{align}
Performing exactly the same steps as in subsection \ref{subsec:PenroseFromWitten} we land up with the analog of the expression  \eqref{FouriertoPenrosestep2},
\begin{align}\label{SUSYFourierstep1}
    \mathbf{J}_s^{a_1\cdots a_{2s}}(x,\theta)=\frac{1}{2^s\text{Vol}(GL(1,\mathbb{R}))}\int\frac{d^2\lambda d^2\Bar{\lambda}}{(2\pi)^3}\lambda^{a_1}\cdots \lambda^{a_{2s}}e^{i\Bar{\lambda}_a\lambda_b x^{ab}}\hat{\mathbf{J}}_s^{+}(\lambda,\Bar{\lambda},\theta).
\end{align}
The integrand of the above expression is only in spinor helicity variables. However, for the supersymmetric scenario, it is preferable to use super-spinor helicity variables \cite{Jain:2023idr} by expressing $\theta$ in the basis of $\lambda$ and $\Bar{\lambda}$. We do this using the Faddeev-Popov method as follows:

\begin{tcolorbox}[colback=white]
\textbf{Claim:}
\begin{equation}\label{GrassmannId}
(\lambda \cdot \bar{\lambda}) \int d\eta\, d\bar{\eta}~
\delta^2\left( \theta^a + \frac{\bar{\eta} \lambda^a + \eta \bar{\lambda}^a}{\lambda \cdot \bar{\lambda}} \right) = 1
\end{equation}

\textbf{Proof:}
\small
\begin{align}
&(\lambda \cdot \bar{\lambda}) \int d\eta\, d\bar{\eta}~
\delta^2\left( \theta^a + \frac{\bar{\eta} \lambda^a + \eta \bar{\lambda}^a}{\lambda \cdot \bar{\lambda}} \right)= (\lambda \cdot \bar{\lambda}) \int d\eta\, d\bar{\eta}~
\left( \theta^1 + \frac{\bar{\eta} \lambda^1 + \eta \bar{\lambda}^1}{\lambda \cdot \bar{\lambda}} \right)
\left( \theta^2 + \frac{\bar{\eta} \lambda^2 + \eta \bar{\lambda}^2}{\lambda \cdot \bar{\lambda}} \right) \notag \\
&= \frac{1}{\lambda \cdot \bar{\lambda}} \int d\eta\, d\bar{\eta}~ \bar{\eta} \eta \left( \lambda^1 \bar{\lambda}^2 - \bar{\lambda}^1 \lambda^2 \right)= \frac{1}{\lambda \cdot \bar{\lambda}} \left( \lambda^1 \bar{\lambda}^2 - \bar{\lambda}^1 \lambda^2 \right) \int d\eta\, d\bar{\eta}~ \bar{\eta} \eta \notag \\
&= \frac{1}{\lambda \cdot \bar{\lambda}} \left( \lambda^1 \bar{\lambda}^2 - \bar{\lambda}^1 \lambda^2 \right) \cdot 1= 1 \qquad \square
\end{align}
\normalsize
\end{tcolorbox}
Inserting \eqref{GrassmannId} in \eqref{SUSYFourierstep1} results in,
\begin{align}\label{SUSYFourierstep2}
    \mathbf{J}_s^{a_1\cdots a_{2s}}(x,\theta)=\frac{1}{2^{2s}\text{Vol}(GL(1,\mathbb{R}))}&\int\frac{d^2\lambda d^2\Bar{\lambda}}{(2\pi)^3}e^{i\Bar{\lambda}_a\lambda_b x^{ab}}\lambda^{a_1}\cdots \lambda^{a_{2s}}(\lambda\cdot \Bar{\lambda})\notag\\&\times \int d\eta d\Bar{\eta}~
\delta^2\left( \theta^a + \frac{\bar{\eta} \lambda^a + \eta \bar{\lambda}^a}{\lambda \cdot \bar{\lambda}} \right) \mathbf{\hat{J}}_s^{+}(\lambda,\Bar{\lambda},\eta,\Bar{\eta}),
\end{align}
where we replaced $\mathbf{\hat{J}}_s^{+}(\lambda,\Bar{\lambda},\theta)$ by $\mathbf{\hat{J}}_s^{+}(\lambda,\Bar{\lambda},\eta,\Bar{\eta})$ on the support of the delta function we have inserted. We now express the latter by writing it as a half-Fourier transform of its super-twistor space counterpart \cite{Bala:2025jbh}:
\begin{superwittenbox}
The relation between the current in super-spinor helicity variables and super-twistor space is as follows:
\begin{align}\label{JsetaetabarinJsetachi}
    &\mathbf{\hat{J}}_s^{+}(\lambda,\Bar{\lambda},\eta,\Bar{\eta})=2\int d\chi e^{-\frac{\chi \Bar{\eta}}{2}}\mathbf{\hat{J}}_s^{+}(\lambda,\Bar{\lambda},\eta,\chi)=2\int d\chi e^{-\frac{\chi \Bar{\eta}}{2}} (\chi-\eta)\hat{\mathbf{J}}_s^{+}(\lambda,\Bar{\lambda},\chi+\eta)\notag\\
    &=2\int d^2\Bar{\mu}~e^{-i\Bar{\lambda}\cdot \Bar{\mu}}\int d\chi e^{-\frac{\chi \Bar{\eta}}{2}} (\chi-\eta)\hat{\mathbf{J}}_s^{+}(\lambda,\Bar{\mu},\chi+\eta).
\end{align}
\end{superwittenbox}
We now substitute  \eqref{JsetaetabarinJsetachi} back into \eqref{SUSYFourierstep2} and change variables from $(\eta,\chi)\to (\xi_+,\xi_-)=(\chi+\eta,\chi-\eta)$\footnote{We will also later make a simple variable change from $\xi_+$ to $\psi$ where $\psi\propto \xi_+$ which is the variable we prefer to work with for the most part.}.This yields,
\begin{align}\label{SUSYFourierstep3}
      \mathbf{J}_s^{a_1\cdots a_{2s}}(x,\theta)&=-\frac{1}{2^{2s}\text{Vol}(GL(1,\mathbb{R}))}\int\frac{d^2\lambda d^2\Bar{\lambda}}{(2\pi)^3}e^{-i\Bar{\lambda}_a(\Bar{\mu}^a-\lambda_b x^{ab})}\lambda^{a_1}\cdots \lambda^{a_{2s}}(\lambda\cdot \Bar{\lambda})\notag\\&\times \int d\xi_+ d\xi_- d\Bar{\eta}\delta^2\left( \theta^a + \frac{\bar{\eta} \lambda^a + (\frac{\xi_+-\xi_-}{2}) \bar{\lambda}^a}{\lambda \cdot \bar{\lambda}} \right)\xi_- e^{-\frac{(\xi_+ +\xi_-)\Bar{\eta}}{4}}\hat{\mathbf{J}}_s^{+}(\lambda,\Bar{\mu},\xi_+)\notag\\
      &=\frac{1}{2^{2s}\text{Vol}(GL(1,\mathbb{R})}\int\frac{d^2\lambda d^2\Bar{\mu}d\xi_+}{(2\pi)^3}\lambda^{a_1}\cdots \lambda^{a_{2s}}\mathbf{J}_s^{+}(\lambda,\Bar{\mu},\xi_+)\mathcal{I},
\end{align}
where,
\begin{align}
\mathcal{I}=\int d^2\Bar{\lambda}(\lambda\cdot \Bar{\lambda})d\Bar{\eta}\delta^2\left( \theta^a + \frac{\bar{\eta} \lambda^a + \frac{\xi_+}{2}\bar{\lambda}^a}{\lambda \cdot \bar{\lambda}} \right) e^{-\frac{\xi_+\Bar{\eta}}{4}}e^{-i\Bar{\lambda}_a(\Bar{\mu}^a-x^{ab}\lambda_b)}
\end{align}
Let us now express the Grassmannian delta function in the basis of $\lambda$ and $\Bar{\lambda}$. This yields,
\begin{align}\label{GrassmannID1}
    \delta^2\left( \theta^a + \frac{\bar{\eta} \lambda^a + \frac{\xi_+}{2}\bar{\lambda}^a}{\lambda \cdot \bar{\lambda}} \right)=\frac{1}{(\lambda\cdot \Bar{\lambda})}\delta(\Bar{\eta}-(\Bar{\lambda}\cdot \theta))\delta(\xi_+ +2
    (\lambda\cdot \theta)).
\end{align}
Using \eqref{GrassmannID1} in \eqref{SUSYFourierstep3} and performing the $\Bar{\eta}$ integral results in,
\begin{align}
    \mathcal{I}=\delta(\xi_+ +2(\lambda\cdot \theta))\int d^2\Bar{\lambda}e^{\frac{(\lambda\cdot \theta)(\Bar{\lambda}\cdot \theta)}{2}}e^{-i\Bar{\lambda}_a(\Bar{\mu}^a-x^{ab}\lambda_b)}=\delta(\xi_+ +2(\lambda\cdot \theta))\delta^2(\Bar{\mu}^a-x^{ab}\lambda_b-\frac{i}{4}\theta^2 \lambda^a).
\end{align}
Plugging this back in \eqref{SUSYFourierstep3}, changing variables from $\xi_+\to \psi=-\frac{e^{-\frac{i\pi}{4}}}{\sqrt{2}}\xi_+$ and performing the integrals over $\Bar{\mu}$ and $\psi$ and using the projective integral formula \eqref{projID1} results in,
\begin{align}\label{SUSYFourierstep4}
    \mathbf{J}_s^{a_1\cdots a_{2s}}(x,\theta)=\int \langle \lambda d\lambda\rangle \lambda^{a_1}\cdots \lambda^{a_{2s}}\hat{\mathbf{J}}_s^{+}(\lambda,\Bar{\mu},\psi)|_{\mathcal{X}},
\end{align}
where $\mathcal{X}$ are precisely the super-incidence relations we derived earlier \eqref{Superincidence}. Therefore, \eqref{SUSYFourierstep4} is indeed the super-Penrose transform. To summarize, we have shown that the ordinary Fourier transform that takes us from position space to momentum space \eqref{superFourierTrans1} coupled with the supersymmetric Witten transform \eqref{JsetaetabarinJsetachi} yields the super-Penrose transform \eqref{SuperPenroseTransform}!

\subsection{The super-conformal generators and invariants}
So far we have discussed the super-Penrose and super-Witten transforms that respectively connect position super-space and super-spinor helicity variables to super-twistor space and discussed the relation between them. We now proceed to setup the super-conformal Ward identities in super-twistor space that we shall use to solve for Wightman functions.

Our supertwistor space carries a natural action of the super conformal group OSp($\mathcal{N}|4;\mathbb{R}$) with generators acting on super-currents as follows \cite{Bala:2025jbh},
\begin{align}\label{supertwistorgeneratorZ/W}  [\mathcal{T}^{\mathcal{A}\mathcal{B}},\mathbf{\hat{J}}_s^{\pm}(\mathcal{Z})]= \mathcal{Z}^{\mathcal{(A}}\frac{\partial}{\partial\mathcal{Z
}_{B]}}\mathbf{\hat{J}}_s^{\pm}(\mathcal{Z}), \,\,\,\,\,[\mathcal{T}^{\mathcal{A}\mathcal{B}},\mathbf{\hat{J}}_s^{\pm}(\mathcal{W})]= \mathcal{W}^{\mathcal{(A}}\frac{\partial}{\partial\mathcal{W}_{B]}}\mathbf{\hat{J}}_s^{\pm}(\mathcal{W}),
\end{align}
where the indices are symmetrized taking grading into account, please see \cite{Bala:2025jbh} for more details. Another important operator that we must consider is the super helicity operator. This operator ensures the correct behaviour of the super-currents under projective rescaling in the super-Penrose transform \eqref{SuperPenroseTransform}. For a $n-$point super-Wightman funcction, we require that for each current, 
\begin{align}\label{superhelicityZ}
        \mathcal{Z}_{i\mathcal{A}}\frac{\partial}{\partial \mathcal{Z}_{i\mathcal{A}}}\langle 0| \cdots \mathbf{\hat{J}_{s_i}^{+}}(\mathcal{Z}_i)\cdots|0\rangle = -2(s_i+1)\langle 0| \cdots \mathbf{\hat{J}_{s_i}^{+}}(\mathcal{Z}_i)\cdots|0\rangle , \notag \\
        \mathcal{Z}_{i\mathcal{A}}\frac{\partial}{\partial \mathcal{Z}_{i\mathcal{A}}}\langle 0| \cdots \mathbf{\hat{J}_{s_i}^{-}}(\mathcal{Z}_i)\cdots|0\rangle = -2(-s_i+1-\frac{\mathcal{N}}{2})\langle 0| \cdots \mathbf{\hat{J}_{s_i}^{+}}(\mathcal{Z}_i)\cdots|0\rangle,
    \end{align}
    with similar formulae for the super-dual twistor space currents that can be obtained via the super-twistor Fourier transform that connects them viz,
    \begin{align}\label{supertwistorFouriertransform}
        \mathbf{\hat{J}}_s^{\pm}(\mathcal{W})=\int d^{4|\mathcal{N}}\mathcal{Z}e^{i\mathcal{Z}\cdot\mathcal{W}}\mathbf{\hat{J}}_s^{\pm}(\mathcal{Z}).
    \end{align}
With the aid of the above equations, one can fully fix the full functional form of two and three point correlators as was done in \cite{Bala:2025jbh}. 
Given a number of Dual/twistors($\mathcal{W/Z}$), natural invariants one can form are the super-symplectic dot products of $\mathcal{Z/W}$ that lead to the set,
\begin{align}\label{supertwistordotprods}
    \{\mathcal{Z}_i\cdot \mathcal{Z}_j=-\mathcal{Z}_i^{\mathcal{A}}\Omega_{\mathcal{A}\mathcal{B}}\mathcal{Z}_j^{\mathcal{B}},\mathcal{W}_i\cdot \mathcal{Z}_j=\mathcal{W}_{i\mathcal{A}}\mathcal{Z}_j^{\mathcal{A}},\mathcal{W}_i\cdot \mathcal{W}_j=\mathcal{W}_{i\mathcal{A}}\Omega^{\mathcal{A}\mathcal{B}}\mathcal{W}_{j\mathcal{B}}\}
\end{align}
where i and j are indices labelling the (dual)super-twistors.

However, apart from these invariants, there is another invariant of $OSp(\mathcal{N}|4;\mathbb{R})$ which is the supersymmetrization of the projective delta function analzyed in the
non-supersymmetric case in \eqref{firstdefdelta4}. As an illustrative example, let us obtain this additional solution involving by considering three point twistors $\mathcal{Z}_1, \mathcal{Z}_2$ and $\mathcal{Z}_3$. Then,
           \begin{align}
                \delta^{4|\mathcal{N}}(\mathcal{Z}_3 + c_{23}\mathcal{Z}_1+c_{31} \mathcal{Z}_2) = \delta^{4}(Z_3 + c_{23} Z_1+c_{31} Z_2) \delta^{N}(\psi_3 + c_{23} \psi_1+c_{31} \psi_2)
           \end{align}
where $c_{23}, c_{31}$ are arbitrary real parameters\footnote{We could have started with three parameters and with the weights integrated over, we can show that one can reduce down to two parameters with a volume factor similar to the non-supersymmetric case as shown in \ref{app:DerivationofDelta4}.}, is also an invariant of the supergroup i.e. it is solved by $\mathcal{T}_{\mathcal{AB}}$ \eqref{supertwistorgeneratorZ/W} as a distribution when integrated against a well behaved test function for example function of Schwartz class. Alternatively, instead of applying the infinitesimal supergenerators one can check the above expression is also invariant under the finite $OSp(\mathcal{N}|4;\mathbb{R})$ transformation similar to what we did for the non-supersymmetric case in section \ref{sec:Geometry}. 

 In order to remove the dependence on these $c_{ij}$ parameters, we integrate it with weights $c_{ij}^{\alpha_{ij}}$ to obtain, 
\begin{align}
\delta^{2|\mathcal{N}}(\mathcal{Z}_1,\mathcal{Z}_2,\mathcal{Z}_3;\alpha_{23},\alpha_{31}) \equiv \int dc_{23}dc_{31} c_{23}^{\alpha_{23}}c_{31}^{\alpha_{31}} \delta^{4|\mathcal{N}}(\mathcal{Z}_3+c_{23}\mathcal{Z}_1 + c_{31} \mathcal{Z}_2),
\end{align}
 where $\alpha_{ij}$ are arbitrary parameters that will be fixed by demanding the correct projective rescaling (i.e. superhelicity) \eqref{superhelicityZ}. Similar to the non-supersymmetric case \eqref{delta4def}, one cannot satisfy projective rescaling just using this $\delta^{2|\mathcal{N}}$ only. We need to multiply this by another  OSp(4$|\mathcal{N};\mathbb{R}$) invariant viz a dot product of super-twistors. Thus, we define\footnote{Using one of the products $\mathcal{Z}_i\cdot \mathcal{Z}_j$ is sufficient at the support of the $\delta^{4|\mathcal{N}}$.},
\begin{simpleboxenv}
    \begin{align}\label{susydelta3}
    &\delta^{3|\mathcal{N}}(\mathcal{Z}_1,\mathcal{Z}_2,\mathcal{Z}_3;\alpha_{12}, \alpha_{23},\alpha_{31})\notag\\&= (-i)^{\alpha}\delta^{[\alpha_{12}]}(\mathcal{Z}_1\cdot \mathcal{Z}_2) \int dc_{23}dc_{31} c_{23}^{\alpha_{23}}c_{31}^{\alpha_{31}} \delta^{4|\mathcal{N}}(\mathcal{Z}_3+c_{23} \mathcal{Z}_1 + c_{31} \mathcal{Z}_2),
\end{align}
\end{simpleboxenv}
It is easy to check that \eqref{susydelta3} is super-conformally invariant ansatz that is solved by $\mathcal{T}^{\mathcal{AB}}$ \eqref{supertwistorgeneratorZ/W} as a distribution with the super-helicity equation \eqref{superhelicityZ} fixing the values of $\alpha_{12}, \alpha_{23}$ and $\alpha_{31}$ when used in a super-Wightman function. 

We can also generalize the above invariant to $n-$ points via,
\begin{align}\label{nptansatzdelta4susy}
    \mathcal{F}(\mathcal{Z}_1,\cdots \mathcal{Z}_n)= \int dc_1\cdots dc_{n-1}~ f(c_1,\cdots,c_{n-1})\delta^{4|\mathcal{N}}(c_1 \mathcal{Z}_1+\cdots +\mathcal{Z}_n).
\end{align}
To erase arbitrariness in the parameters $c_i$, we have integrated them on the support of a function $f(c_1,c_2,\cdots c_{n-1})$. For the projectiveness of this quantity and to be well defined in the super-Penrose transform, this function will be constrained by the super-helicity identity at each point viz \eqref{superhelicityZ}.
 
 Now, armed with these invariants, we shall discuss the construction of super-correlators in the next section.

\section{Supersymmetric Wightman functions in Twistor space}\label{sec:SUSYWightmanTwistor}
We first discuss Wightman functions of super-currents and how the projective super-delta function \eqref{susydelta3} figures in the results. Just as in the non-supersymmetric case of section \ref{Conformal delta3 in twistor space}, where the $\delta^{3}$ solutions are necessary, we shall find that that their supersymmetric counterparts that we have constructed \eqref{susydelta3} play an analogous role. 
We then extend the formalism to include super-scalars. Further, we shall see the requirement of introducing one more Grassmann coordinate for super-correlators involving scalars. However, we shall see that the super-penrose transform for super-scalars instructs us to integrate out this extra parameter, resulting in drastic simplifications. We restrict our attention to $\mathcal{N}=1$ theories although it should not be too difficult to generalize to arbitrary $\mathcal{N}$.

\subsection{Two and three point functions involving only conserved currents}
Starting with two point functions of super-currents we have the unique results \cite{Bala:2025jbh},
\begin{align}\label{susyJsJsTwopoint}
    \langle0|\mathbf{\hat{J}_{s}^{+}}(\mathcal{Z}_1)\mathbf{\hat{J}_{s}^{+}}(\mathcal{Z}_2)|0\rangle = \frac{1}{(\mathcal{Z}_1\cdot\mathcal{Z}_2)^{2s+2}}, ~~ \langle0|\mathbf{\hat{J}_{s}^{-}}(\mathcal{Z}_1)\mathbf{\hat{J}_{s}^{-}}(\mathcal{Z}_2)|0\rangle = \frac{1}{(\mathcal{Z}_1\cdot\mathcal{Z}_2)^{2(-s+\frac{1}{2})+2}}.
\end{align}
At the level of three points, the authors of \cite{Bala:2025jbh} show that one obtains \textit{homogeneous} super-correlators for net-integer spin and \textit{non-homogeneous} super-correlators for net half-integer spin.
Lets consider three integer spin supercurrents with all + helicity. Then, the correlator made out of super symplectic twistor dot products \eqref{supertwistordotprods} is given by \cite{Bala:2025jbh}, 
\begin{align}\label{3deltaSusydot}
    \langle0|\mathbf{\hat{J}_{s_1}^+ }(\mathcal{Z}_1)\mathbf{\hat{J}_{s_2}^+ }(\mathcal{Z}_2)\mathbf{\hat{J}_{s_3}^+ }(\mathcal{Z}_3)|0\rangle_h=  (-i)^{s_T} \delta^{[s_1+s_2-s_3]}(\mathcal{Z}_1\cdot \mathcal{Z}_2)\delta^{[s_2+s_3-s_1]}(\mathcal{Z}_2\cdot \mathcal{Z}_3)\delta^{[s_3+s_1-s_2]}(\mathcal{Z}_3 \cdot \mathcal{Z}_1).
\end{align}
For three half-integer super-currents, the natural answer is given in dual super twistor variables. We have,
\small
\begin{align}\label{3deltaSusydotNHhalfint}
    \langle0|\mathbf{\hat{J}_{s_1}^+ }(\mathcal{W}_1)\mathbf{\hat{J}_{s_2}^+ }(\mathcal{W}_2)\mathbf{\hat{J}_{s_3}^+ }(\mathcal{W}_3)|0\rangle_{nh} =  (-i)^{s_T} \delta^{[s_1+s_2-s_3+\frac{\mathcal{N}}{2}]}(\mathcal{Z}_1\cdot \mathcal{Z}_2)\delta^{[s_2+s_3-s_1+\frac{\mathcal{N}}{2}]}(\mathcal{Z}_2\cdot \mathcal{Z}_3)\delta^{[s_3+s_1-s_2+\frac{\mathcal{N}}{2}]}(\mathcal{Z}_3 \cdot \mathcal{Z}_1).
\end{align}
\normalsize
However, it is desirable to also have the expression for this correlator in super-twistor variables.
To this end, one can use \eqref{susydelta3} to construct this three point function. The three point function in terms of the projective super-delta function \footnote{One can also obtain this solution using the super Fourier transform \eqref{supertwistorFouriertransform} of \eqref{3deltaSusydotNHhalfint}.} is given by,

\begin{align}\label{delta4+++}
    &\langle0|\mathbf{\hat{J}_{s_1}^+ }(\mathcal{Z}_1)\mathbf{\hat{J}_{s_2}^+ }(\mathcal{Z}_2)\mathbf{\hat{J}_{s_3}^+ }(\mathcal{Z}_3)|0\rangle_{nh}\notag\\&= (-i)^{s_T}\delta^{[s_1+s_2+s_3]}(\mathcal{Z}_1\cdot\mathcal{Z}_2) \int dc_{23} dc_{31} c_{23}^{-s_2-s_3+s_1} c_{31}^{-s_3-s_1+s_2} \delta^{4|\mathcal{N}}(\mathcal{Z}_3^{\mathcal{A}}+ c_{31} \mathcal{Z}_2^{\mathcal{A}}+c_{23}\mathcal{Z}_1^{\mathcal{A}})
\end{align}

Therefore, \eqref{3deltaSusydot} and \eqref{delta4+++} are the homogeneous and non-homogeneous solution respectively valid for net-integer and net-half integer super-current three point correlators. Thus, similar to the non-supersymmetric case of section \ref{sec:WightmanConserved3pt}, we conclude from here that if we want to express the supercorrelator using only $\mathcal{Z}$ variables, one has to use $\delta^{3|\mathcal{N}}$ invariants. 

\subsection{Parity odd super-conformal Wightman functions}
Consider the supersymmetric generalization of the epsilon transform \eqref{EPTspinornotation}\footnote{The result in \eqref{EPTspinornotation}is given for spin-1 but the result is a simple generalization to arbitrary integer spin. One just needs to add extra indices to the formula and symmetrize them.}. For super-currents the epsilon transform is a simple generalization of the same and reads,
\begin{align}\label{superspacecurrentEPT}
    (\epsilon\cdot \mathbf{J}_s(x,\theta))^{a_1\cdots a_{2s}}=-i\int \frac{d^3 y}{|y-x|^2}\frac{\partial}{\partial y^{b}_{(a_1}}\mathbf{J}^{a_2\cdots a_{2s})b}_{s}(y,\theta)
\end{align}
The super-twistor space epsilon transform can then therefore be derived from \eqref{superspacecurrentEPT} and the super-Penrose transform \eqref{SuperPenroseTransform}.
\begin{tcolorbox}[
    colback=blue!5!white,
    colframe=blue!75!black,
    title=Super Epsilon Transform in Twistor Space,
    sharp corners=south,
    boxrule=0.8pt,
    arc=4pt,
    fonttitle=\bfseries
]
The Super-epsilon transform in twistor space for a spin-s current is given by,
\begin{align}\label{supertwistorEPTspins}
    (\epsilon \cdot \hat{\mathbf{J}}_s)^+(\mathcal{Z})= 
    -i \int \frac{d^3 y}{y^2} \,
    \mathcal{Z}^A \mathcal{I}_{AB} \frac{\partial}{\partial \mathcal{Z}_B}
    \hat{\mathbf{J}}_s^+(\lambda, \bar{\mu},\psi) \bigg|_{\bar{\mu}^a \to \bar{\mu}^a + y^{ab} \lambda_b},
\end{align}
where $\mathcal{I}_{AB}$ is the super-infinity twistor \eqref{superinfinitytwistor} to be defined shortly .
The parity odd super-Penrose transform is thus given by,
\begin{align}\label{parityoddSuperPenrose}
    (\epsilon\cdot \mathbf{J}_s)^{a_1\cdots a_{2s}}(x,\theta)=\int \langle \lambda d\lambda\rangle \lambda^{a_1}\cdots \lambda^{a_{2s}}(\epsilon\cdot \hat{\mathbf{J}}_s)^{+}(\mathcal{Z})|_{\mathcal{X}},
\end{align}
where $\mathcal{X}$ denote the super-incidence relations.
\end{tcolorbox}
 We have also defined the super-infinity twistor that is equal to,
\begin{align}\label{superinfinitytwistor}
    \mathcal{I}_{\mathcal{A}\mathcal{B}}=I_{AB}\oplus 0_{\mathcal{N}\times \mathcal{N}}.
\end{align}
Similarly, the parity odd super-Witten transform \eqref{wittentransparityodd} is the exact analog of its non-supersymmetric counterpart \eqref{wittentransparityodd} obtained using \eqref{JsetaetabarinJsetachi} and the momentum space version of the super-epsilon transform \eqref{superspacecurrentEPT}.
\begin{parityoddsuperwittenbox}
\begin{align}\label{JsetaetabarinJsetachiODD}
    &i\text{Sign}(\sqrt{p^2})(\epsilon\cdot\mathbf{\hat{J}}_s)^{+}(\lambda,\Bar{\lambda},\eta,\Bar{\eta})=2\int d^2\Bar{\mu}~e^{-i\Bar{\lambda}\cdot \Bar{\mu}}\int d\chi e^{-\frac{\chi \Bar{\eta}}{2}} (\chi-\eta)\mathbf{\hat{J}}_s^{+}(\lambda,\Bar{\mu},\chi+\eta),
\end{align}
where the relation between the Grassmann variables $\eta,\chi$ to the ones we use more frequently is given below \eqref{JsetaetabarinJsetachi}.
\end{parityoddsuperwittenbox}

Using \eqref{supertwistorEPTspins} and the two point even expression \eqref{susyJsJsTwopoint} we obtain,
\begin{align}\label{twopointoddSUSY}
    \langle 0|\mathbf{\hat{J}}_{s_1}^{+}(\mathcal{Z}_1)\mathbf{\hat{J}}_{s_2}^{+}(\mathcal{Z}_2)|0\rangle=\text{Sgn}(\mathcal{Z}_1\mathcal{I}\mathcal{Z}_2)\delta^{[2s+1]}(\mathcal{Z}_1\cdot \mathcal{Z}_2).
\end{align}
Similarly, using the expression of the parity even super three point function \eqref{3deltaSusydot} and the super-epsilon transform \eqref{supertwistorEPTspins} results in,
\small
\begin{align}\label{SUSY3pointODD}
  &\langle 0|\mathbf{\hat{J}}_{s_1}^{+}(\mathcal{Z}_1)\mathbf{\hat{J}}_{s_2}^{+}(\mathcal{Z}_2)\mathbf{\hat{J}}_{s_3}^{+}(\mathcal{Z}_3)|0\rangle\notag\\&=\frac{i}{2}\int dc_{12}dc_{23}dc_{31}c_{12}^{s_1+s_2-s_3}c_{23}^{s_2+s_3-s_1}c_{31}^{s_3+s_1-s_2}\text{Sgn}\big(c_{12}\langle \mathcal{Z}_1\mathcal{I}\mathcal{Z}_2\rangle+c_{31}\langle \mathcal{Z}_3 \mathcal{I}\mathcal{Z}_2\rangle\big)e^{-ic_{12}\mathcal{Z}_1\cdot \mathcal{Z}_2-ic_{23}\mathcal{Z}_2\cdot \mathcal{Z}_3-ic_{31}\mathcal{Z}_3\cdot \mathcal{Z}_1}\bigg|_{\mathcal{X}}.
\end{align}
\normalsize
Very nicely, we see that these results are obtained from something as simple as replacing twistors by super-twistors in the non-supersymmetric results \eqref{twopointodd}, \eqref{Js1s2s3TwistorNEW}. Moreover, we bring attention to the fact that these expressions also involve the super-infinity twistor \eqref{superinfinitytwistor}. One can check that these expressions are solved by the super-conformal generators \eqref{supertwistorgeneratorZ/W} as a distribution.  

\subsection{Extensions to super scalars}
The subject of the final part of this section shall be super-scalar operators with scaling dimension one.
In position space, the scalar super-field is given by \cite{Nizami:2013tpa},
\begin{align}\label{J0posspace}
    \mathbf{J}_0(x,\theta)=O_1(x)+\theta_a O_{1/2}^{a}(x)+\theta^2 O_2(x).
\end{align}
We have seen previously that super-twistor space for conserved super-currents was spanned by a single super-twistor coordinates $\mathcal{Z}^{\mathcal{A}}=(Z^A,\psi)$. However, for scalars, an additional Grassmann coordinate called $\psi_-$ is required to describe them. This can be understood from the structure of the scalar multiplet (which one can obtain by a Witten transform of the expression in \cite{Jain:2023idr} and some simple variable changes):
\begin{align}\label{scalarmultiplet}
    \mathbf{\hat{J}}_0(\mathcal{Z},\psi_-)=\frac{e^{\frac{i\pi}{4}}}{2\sqrt{2}}(\psi+\psi_-)\hat{O}_1(Z)+\frac{1}{2}\hat{O}_{1/2}^{-}(Z)+\frac{i}{2}\psi_- \psi \hat{O}_{1/2}^{+}(Z)-\sqrt{2}e^{\frac{i\pi}{4}}(\psi-\psi_-)\hat{O}_2(Z).
\end{align}

For conserved super-currents, the Penrose transform was given by \eqref{SuperPenroseTransform}. For the super-scalar \eqref{J0posspace}, we propose that it is related to \eqref{scalarmultiplet} as follows:
\begin{superpenroseboxscalar}
\begin{align}\label{SuperPenroseTransformScalar}
    \mathbf{J}_0(x,\theta)=\int \langle \lambda d\lambda\rangle\int e^{-\frac{i\pi}{4}}d\psi_-~\mathbf{\hat{J}}_0(\mathcal{Z},\psi_-)|_{\mathcal{X}},
\end{align}
where $\mathcal{X}$ are the super-incidence relations \eqref{Superincidence}.
A few words are in order. From \eqref{scalarmultiplet}, we can see that $\psi,\psi_-$ both have helicity $-\frac{1}{2}$. Therefore, $\mathbf{\hat{J}}_0(\mathcal{Z})$ behaves like a $s=-\frac{1}{2}$ current (and not like a scalar with  $s=0$!) and satisfies,
\begin{align}\label{J0projective}
    \mathbf{\hat{J}}_0(r\mathcal{Z},r\psi_-)=\frac{1}{r}\mathbf{\hat{J}}_0(\mathcal{Z},\psi_-).
\end{align}
Then under $\lambda\to r\lambda,\psi_-\to r\psi_-$, \eqref{SuperPenroseTransformScalar}, the measure transforms as $\langle\lambda d\lambda\rangle d\psi_-\to r^2 \langle \lambda d\lambda\rangle d(r\psi_-)=r^2 \langle \lambda d\lambda\rangle \frac{1}{r}d(\psi_-)=r \langle \lambda d\lambda\rangle d(\psi_-)$. This is precisely canceled out by the transformation of $J_0$ \eqref{J0projective} and thus the integral is invariant under projective rescalings.
\end{superpenroseboxscalar}
Another point is that the super-field \eqref{J0posspace} has a scaling dimension equal to one, which implies that the super-twistor space scalar is dimensionless. Note that although the super-twistor component expansion \eqref{scalarmultiplet} contains both positive and negative helicity components of the spin-1/2 operator, in the context of the super-Penrose transform \eqref{SuperPenroseTransformScalar} only the positive helicity operator contributes. This is yet again a manifestation of the fact that equal in magnitude and opposite in sign helicities in a Penrose transform contribute equally like we discussed in the paragraph above \eqref{PenroseTransformNonConservedSPins}.
Next, the super-Witten transform for scalars is identical to the spinning case \eqref{JsetaetabarinJsetachi} with $s=0$. We have,
\\
\begin{superwittenboxscalar}
\begin{align}\label{J0etaetabarinJsetachi}
    &\mathbf{\hat{J}}_0(\lambda,\Bar{\lambda},\eta,\Bar{\eta})=2\int d\chi e^{-\frac{\chi \Bar{\eta}}{2}}\mathbf{\hat{J}}_0(\lambda,\Bar{\lambda},\eta,\chi)=2\int d\chi e^{-\frac{\chi \Bar{\eta}}{2}} \hat{\mathbf{J}}_0^{+}(\lambda,\Bar{\lambda},\chi+\eta,\chi-\eta)\notag\\
    &=2\int d^2\Bar{\mu}~e^{-i\Bar{\lambda}\cdot \Bar{\mu}}\int d\chi e^{-\frac{\chi \Bar{\eta}}{2}} \mathbf{\hat{J}}_0^{+}(\lambda,\Bar{\mu},\chi+\eta,\chi-\eta).
\end{align}
The main difference from the spinning case \eqref{JsetaetabarinJsetachi} is that the scalar super-field \eqref{scalarmultiplet} depends on the difference $\chi-\eta\sim \psi_-$ in a non-trivial way unlike the spinning case \eqref{JsetaetabarinJsetachi} where it appears as an overall multiplicative quantity.
\end{superwittenboxscalar}
\subsubsection{Two point functions}
Lets start with two point functions. An ansatz consistent with projective rescaling and dimensional analysis is given by,
\begin{simpleboxenv}
\begin{align}
    \langle 0|\mathbf{\hat{J}}_0(\mathcal{Z}_1,\psi_{1-})\mathbf{\hat{J}}_0(\mathcal{Z}_2,\psi_{2-}|0\rangle=\frac{c_{\frac{1}{2}}}{(\mathcal{Z}_1\cdot \mathcal{Z}_2)}+\frac{c_{\Delta=1}\psi_{1-}\psi_{2-}}{(\mathcal{Z}_1\cdot \mathcal{Z}_2)^2}.
\end{align}
\end{simpleboxenv}
However, the supersymmetric Penrose transform \eqref{SuperPenroseTransformScalar} for scalars instructs us to integrate over the $\psi_-$ coordinates thus yielding the simpler result,
\begin{align}\label{scalarsuperpenrose2point}
    \langle 0|\mathbf{J}_0(x_1,\theta_1)\mathbf{J}_0(x_2,\theta_2)|0\rangle&=\int \langle \lambda_1d\lambda_1\rangle\langle\lambda_2d\lambda_2\rangle\int d\psi_{1-}d\psi_{2-}\bigg(\frac{c_{\frac{1}{2}}}{(\mathcal{Z}_1\cdot \mathcal{Z}_2)}+\frac{c_{\Delta=1}\psi_{1-}\psi_{2-}}{(\mathcal{Z}_1\cdot \mathcal{Z}_2)^2}\bigg)\notag\\
    &=\int \langle \lambda_1d\lambda_1\rangle\langle\lambda_2d\lambda_2\rangle\frac{c_{\Delta=1}}{(\mathcal{Z}_1\cdot \mathcal{Z}_2)^{2}}.
\end{align}
which is simply obtained from the spinning result \eqref{susyJsJsTwopoint} by setting $s=0$.
Using the super incidence relations \eqref{Superincidence} and performing a Grassmann expansion results in,
\begin{align}\label{supertwistorexpansion2pointscalar}
    &\langle 0|\mathbf{J}_0(x_1,\theta_1)\mathbf{J}_0(x_2,\theta_2)|0\rangle\notag\\
    &=\int \langle \lambda_1d\lambda_1\rangle\langle\lambda_2d\lambda_2\rangle \frac{i^{2s+2}}{(Z_1\cdot Z_2)_0^{2}}-4i\theta_{1a}\theta_{2b}\int \langle \lambda_1d\lambda_1\rangle\langle\lambda_2d\lambda_2\rangle \frac{i^{2}\lambda_1^a\lambda_2^b}{(Z_1\cdot Z_2)_0^{3}}\notag\\
    &-\frac{i}{2}(\theta_1^2+\theta_2^2)\int \langle \lambda_1d\lambda_1\rangle\langle\lambda_2d\lambda_2\rangle \frac{\langle 1 2\rangle}{(Z_1\cdot Z_2)_0^{3}}-\frac{3}{8}\theta_1^2\theta_2^2\int \langle \lambda_1d\lambda_1\rangle\langle\lambda_2d\lambda_2\rangle \frac{\langle 1 2\rangle^2}{(Z_1\cdot Z_2)_0^{4}}.
\end{align}

The first integral is the $\langle 0|O_1(x_1)O_1(x_2)|0\rangle$, the second one is $\langle O_{1/2}^{a}(x_1)O_{1/2}^{b}(x_2)\rangle$ whereas the fourth integral is $\langle 0|O_2O_2|0\rangle$. Remarkably, the Supersymmetric Penrose transform has reproduced for us twistor space expression for $\langle O_2(x_1)O_2(x_2)\rangle$ which we earlier had to obtain via a Legendre transform \eqref{O2O2O2twistor}. Thus, we see that the infinity twistor is naturally accommodated by the super-incidence relations \eqref{Superincidence}. The third integral in \eqref{supertwistorexpansion2pointscalar} is zero since it is odd under the exchange of $\lambda_1$ and $\lambda_2$. Thus, we have obtained a result that is perfectly consistent with the superfield expansion \eqref{scalarmultiplet}. However, we are neglecting the potential contact term contributions that arise from $\langle 0|O_1O_2|0\rangle$ as well as the parity odd contribution to $\langle 0|O_{1/2}O_{1/2}|0\rangle$. Let us now show how to obtain the supersymmetric version of the same.

\subsubsection{Contact terms}
 It is therefore also of interest to study and classify superconformal contact terms as pointed out in \cite{Nakayama:2019mpz}. We already analyzed non-supersymmetric two point contact terms that occur in the $\langle O_2O_1\rangle$ two point function \eqref{O2O1Twistor} and the parity odd two point functions \eqref{twopointodd}. Given the fact that the scalar multiplet \eqref{scalarmultiplet} contains $O_1,O_2$ as well as a spin-1/2 operator, let us see if supersymmetry allows for these contact term contributions. An ansatz that we can write down taking inspiration from \eqref{O2O1Twistor} and \eqref{twopointodd} is as follows:
\begin{simpleboxenv}
\begin{align}\label{susytwopointcontact}
    \langle 0|\mathbf{\hat{J}}_0(\mathcal{Z}_1,\psi_{1-})\mathbf{\hat{J}}_0(\mathcal{Z}_2,\psi_{2-}|0\rangle= \psi_{1-}\psi_{2-} \text{Sgn}(\langle \mathcal{Z}_1 \mathcal{I}\mathcal{Z}_2\rangle)\delta^{[1]}(\mathcal{Z}_1\cdot\mathcal{Z}_2).
\end{align}
\end{simpleboxenv}
The above ansatz has correct projective properties \eqref{J0projective} and dimensionality.
Plugging this into the super-scalar super-Penrose transform \eqref{SuperPenroseTransformScalar} results in,
\begin{align}\label{J0J0contactsuperpenrose}
    \langle 0|\mathbf{J}_0(x_1,\theta_1)\mathbf{J}_0(x_2,\theta_2)|0\rangle_{\text{contact}}=\int \langle\lambda_1 d\lambda_1\rangle\langle\lambda_2 d\lambda_2\rangle \text{Sgn}(\langle \mathcal{Z}_1 \mathcal{I}\mathcal{Z}_2\rangle)\delta^{[1]}(\mathcal{Z}_1\cdot\mathcal{Z}_2)|_{\mathcal{X}}.
\end{align}
Using the super-incidence relations \eqref{Superincidence}, expanding \eqref{J0J0contactsuperpenrose} into components like in \eqref{supertwistorexpansion2pointscalar} and matching with the position space super-field expansion obtained using \eqref{J0posspace} shows that the correct two point contact terms \eqref{O2O1Twistor} and \eqref{twopointodd} are reproduced correctly. In fact, they combine very nicely thanks to supersymmetry to become the following supersymmetric contact term,
\begin{align}
    \langle 0|\mathbf{J}_0(x_1,\theta_1)\mathbf{J}_0(x_2,\theta_2)|0\rangle_{\text{contact}}=\delta^2(\theta_1-\theta_2)\delta^3(x_1-x_2).
\end{align}
We leave to the future, a more complete and systematic analysis of super-symmetric contact terms and their applications.
\subsubsection{Three point functions}
Let us now move onto three point functions involving scalar super-fields. First, consider a supercorrelator consisting of one scalar in $\mathcal{N}=1$ and two conserved currents with + helicity: The most general ansatz consistent with \eqref{SuperPenroseTransform} and \eqref{SuperPenroseTransformScalar} is,
\begin{simpleboxenv}
  \begin{align}\label{susyJs1Js2J0}
      &\langle 0| \hat{\mathbf{J}}^+_{s_1}(\mathcal{Z}_1) \mathbf{\hat{J}}^+_{s_2}(\mathcal{Z}_2)\mathbf{\hat{J}}_0(\mathcal{Z}_3,\psi_{3-})| 0\rangle = \alpha_1 \psi_{3-}\delta^{[s_1+s_2]}(\mathcal{Z}_1 \cdot \mathcal{Z}_2 )\delta^{[-s_1+s_2]}(\mathcal{Z}_2 \cdot \mathcal{Z}_3 )\delta^{[s_1-s_2]}(\mathcal{Z}_3 \cdot \mathcal{Z}_1 ) \notag \\  
      & + \alpha_2 \int dc_{12} dc_{23} dc_{31} c_{12}^{s_1+s_2} c_{23}^{s_1-s_2}c_{31}^{-s_1+s_2}  e^{i c_{12} \mathcal{Z}_1\cdot\mathcal{Z}_2} \delta^{4|1}(\mathcal{Z}_3^{\mathcal{A}}+ c_{23} \mathcal{Z}_1^{\mathcal{A}} + c_{31} \mathcal{Z}_2^{\mathcal{A}}),
  \end{align}
  \end{simpleboxenv}
  where $\alpha_1$ and $\alpha_2$ are coefficients that contain information about the OPE coefficients and can be read off from the super-field expansions of the scalar and currents. The important point to note is that the we get a linear combination the $\delta\delta\delta$ and $\delta^4$ solutions which are individually super-conformal invariant. However, in the context of the super-Penrose transform \eqref{SuperPenroseTransformScalar}, one should integrate out $\psi_{3-}$ and therefore only the second term in \eqref{susyJs1Js2J0} is important to keep.

The results with two scalars and one conserved current with + helicity is given by,
\begin{simpleboxenv}
\begin{align}
     \langle0| \hat{\mathbf{J}}^+_{s_1}&(\mathcal{Z}_1)\hat{\mathbf{J}}_0(\mathcal{Z}_2) \hat{\mathbf{J}}_0(\mathcal{Z}_3)| 0\rangle = \alpha_1 \delta^{[s_1]}(\mathcal{Z}_1\cdot\mathcal{Z}_2)\delta^{[-1-s_1]}(\mathcal{Z}_2\cdot\mathcal{Z}_3)\delta^{[s_1]}(\mathcal{Z}_3\cdot\mathcal{Z}_1) \notag \\
    & +\alpha_2 \psi_{2-}\psi_{3-} \delta^{[s_1]}(\mathcal{Z}_1\cdot\mathcal{Z}_2)\delta^{[-s_1]}(\mathcal{Z}_2\cdot\mathcal{Z}_3)\delta^{[s_1]}(\mathcal{Z}_3\cdot\mathcal{Z}_1) \notag \\
    & + \alpha_3 \psi_{2-} \int dc_{12}dc_{23}dc_{31} c_{12}^{s_1} c_{23}^{s_1}c_{31}^{-s_1} e^{i c_{12} \mathcal{Z}_1\cdot\mathcal{Z}_2 } \delta^4(c_{23}\mathcal{Z}_1^{\mathcal{A}}+ c_{31} \mathcal{Z}_2^{\mathcal{A}} + \mathcal{Z}_3^{\mathcal{A}}  ) \notag \\
     &+ \alpha_4 \psi_{3-} \int dc_{12}dc_{23} dc_{31} c_{12}^{-1-s_1}c_{23}^{s_1}c_{31}^{-s_1} e^{i c_{23} \mathcal{Z}_2\cdot\mathcal{Z}_3} \delta^4(-\mathcal{Z}_1^{\mathcal{A}} +c_{12}\mathcal{Z}_2^{\mathcal{A}} - c_{31} \mathcal{Z}_3^{\mathcal{A}}),
\end{align}
\end{simpleboxenv}
where the $\alpha_i$ are related to the OPE coefficients. However, the scalar super-Penrose transform \eqref{SuperPenroseTransformScalar} tells us that only the coefficient of $\alpha_2$ suffices to recover the super-position space result.
Similarily, the results with three scalars can be obtained in a similar way and is given by- 
\begin{simpleboxenv}
\begin{align}\label{scalarsusy3pt}
   \langle 0|\hat{\mathbf{J}}_0(\mathcal{Z}_1) &\hat{\mathbf{J}}_0(\mathcal{Z}_2)\hat{\mathbf{J}}_0(\mathcal{Z}_3) =  \alpha_1 \int dc_{12}dc_{23}dc_{31} e^{i c_{12} \mathcal{Z}_1\cdot \mathcal{Z}_2} \delta^{4|1}(c_{23} \mathcal{Z}_2^{\mathcal{A}}- c_{31} \mathcal{Z}_1^{\mathcal{A}}+\mathcal{Z}_3^{\mathcal{A}}) \notag \\
   + & \alpha_2  \psi_{1-} \delta(\mathcal{Z}_1\cdot\mathcal{Z}_2)\delta^{[-1]}(\mathcal{Z}_2\cdot\mathcal{Z}_3)\delta(\mathcal{Z}_3\cdot\mathcal{Z}_1) \notag \\
    + & \alpha_3  \psi_{2-} \delta(\mathcal{Z}_1\cdot\mathcal{Z}_2)\delta(\mathcal{Z}_2\cdot\mathcal{Z}_3)\delta^{[-1]}(\mathcal{Z}_3\cdot\mathcal{Z}_1) \notag \\
  + & \alpha_4  \psi_{3-} \delta(\mathcal{Z}_1\cdot\mathcal{Z}_2)\delta^{[-1]}(\mathcal{Z}_2\cdot\mathcal{Z}_3)\delta(\mathcal{Z}_3\cdot\mathcal{Z}_1) \notag \\
    + &\alpha_5  \psi_{1-}\psi_{2-} \int dc_{12}dc_{23 }dc_{31} e^{i c_{12} \mathcal{Z}_1\cdot \mathcal{Z}_2} \delta^{4|1}(c_{23} \mathcal{Z}_1^{\mathcal{A}} + c_{31} \mathcal{Z}_2^{\mathcal{A}} + \mathcal{Z}_3^{\mathcal{A}}) \notag \\
    + &\alpha_6  \psi_{1-}\psi_{3-} \int dc_{12}dc_{23 }dc_{31} e^{i c_{12} \mathcal{Z}_1\cdot \mathcal{Z}_2} c_{31}^{-1} \delta^{4|1}(c_{23} \mathcal{Z}_1^{\mathcal{A}} + c_{31} \mathcal{Z}_2^{\mathcal{A}} + \mathcal{Z}_3^{\mathcal{A}}) \notag \\
    + &\alpha_7  \psi_{2-}\psi_{3-} \int dc_{12}dc_{23 }dc_{31} e^{i c_{12} \mathcal{Z}_1\cdot \mathcal{Z}_2} c_{23}^{-1} \delta^{4|1}(c_{23} \mathcal{Z}_1^{\mathcal{A}} + c_{31} \mathcal{Z}_2^{\mathcal{A}} + \mathcal{Z}_3^{\mathcal{A}}) \notag \\
    + & \alpha_8 \psi_{1-}\psi_{2-}\psi_{3-} \delta(\mathcal{Z}_1\cdot \mathcal{Z}_2)\delta(\mathcal{Z}_2\cdot \mathcal{Z}_3)\delta(\mathcal{Z}_3\cdot\mathcal{Z}_1).
\end{align}
\end{simpleboxenv}
The eight $\alpha_i$ (which can be reduced by imposing permutation symmetry to $4$) encode the information of the OPE coefficients. Similar to the previous cases, the only term that contributes to the super-Penrose transform is the coefficient of $\alpha_8$. However, one must keep in mind that the super-Witten transform \eqref{JsetaetabarinJsetachi} requires all eight terms to reproduce the correct Grassmann spinor helicity expression.

This ends our discussion on super-scalars. The main message we want to convey is that although the expressions for three point functions involves many terms like in \eqref{scalarsusy3pt}, only one term contributes to the super-Penrose transform \eqref{SuperPenroseTransformScalar}. It would however be interesting to see whether the other terms have any role to play. Similar to the non-supersymmetric case of subsection \ref{subsec:genscalar} and \ref{subsec:nonconsTwistor}, one can derive and develop the super-Penrose transform for general $\Delta$ super-fields as well as non-conserved operators. We leave such an exercise to the future.

\section{Discussion and future directions}\label{sec:Discussion}
In this paper we have discussed and constructed Sp$(4)$ invariants in twistor space for 3d CFT. We find two main classes of invariants: (1) Those that are made out of symplectic dot products of twistors and projective delta functions enforcing collinearity of twistors, (2) Those that also involve the infinity twistor of $\mathbb{R}^{2,1}$. We discussed how they occur in conformal two and three point Wightman functions of conserved currents and scalars and in particular, how the latter invariants allow us to extend the twistor space construction for arbitrary $\Delta$ scalars, generic spinning primaries as well as parity odd Wightman functions. We then derived the supersymmetric Penrose transform and discussed its relation to the supersymmetric Witten transform via the Fourier transform and employing the Faddeev Popov method. Finally, we constructed the OSp$(\mathcal{N}|4)$ invariants and some prototypical examples of Wightman functions involving scalar super-fields for $\mathcal{N}=1$ theories. One of the main messages of this paper is that to accomodate generic primary operators in twistor space, the infinity twistor of $\mathbb{R}^{2,1}$ must be accommodated into the analysis. Although this may seem paradoxial as the infinity twistor breaks conformal invariance down to its Poincare subgroup, we have shown that the representation in which generic operators transform in twistor space is non-local and naturally involves the infinity twistor and with these generators, conformal invariance is indeed present. We have presented and derived many different versions of the Penrose and Witten transforms in this paper. They are summarized for the convenience of the reader in table \ref{tab:penrose-witten}.

\begin{table}[h!]
\centering
\renewcommand{\arraystretch}{1.4} 
\begin{tabular}{|l|l|l|}
\hline
\textbf{Operator} & \textbf{Penrose Transform} & \textbf{Witten Transform} \\ \hline
Conserved Currents & \eqref{PenroseTransform} & \eqref{WittenTransform} \\ \hline
Dimension $\Delta$ Scalars & \eqref{ODeltaPenrose} &  \eqref{ODeltaWitten}\\ \hline
Epsilon transformed Current & \eqref{parityoddPenrosetransform1}& \eqref{wittentransparityodd}\\ \hline
Non Conserved Current & \eqref{PenroseTransformNonConservedSPins} & \eqref{nonconsWittenTransformspins}\\ \hline
Conserved Super-Current&\eqref{SuperPenroseTransform}  & \eqref{JsetaetabarinJsetachi}\\ \hline
Epsilon transformed Super-current&\eqref{supertwistorEPTspins}  & \eqref{JsetaetabarinJsetachiODD}\\ \hline
$\Delta=1$ Super-Scalar& \eqref{SuperPenroseTransformScalar}&\eqref{J0etaetabarinJsetachi}\\
\hline
\end{tabular}
\caption{The Penrose and Witten transforms for various operators.}
\label{tab:penrose-witten}
\end{table}

There are a number of interesting future directions, some of which we discuss below.

\subsection*{Future directions}
One immediate problem is to use the formalism of this paper to bootstrap twistor space three point correlators of involving at least one generic primary operator. This should also help setting up the higher point conformal bootstrap, particularly for spinning correlators since twistor space seems like an appropriate stage to study spinning correlators as they are as simple as their scalar counterparts\footnote{If anything, conserved current correlators seem simpler than their scalar counterparts in twistor space!}. It would be interesting to perform the twistor space version of the analysis of \cite{Costa:2011dw,Costa:2011mg} as a starting point. Developing the machinery of Weight shifting and spin raising operators \cite{Karateev:2017jgd,Baumann:2019oyu} in twistor space also presents an interesting avenue to explore.  Further, as discussed earlier, it would be interesting to impose the constraints of dual conformal and Yangian symmetry given the Sp(4) and OSp($\mathcal{N}|4$) projective delta functions and study their implications. In the context of four dimensional (supersymmetric) scattering amplitudes in $\mathcal{N}=4$ super Yang-Mills theory, it turns out that there is a unique Yangian invariant quantity that one can construct \cite{Drummond:2010qh,Drummond:2010uq,Korchemsky:2010ut,Elvang:2013cua}. We hope to return in the future to perform an analogous analysis in the context of $CFT_3$ and study the implications of Yangian invariance. Another important direction to pursue is to develop the computation of the super-Penrose transform to obtain the super-position space building blocks which shall be the subject of \cite{Mazumdar:2025new}. It is also desirable to extend our formalism to extended supersymmetry and also accommodate non conserved currents. The latter would also be interesting from the point of view of studying BPS states in CFT. Finally, the recent paper \cite{Note:2025djf} discusses the presence of sign functions in twistor space and how they are artifacts of working in Klein space. The author shows that these factors are absent in the complexified twistor space and thus it begs the question whether the ubiquitous sign functions that occur in the results of this paper such as in the parity odd correlators of subsection \ref{subsec:parityoddsubsec} are also artifacts of working in the real twistor space associated to $\mathbb{R}^{2,1}$.

\acknowledgments
We would like to thank Mohd Ali, Nipun Bhave, Sachin Jain, Deep Mazumdar, Vibhor Singh and Brijesh Thakkar for useful discussions. We would especially like to thank Sachin Jain for his insightful comments on several subtle points. AB acknowledges a UGC-NET fellowship.

\appendix

\section{Notation and Conventions}\label{app:Notation}
We work in $\mathbb{R}^{2,1}$ with the metric,
\begin{align}
    ds^2=-dt^2+dx^2+dz^2.
\end{align}
Vector indices are raised and lowered using the metric $\eta_{\mu\nu}$ and its inverse via,
\begin{align}
    v^\mu=\eta^{\mu\nu}v_\nu, v_{\mu}=\eta_{\mu\nu}v^\nu.
\end{align}
We can trade a vector for a $SL(2,\mathbb{R})$ bi-spinor using the Pauli matrices via,
\begin{align}
    x_{ab}=x^\mu (\sigma_\mu)_{ab}.
\end{align}
$SL(2,\mathbb{R})$ spinor indices are raised and lowered using the 2d Levi-Civita symbol via,
\begin{align}
    \lambda_a=\epsilon_{ba}\lambda^b,\lambda^a=\epsilon^{ab}\lambda_b.
\end{align}

The explicit forms of these matrices are given by,
\begin{align}
    (\sigma^x)^a_b=\begin{pmatrix}
        0&&1\\
        1&&0
    \end{pmatrix},(\sigma^t)^a_b=\begin{pmatrix}
        0&&-1\\
        1&&0
    \end{pmatrix},(\sigma^z)^a_b=\begin{pmatrix}
        1&&0\\
        0&&-1
    \end{pmatrix},\epsilon_{ab}=\epsilon^{ab}=\begin{pmatrix}
        0&&-1\\
        1&&0
    \end{pmatrix}.
\end{align}
For $SL(2,\mathbb{R})$ spinor dot products we use the notation,
\begin{align}
    \lambda_a\rho^a=\lambda\cdot \rho.
\end{align}
In twistor space, we raise and lower indices using the Sp$(4)$ invariant\footnote{ We often denote the group Sp$(4;\mathbb{R})$ simply by Sp$(4)$ for convinience.}$\Omega$ \eqref{Omega} via,
\begin{align}
    Z^A=\Omega^{AB}Z_B,Z_B=\Omega_{BA}Z^B.
\end{align}
We also define the symplectic dot products and products involving the infinity twistor as follows:
\begin{align}
        Z_i\cdot Z_j=-Z_i^A \Omega_{AB} Z_j^B, \langle Z_i I Z_j\rangle=Z_i^A I_{AB}Z_j^B.
\end{align}

The contractions of supertwistors $\mathcal{Z}^\mathcal{A}$ and $\mathcal{W}_\mathcal{A}$ are done using the invariant tensor $\Omega_{\mathcal{AB}}$ of the $OSp(\mathcal{N}|4;\mathbb{R})$ super group given by,
\begin{align}
\Omega_{\mathcal{A}\mathcal{B}}\equiv\Omega^{\mathcal{A}\mathcal{B}}=\begin{pmatrix}
        0&\delta_a^{b}&0\\
        -\delta^b_{a} & 0&0\\
        0&0&\mathbb{I}_{\mathcal{N}\cross \mathcal{N}}
    \end{pmatrix}.
\end{align}
The contraction of the invariant tensor is given by,
\begin{align}
\Omega_{\mathcal{A}\mathcal{B}}\Omega^{\mathcal{A}\mathcal{C}}= \delta^{\mathcal{C}}_{\mathcal{B}}.
\end{align} 
In terms of the operators that we use in this paper, we follow the following notation. $J_s^{\pm}(\lambda,\Bar{\lambda})$ denotes the $\pm$ helicity component of the momentum space current $J_s^{a_1\cdots a_{2s}}(p^\mu)$ which are obtained by contracting with the polarization spinors $\eqref{SHvariables}$ like in \eqref{Jspm}. We denote the currents rescaled with $\frac{1}{|p|^{s-1}}$ with a hat like $\hat{J}_s^{\pm}(\lambda,\Bar{\lambda})$. Since the Witten transform \eqref{WittenTransform} makes use of these rescaled currents we define the twistor space currents with a hat as in $\hat{J}_s^{\pm}(\lambda,\Bar{\mu})=\hat{J}_s^{\pm}(Z)$. For the supersymmetric setting, we follow the same notation as above but with the super-operators denoted in bold like $\hat{\mathbf{J}}_s^{\pm}(\mathcal{Z})$. In dealing with Wightman functions, we often ignore overall coefficients since our aim is only to obtain the functional form of the results and the overall coefficients can be restored later.

 We also frequently use the integral identity,
 \begin{align}
     \int_{-\infty}^{\infty} \frac{dx}{x}e^{ia x}\propto \text{Sgn}(a),
 \end{align}
 and related ones which are to be interpreted in a regularized sense like in \cite{Mason:2009sa}.

 Finally, the notation $|_X$ denotes imposing the incidence relation \eqref{IncidenceRelation} whereas $|_{\mathcal{X}}$ denotes imposing the super-incidence relations \eqref{Superincidence}.

\section{Dimensional reduction of Twistor space: $SL(4,\mathbb{R})\to \text{Sp}(4,\mathbb{R})$}\label{app:4dto3d}
In this appendix, we shall start with the generators of $SL(4,\mathbb{R})$, which is the 4d conformal group and dimensionally reduce to three dimensions and obtain the Sp$(4)$ generators \eqref{TABZ}. For $U\in SL(4,\mathbb{R})$ we have,
\begin{align}
    \text{Det}[U]=1.
\end{align}
Exponentiating $U=e^{i\alpha G}$ with $G\in \mathfrak{sl}(4,\mathbb{R})$ we obtain,
\begin{align}
    \text{Det}[e^{i\alpha G}]=e^{i\alpha \text{Tr}(G)}=1\implies \text{Tr}(G)=0.
\end{align}
Let us now construct a projection of $\mathfrak{sl}(4,\mathbb{R})$ onto the subalgebra $\mathfrak{sp}(4,\mathbb{R})$ using the symplectic form $\Omega$. In index notation we have\footnote{Note that this is very reminiscent of how spinor helicity variables are dimensionally reduced like from Klein space to $\mathbb{R}^{2,1}$, see \cite{Bala:2025gmz}.},
\begin{align}\label{3dTfrom4dG}
    T^A_B=\Omega_{EB}\big(\Omega^{C(A}G^{E)}_C\big)=\Omega_{EB}T^{AE}.
\end{align}
Let us exponentiate this quantity and check whether it satisfies the Sp$(4)$ group property viz,
\begin{align}\label{sp4condition}
    M\in Sp(4)\implies M^T \Omega M=\Omega\implies T^T \Omega+\Omega T=0~\text{for}~T\in \mathfrak{sp}(4,\mathbb{R}).
\end{align}
Putting $M=e^{i\alpha T}$ and using the relation \eqref{3dTfrom4dG}, it is easy to see \eqref{sp4condition} is indeed satisfied. Let us now see how to derive the particular representation of Sp$(4,\mathbb{R})$ that acts on twistor space from the twistor space representation of the generators of $SL(4,\mathbb{R})$. The $SL(4,\mathbb{R})$ generators that act on twistor space associated to $\mathbb{R}^{2,2}$ \footnote{For instance, these can be found by starting with the $GL(4,\mathbb{R})$ generators in \cite{Adamo:2017qyl} and subtracting out the trace part}. are given by,
\begin{align}
    G^A_B=Z^A\frac{\partial}{\partial Z^B}-\frac{1}{4}\delta^{A}_{B}Z^C\frac{\partial}{\partial Z^C}.
\end{align}
They obey the $\mathfrak{sl}(4,\mathbb{R})$ algebra viz,
\begin{align}
[G^A_B,G^C_D]=\delta^{C}_{B}G^{A}_{D}-\delta^{A}_{D}G^{C}_{B}.
\end{align}
Using the definition \eqref{3dTfrom4dG} with,
\begin{align}
T^{AB}=\Omega^{C(A}T^{B)}_C=Z^{(A}\frac{\partial}{\partial Z_{B)}},
\end{align}
it is easy to show that,
\begin{align}
    [T^{AB},T^{CD}]\propto \Omega^{AC}T^{BD}+\Omega^{AD}T^{BC}+\Omega^{BC}T^{AD}+\Omega^{BD}T^{AC},
\end{align}
which is exactly the Lie-algebra of $\mathfrak{sp}(4,\mathbb{R})$.

What we see via the above representation theory exercise is that the twistor space associated to $\mathbb{R}^{2,1}$ is a subspace of the twistor space of the twistor space associated to $\mathbb{R}^{2,2}$. Both of these are subsets of $\mathbb{RP}^{3}$ where the latter allows for general volume preserving transformations whereas the former restricts to those transformations that preserve the symplectic form.

\section{Projective VS non-projective integrals}\label{app:ProjectiveVSnonproj}
In this section, we derive two useful formulae that relate projective integrals to ordinary nonprojective integrals by quotienting by $\text{Vol}(GL(1,\mathbb{R})$. This volume factor is given by the following integral,
\begin{align}\label{volumeGL1R}
    \text{Vol}(GL(1,\mathbb{R})=\int_{-\infty}^{\infty}\frac{dc}{|c|}.
\end{align}
\subsection{From $\mathbb{RP}^1$ to $\mathbb{R}^2$}
Consider the following quantity
\begin{align}\label{Penrosetypeintegral}
    \int \langle\lambda d\lambda\rangle f(\lambda),
\end{align}
such that,
\begin{align}\label{fprop1}
    f(r \lambda)=\frac{1}{r^2}f(\lambda).
\end{align}
The Penrose transform \eqref{PenroseTransform} is an example of such an integral with $f(\lambda)\sim \lambda^{a_1}\cdots \lambda^{a_{2s}}J_s^{+}(\lambda,\Bar{\mu})|_X$. The integral in \eqref{Penrosetypeintegral} is applied in all of $\mathbb{RP}^1$. For instance we can locally choose $\lambda_a=(\lambda_1,\lambda_2)=(1,\xi)$ such that \eqref{Penrosetypeintegral} becomes,
\begin{align}\label{projectiveint1step2}
    \int \langle \lambda d\lambda\rangle f(\lambda_1,\lambda_2)=\int_{-\infty}^{\infty}d\xi f(1,\xi).
\end{align}
We now want to show that,
\begin{align}\label{claim1step1}
    \int \langle\lambda d\lambda\rangle f(\lambda)=\frac{1}{\text{Vol}(GL(1,\mathbb{R}))}\int d^2\lambda f(\lambda).
\end{align}
Lets write out the right hand-side of \eqref{claim1step1} explicitly:
\begin{align}
    \frac{1}{\text{Vol}(GL(1,\mathbb{R}))}\int d^2\lambda f(\lambda)=\frac{1}{\text{Vol}(GL(1,\mathbb{R}))}\int d\lambda_1\wedge d\lambda_2 f(\lambda_1,\lambda_2).
\end{align}
Now perform a variable change,
\begin{align}
    \lambda_1=\rho_1,\lambda_2=\rho_1 \rho_2\implies d\lambda_1\wedge d\lambda_2=|\rho_1|d\rho_1\wedge d\rho_2.
\end{align}
We thus have,
\begin{align}
    \frac{1}{\text{Vol}(GL(1,\mathbb{R}))}\int d\lambda_1\wedge d\lambda_2 f(\lambda_1,\lambda_2)&=\frac{1}{\text{Vol}(GL(1,\mathbb{R}))}\int d\rho_1\wedge d\rho_2 |\rho_1| f(\rho_1,\rho_1\rho_2)\notag\\
    &=\frac{1}{\text{Vol}(GL(1,\mathbb{R}))}\int d\rho_1\wedge d\rho_2 |\rho_1| \frac{1}{\rho_1^2}f(1,\rho_2)
\end{align}
where we used the projective property of the function $f$ \eqref{fprop1}. The integral now factorizes and the one above $\rho_1$ completely cancels out the volume factor \eqref{volumeGL1R}. Relabeling $\rho_2=\xi$ then yields \eqref{projectiveint1step2} thus proving \eqref{claim1step1}.

\subsection{From $\mathbb{R}^3$ to $\mathbb{R}^2\times \mathbb{R}^2$}
We now consider the following integral,
\begin{align}
    \int d^3 x f(x).
\end{align}
We now write $x$ as a bispinor by contracting with the Pauli matrices,
\begin{align}
    x^{ab}=\frac{\lambda^a \rho^b+\lambda^b \rho^a}{2}.
\end{align}
By definition we have a redundancy,
\begin{align}\label{xinlambdarhored}
    \lambda\to \frac{1}{r}\lambda, \rho\to r \rho\implies x\to x.
\end{align}
We want to show that,
\begin{align}\label{integralclaim2}
    \int d^3 x f(x)=\frac{1}{4\text{Vol}(GL(1,\mathbb{R}))}\int d^2\lambda d^2\rho |\lambda\cdot \rho| f(\lambda,\rho),
\end{align}
where the redundancy \eqref{xinlambdarhored} implies that,
\begin{align}\label{rescalingflambdarho}
    f(\frac{\lambda}{r},r\rho)=f(\lambda,\rho).
\end{align}
To prove this statement, let us start with the RHS of \eqref{integralclaim2}. Writing it out explicitly we have,
\small
\begin{align}\label{lambdarhointstep2}
    \frac{1}{2\text{Vol}(GL(1,\mathbb{R}))}\int d^2\lambda d^2\rho |\lambda\cdot \rho| f(\lambda,\rho)=\frac{1}{\text{Vol}(GL(1,\mathbb{R}))}\int d\lambda_1\wedge d\lambda_2\wedge d\rho_1\wedge d\rho_2|\lambda_1\rho_2-\lambda_2\rho_1| f(\lambda_1,\lambda_2,\rho_1,\rho_2).
\end{align}
\normalsize
We now make a variable change $\lambda_1=\frac{\chi_1}{\rho_2},\lambda_2=\frac{\chi_2}{\rho_2},\rho_1=\nu_1 \rho_2, \rho_2=\nu_2$. \eqref{lambdarhointstep2} along with the rescaling property \eqref{rescalingflambdarho} becomes,
\begin{align}\label{lambdarhointstep3}
    &\frac{1}{\text{Vol}(GL(1,\mathbb{R}))}\int\frac{1}{|\nu_2|} d\chi_1\wedge d\chi_2\wedge d\nu_1\wedge d\nu_2|\chi_1-\chi_2 \nu_1| f(\chi_1,\chi_2,\nu_1,1)\notag\\
    &=\int d\chi_1\wedge d\chi_2\wedge d\nu_1 |\chi_1-\chi_2\nu_1|f(\chi_1,\chi_2,\nu_1,1).
\end{align}
Let us now deal with the LHS of \eqref{integralclaim2} and prove that it is also equal to \eqref{lambdarhointstep3}. We have,
\begin{align}
    &\int d^3 x f(x)=\frac{1}{2}\int dx^{11}\wedge dx^{12}\wedge dx^{22} f(x_{11},x_{12},x_{22})\notag\\
    &\frac{1}{4}\int d\lambda_1\wedge d\lambda_2\wedge d\rho_1|\lambda_1-\lambda_2\rho_1|f(\lambda_1,\lambda_2,\rho_1,1).
\end{align}
To go from the first line to the second we used the relation between $x$ and $\lambda,\rho$ viz \eqref{xinlambdarhored} and chose to set $\rho_2=1$. Relabeling $\lambda_1\to \chi_1,\lambda_2\to \chi_2$ and $\rho_1\to \nu_1$ yields \eqref{lambdarhointstep3}, thus proving the formula \eqref{integralclaim2}.

\section{Derivation of three point Sp$(4)$ projective delta function}\label{app:DerivationofDelta4}
Consider three twistors $Z_1,Z_2,Z_3$. The object,
\begin{align}
    \delta^4(c_{23}Z_1+c_{31}Z_2+c_{12}Z_3),
\end{align}
is invariant under an $Sp(4)$ transformation \eqref{TABZ}. To see this, consider a finite $Sp(4)$ transformation $M\in Sp(4)$. The delta function transforms as follows:
\begin{align}\label{delta4Sp4invariance3ptsa}
    \delta^4(c_{23}Z_1+c_{31}Z_2+c_{12}Z_3)&\to  \delta^4(c_{23}M Z_1+c_{31}M Z_2+c_{12}M Z_3)=\frac{1}{\text{Det}(M)}\delta^4(c_{23}Z_1+c_{31}Z_2+c_{12}Z_3)\notag\\&=\delta^4(c_{23}Z_1+c_{31}Z_2+c_{12}Z_3), 
\end{align}
since $\text{Det}(M)=1$ for $Sp(4)$ transformations\footnote{An $Sp(4)$ transformation $M$ satisfies $M^T \Omega M=\Omega$ where $\Omega$ is the symplectic form. Taking determinant on both sides yields $\text{Det}(M)^2=1$ which implies $\text{Det}(M)=1$ for the connected component of the symplectic group.}.
To remove the dependence on the arbitrary real parameters $c_{ij}$ we integrate over them with weights $f(c_{12},c_{23},c_{31})$ to obtain
\begin{align}\label{delta4ansatzstep1a}
    \int dc_{12}dc_{23}dc_{31}f(c_{12},c_{23},c_{31})\delta^4(c_{23}Z_1+c_{31}Z_2+c_{12}Z_3).
\end{align}
Since we are interested in constructing a function that has nice projective properties with respect to all three twistors\footnote{More precisely, this quantity should be rescaled only by an overall amount when any among $Z_1,Z_2,Z_3$ are rescaled. With some hindsight, we also disallow sign factors of these parameters as they lead to the wrong spinor helicity variables expression after a half-Fourier transform.}, we see that we require the function $f(c_{12},c_{23},c_{31})$ to be separable.
\begin{align}\label{fansatzstep1a}
    f(c_{12},c_{23},c_{31})\propto c_{12}^{\beta}c_{23}^{\alpha_{23}}c_{31}^{\alpha_{31}}.
\end{align}
Substituting \eqref{fansatzstep1a} in \eqref{delta4ansatzstep1a} yields,
\begin{align}
    \int dc_{12}dc_{23}dc_{31}c_{12}^{\beta}c_{23}^{\alpha_{23}}c_{31}^{\alpha_{31}}\delta^4(c_{23}Z_1+c_{31}Z_2+c_{12}Z_3).
\end{align}
However, this integral as it stands is ill-defined since one can re-define $c_{23}$ and $c_{31}$ to pull out an overall factor of $\int dc_{12}c_{12}^{\beta+\alpha_{23}+\alpha_{31}-2}$ which is either zero or infinity. We thus divide by this factor and consider,
\begin{align}\label{delta4ansatzstep2a}
    \int dc_{23}dc_{31}c_{23}^{\alpha_{23}}c_{31}^{\alpha_{31}}\delta^4(c_{23}Z_1+c_{31}Z_2+Z_3).
\end{align}
Under rescaling $Z_1,Z_2$ and $Z_3$ by amounts $r_1,r_2$ and $r_3$ we obtain,
\small
\begin{align}\label{delta4ansatzstep3a}
    \int dc_{23}dc_{31}c_{23}^{\alpha_{23}}c_{31}^{\alpha_{31}}\delta^4(c_{23}Z_1+c_{31}Z_2+Z_3)\to \frac{\text{Sgn}(r_1)\text{Sgn}(r_2)}{r_1^{1+\alpha_{23}}r_2^{1+\alpha_{31}}r_3^{2-\alpha_{23}-\alpha_{31}}}\int dc_{23}dc_{31}c_{23}^{\alpha_{23}}c_{31}^{\alpha_{31}}\delta^4(c_{23}Z_1+c_{31}Z_2+Z_3).
\end{align}
\normalsize
We thus see that if $r_1$ or $r_2$ are negative, we do not get nice scaling properties. Thus, we shall multiply \eqref{delta4ansatzstep2a} by a delta function enforcing $Z_1\cdot Z_2=0$ which is a conformally invariant constraint as this dot product is conformally invariant, please see \eqref{twistordotprods}. Before we do this let us note that the four dimensional delta function in \eqref{delta4ansatzstep2a} enforces that $Z_3$ lies in the plane spanned by $Z_1$ and $Z_2$. Thus only the inner produce $Z_1\cdot Z_2$ is independent. Therefore we consider
\footnote{This ansatz is also consistent with the relationship between functions in dual twistor space and twistor space as discussed in appendix \ref{app:TwistortoDual}. Moreover, such an ansatz ensures nice projective properties as we shall soon see.},
\small
\begin{align}\label{delta4defa}
&\delta^3(Z_1,Z_2,Z_3;\alpha_{12},\alpha_{23},\alpha_{31})=(-i)^{-\alpha_{12}-\alpha_{23}-\alpha_{31}}\delta^{[-\alpha_{12}-\alpha_{23}-\alpha_{31}]}(Z_1\cdot Z_2)\int dc_{23}dc_{31}c_{23}^{\alpha_{23}}c_{31}^{\alpha_{31}}\delta^4(c_{23}Z_1+c_{31}Z_2+Z_3)\notag\\&=\int dc_{12}dc_{23}dc_{31}c_{12}^{\alpha_{12}}c_{23}^{\alpha_{23}}c_{31}^{\alpha_{31}}\delta^4(c_{23} Z_1+c_{31} Z_2+c_{12} Z_3)e^{\frac{i}{3}\big(\frac{Z_1\cdot Z_2}{c_{12}}+\frac{Z_2\cdot Z_3}{c_{23}}+\frac{Z_3\cdot Z_1}{c_{31}}\big)},
\end{align}
\normalsize
which completes the deriavation of \eqref{delta4def}.
\section{From twistor space to dual twistor space}\label{app:TwistortoDual}
In this appendix, we perform an explicit Fourier transform from the non-homogeneous $(+++)$ helicity written in dual twistor space $W$ to twistor space $Z$. This will yield the projective $\mathbb{RP}^3$ delta function solution of section \ref{sec:Geometry} from the result of the ``Three delta product"  solution \eqref{twistorspace3points}. Our starting point is,
\small
\begin{align}
    \langle 0|\tilde{J}^{+}_{s_1}(W_1)\tilde{J}^{+}_{s_2}(W_2)\tilde{J}^{+}_{s_3}(W_3)|0\rangle_{nh}=i^{-s_1-s_2-s_3} \delta^{[-s_1-s_2+s_3]}(W_1\cdot W_2)\delta^{[-s_2-s_3+s_1]}(W_2\cdot W_3)\delta^{[-s_3-s_1+s_2]}(W_3\cdot W_1),
\end{align}
\normalsize     
Performing a half-Fourier transform we get,
\small
\begin{align}\label{delta4WWWtoZZZ}
    &\langle 0|\tilde{J}^{+}_{s_1}(Z_1)\tilde{J}^{+}_{s_2}(Z_2)\tilde{J}^{+}_{s_3}(Z_3)|0\rangle_{nh}=\int d^4 W_1 d^4 W_2 d^4 W_3 e^{iZ_1\cdot W_1+iZ_2\cdot W_2+i Z_3\cdot W_3}\langle 0|\tilde{J}^{+}_{s_1}(W_1)\tilde{J}^{+}_{s_2}(W_2)\tilde{J}^{+}_{s_3}(W_3)|0\rangle_{nh}\notag\\
    &=\int dc_{12}dc_{23}dc_{31}c_{12}^{-s_1-s_2+s_3}c_{23}^{-s_2-s_3+s_1}c_{31}^{-s_3-s_1+s_2}\notag\\&~~\times \int d^4W_1d^4W_2d^4W_3e^{i(Z_1\cdot W_1+Z_2\cdot W_2+Z_3\cdot W_3-c_{12}W_1\cdot W_2-c_{23}W_2\cdot W_3-c_{31}W_3\cdot W_1)}\notag\\
    &=\int dc_{12}dc_{23}dc_{31}c_{12}^{-s_1-s_2+s_3}c_{23}^{-s_2-s_3+s_1}c_{31}^{-s_3-s_1+s_2}\int d^4 W_2 d^4 W_3 e^{i(Z_2\cdot W_2+Z_3\cdot W_3-c_{23}Z_2\cdot Z_3)}\delta^4(Z_1^A+c_{12}W_2^A-c_{31}W_3^A)\notag\\
    &=\int dc_{12}dc_{23}dc_{31}c_{12}^{-s_1-s_2+s_3-4}c_{23}^{-s_2-s_3+s_1}c_{31}^{-s_3-s_1+s_2}e^{i\frac{Z_1\cdot Z_2}{c_{12}}}\int d^4 W_3 e^{i(Z_3\cdot W_3+\frac{c_{23}}{c_{12}}Z_1\cdot W_3+\frac{c_{31}}{c_{12}}Z_2\cdot W_3)}\notag\\
    &=\int dc_{12}dc_{23}dc_{31}c_{12}^{-s_1-s_2+s_3}c_{23}^{-s_2-s_3+s_1}c_{31}^{-s_3-s_1+s_2}e^{i\frac{Z_1\cdot Z_2}{c_{12}}}\delta^4(c_{12}Z_3^A+c_{23}Z_1^A+c_{31}Z_2^A)\notag\\
    &:=\delta(Z_1,Z_2,Z_3;-s_2-s_3+s_1,-s_3-s_1+s_2,-s_1-s_2+s_3),
\end{align}
\normalsize
where we have defined the shorthand notation,
\begin{align}
    \delta(Z_1,Z_2,Z_3;n_{23},n_{31},n_{12})=\int dc_{12}dc_{23}dc_{31}c_{12}^{n_{12}}c_{23}^{n_{23}}c_{31}^{n_{31}}e^{i\frac{Z_1\cdot Z_2}{c_{12}}}\delta^4(c_{12}Z_3^A+c_{23}Z_1^A+c_{31}Z_2^A)
\end{align}
Further, on the support of the delta function, \eqref{delta4WWWtoZZZ} can be written in a more symmetric form viz,
\begin{align}
    &\langle 0|\tilde{J}^{+}_{s_1}(Z_1)\tilde{J}^{+}_{s_2}(Z_2)\tilde{J}^{+}_{s_3}(Z_3)|0\rangle_{nh}\notag\\&=\int dc_{12}dc_{23}dc_{31}c_{12}^{-s_1-s_2+s_3}c_{23}^{-s_2-s_3+s_1}c_{31}^{-s_3-s_1+s_2}\delta^4(c_{12}Z_3^A+c_{23}Z_1^A+c_{31}Z_2^A)e^{\frac{i}{3}\big(\frac{Z_1\cdot Z_2}{c_{12}}+\frac{Z_2\cdot Z_3}{c_{23}}+\frac{Z_3\cdot Z_1}{c_{31}}\big)}.
\end{align}

\section{Twistor space and dual twistor space three point correlators}\label{app:TwistorSpaceResults}
Here we present the expressions for parity even three point Wightman functions of conserved current in Twistor variables and dual Twistor variables for all helicities. The expressions were obtained by solving the conformal Ward identities \eqref{TABZ},\eqref{TABW} and the helicity identity \eqref{helicityZandW}.

\subsection{Twistor space}\label{subapp:AllTwistor}
The results for correlators in twistor variables($Z$) is given as,
\begin{align}
    &\langle 0|J_{s_1}^{-}(Z_1)J_{s_2}^{-}(Z_2)J_{s_3}^{-}(Z_3)|0\rangle \notag\\
    &= (c^{(h)}_{s_1 s_2 s_3} ) \delta^3(Z_1,Z_2,Z_3;\alpha, \beta, \gamma) + c_{s_1 s_2 s_3}^{(nh)} (-i)^{-s_T}\delta^{-\alpha}(Z_1\cdot Z_2)\delta^{-\beta}(Z_2\cdot Z_3)\delta^{-\gamma}(Z_3\cdot Z_1) 
\end{align}
\begin{align} 
    &\langle 0|J_{s_1}^{+}(Z_1)J_{s_2}^{+}(Z_2)J_{s_3}^{-}(Z_3)|0\rangle \notag \\ 
    &= (c^{(h)}_{s_1 s_2 s_3} ) (-1)^{\gamma} \delta^3(Z_1,Z_2,Z_3;-s_T, \gamma, \beta) + c^{(nh)}_{s_1 s_2 s_3} (-i)^{\alpha}\delta^{s_T}(Z_1\cdot Z_2)\delta^{-\beta}(Z_3\cdot Z_1)\delta^{-\gamma}(Z_2\cdot Z_3)
\end{align}
\begin{align} 
    &\langle 0|J_{s_1}^{+}(Z_1)J_{s_2}^{-}(Z_2)J_{s_3}^{+}(Z_3)|0\rangle \notag \\ 
    &= (c^{(h)}_{s_1 s_2 s_3} ) (-1)^{\beta} \delta^3(Z_1+Z_2+Z_3;\beta,\alpha,-s_T) + c^{(nh)}_{s_1 s_2 s_3} (-i)^{\gamma}\delta^{-\beta}(Z_1\cdot Z_2)\delta^{-\alpha}(Z_2\cdot Z_3)\delta^{s_T}(Z_3\cdot Z_1) 
\end{align}
\begin{align} 
    &\langle 0|J_{s_1}^{-}(Z_1)J_{s_2}^{+}(Z_2)J_{s_3}^{+}(Z_3)|0\rangle \notag \\ 
    &= (c^{(h)}_{s_1 s_2 s_3} ) \delta^3(Z_1,Z_2,Z_3;\gamma, -s_T, \alpha) + c^{(nh)}_{s_1 s_2 s_3} (i)^{\beta}\delta^{[-\gamma]}(Z_1\cdot Z_2)\delta^{[s_T]}(Z_2\cdot Z_3) \delta^{-\alpha}(Z_3 \cdot Z_1)
\end{align}
\begin{align} 
    &\langle 0|J_{s_1}^{-}(Z_1)J_{s_2}^{-}(Z_2)J_{s_3}^{+}(Z_3)|0\rangle \notag \\ 
    &= (c^{(h)}_{s_1 s_2 s_3} )(-i)^{-\alpha}\delta^{[-s_T]}(Z_1\cdot Z_2)\delta^{[\gamma]}(Z_2\cdot Z_3)\delta^{[\beta]}(Z_3\cdot Z_1) + c_{s_1 s_2 s_3}^{(nh)} (-1)^{\gamma} \delta^3(Z_1,Z_2,Z_3;s_T,-\gamma,-\beta)
\end{align}
\begin{align} 
    &\langle 0|J_{s_1}^{-}(Z_1)J_{s_2}^{+}(Z_2)J_{s_3}^{-}(Z_3)|0\rangle \notag \\ 
    &= (c^{(h)}_{s_1 s_2 s_3} )(-i)^{-\gamma}\delta^{[\beta]}(Z_1\cdot Z_2)\delta^{[\alpha]}(Z_2\cdot Z_3)\delta^{[-s_T]}(Z_3\cdot Z_1) + c_{s_1 s_2 s_3}^{(nh)} (-1)^{\beta} \delta^3(Z_1,Z_2,Z_3;-\beta,-\alpha,s_T)
\end{align}
\begin{align} 
    &\langle 0|J_{s_1}^{+}(Z_1)J_{s_2}^{-}(Z_2)J_{s_3}^{-}(Z_3)|0\rangle \notag \\ 
    &= (c^{(h)}_{s_1 s_2 s_3} )
    (-i)^{-\beta}\delta^{[\gamma]}(Z_1\cdot Z_2)\delta^{[-s_T]}(Z_2\cdot Z_3)\delta^{[\alpha]}(Z_3\cdot Z_1) + c_{s_1 s_2 s_3}^{(nh)} (-1)^{\alpha} \delta^3(Z_1,Z_2,Z_3;-\gamma,s_T,-\alpha)
\end{align}
\begin{align} 
    &\langle 0|J_{s_1}^{+}(Z_1)J_{s_2}^{+}(Z_2)J_{s_3}^{+}(Z_3)|0\rangle \notag \\ 
    &=(c^{(h)}_{s_1 s_2 s_3} ) 
     (i)^{s_T}\delta^{[\alpha]}(Z_1\cdot Z_2)\delta^{[\beta]}(Z_2\cdot Z_3)\delta^{[\gamma]}(Z_3\cdot Z_1) + c_{s_1 s_2 s_3}^{(nh)}  \delta^3(Z_1,Z_2,Z_3;-\alpha,-\beta,-\gamma)
\end{align}

\subsection{Dual- Twistor space}\label{subapp:AllDualtwistor}
The dual-twistor space counterparts of the above expressions in all eight helicity configurations are given by,
\begin{align}
    &\langle 0|J_{s_1}^{-}(W_1)J_{s_2}^{-}(W_2)J_{s_3}^{-}(W_3)|0\rangle \notag\\
    & = (c^{(h)}_{s_1 s_2 s_3} )(-i)^{s_1+s_2+s_3}\delta^{[\alpha]}(W_1\cdot W_2)\delta^{\beta}(W_2\cdot W_3)\delta^{\gamma}(W_3\cdot W_1)  \notag \\
    & +c^{(nh)}_{s_1 s_2 s_3}(-i)^{s_T}\delta^{[s_T]}(W_1\cdot W_2)\int \frac{dc_{23} dc_{31}}{(2\pi)^2}c_{23}^{-\beta}c_{31}^{-\gamma}\delta^4(W_3^{A} + c_{31}W_2^A + c_{23}W_1^A)
\end{align}
\begin{align} 
    &\langle 0|J_{s_1}^{+}(W_1)J_{s_2}^{+}(W_2)J_{s_3}^{-}(W_3)|0\rangle \notag \\ 
    &= (c^{(h)}_{s_1 s_2 s_3} )(-i)^{-\alpha}\delta^{-s_T}(W_1\cdot W_2)\delta^{\beta}(W_2\cdot W_3)\delta^{\gamma}(W_3\cdot W_1) \notag \\
    &+ c^{(nh)}_{s_1 s_2 s_3}\int \frac{dc_{12}dc_{23}dc_{31}}{(2\pi)^{3}}c_{12}^{-\alpha}c_{23}^{-\beta}c_{31}^{-\gamma}e^{i c_{12} W_1\cdot W_2}\delta^4(W_3
    ^A + c_{23} W_2^A - c_{31}W_1^A)
\end{align}
\begin{align} 
    &\langle 0|J_{s_1}^{+}(W_1)J_{s_2}^{-}(W_2)J_{s_3}^{+}(W_3)|0\rangle \notag \\ 
    &= (c^{(h)}_{s_1 s_2 s_3} )(-i)^{-\gamma}\delta^{\beta}(W_1\cdot W_2)\delta^{\alpha}(W_2\cdot W_3)\delta^{-s_T}(W_3\cdot W_1) \notag \\
    &+ c^{(nh)}_{s_1 s_2 s_3}\int \frac{dc_{12}dc_{23}dc_{31}}{(2\pi)^{3}}c_{12}^{-\alpha}c_{23}^{-\beta}c_{31}^{-\gamma}e^{i c_{31} W_3\cdot W_1}\delta^4(W_2
    ^A + c_{12} W_1^A - c_{23}W_3^A)
\end{align}
\begin{align} 
    &\langle 0|J_{s_1}^{-}(W_1)J_{s_2}^{+}(W_2)J_{s_3}^{+}(W_3)|0\rangle \notag \\ 
    &= (c^{(h)}_{s_1 s_2 s_3} )(-i)^{-\beta}\delta^{[\gamma]}(W_1\cdot W_2)\delta^{[-s_T]}(W_2\cdot W_3) \delta^{\alpha}(W_3 \cdot W_1) \notag \\
    &+ c^{(nh)}_{s_1 s_2 s_3}\int \frac{dc_{12}dc_{23}dc_{31}}{(2\pi)^{3}}c_{12}^{-\alpha}c_{23}^{-\beta}c_{31}^{-\gamma}e^{i c_{23} W_2\cdot W_3}\delta^4(W_1
    ^A - c_{12} W_2^A + c_{31}W_3^A)
\end{align}
\begin{align} 
    &\langle 0|J_{s_1}^{-}(W_1)J_{s_2}^{-}(W_2)J_{s_3}^{+}(W_3)|0\rangle \notag \\ 
    &= (c^{(h)}_{s_1 s_2 s_3} )(-1)^{s_T}(-i)^{\alpha}\delta^{\alpha}(W_1 \cdot W_2)\int dc_{23}dc_{31} c_{23}^{\beta}c_{31}^{\gamma}\delta^4(W_3^A+c_{23}W_2^A-c_{31}W_1^A)  \notag \\
    &+ c^{(nh)}_{s_1 s_2 s_3}(i)^{\alpha}\delta^{[s_T]}(W_1\cdot W_2)\delta^{[-\gamma]}(W_2\cdot W_3)\delta^{[-\beta]}(W_3\cdot W_1)
\end{align}
\begin{align} 
    &\langle 0|J_{s_1}^{-}(W_1)J_{s_2}^{+}(W_2)J_{s_3}^{-}(W_3)|0\rangle \notag \\ 
    &= (c^{(h)}_{s_1 s_2 s_3} )(-1)^{s_T}(-i)^{\gamma}\delta^{\gamma}(Z_3\cdot Z_1)\int dc_{12}dc_{23} c_{12}^{\alpha} c_{23}^{\beta} \delta^4(W_2^A+c_{12}W_1^A-c_{23} W_3^A) \notag \\
    &+ c^{(nh)}_{s_1 s_2 s_3}(i)^{\gamma}\delta^{[-\beta]}(W_1\cdot W_2)\delta^{[-\alpha]}(W_2\cdot W_3)\delta^{[s_T]}(W_3\cdot W_1)
\end{align}
\begin{align} 
    &\langle 0|J_{s_1}^{+}(W_1)J_{s_2}^{-}(W_2)J_{s_3}^{-}(W_3)|0\rangle \notag \\ 
    &= (c^{(h)}_{s_1 s_2 s_3} )(-1)^{s_T}(-i)^{\beta}\delta^{\beta}(W_2\cdot W_3)\int dc_{12}dc_{31} c_{12}^{\alpha} c_{31}^{\gamma} \delta^4(-W_1^A+c_{12}W_2^A-c_{31} W_3^A) \notag \\
    &+ c^{(nh)}_{s_1 s_2 s_3}(-i)^{\beta}\delta^{[-\gamma]}(W_1\cdot W_2)\delta^{[s_T]}(W_1\cdot W_2)\delta^{[-\alpha]}(W_3\cdot W_1)
\end{align}
\begin{align} 
    &\langle 0|J_{s_1}^{+}(W_1)J_{s_2}^{+}(W_2)J_{s_3}^{+}(W_3)|0\rangle \notag \\ 
    &= (c^{(h)}_{s_1 s_2 s_3} ) (i)^{-s_T}\delta^{[-s_T]}(W_1\cdot W_2)\int dc_{23}dc_{31} c_{23}^{\beta} c_{31}^{\gamma} \delta^4(W_3^A+c_{31}W_2^A+c_{23} W_1^A) \notag \\
    &+ c^{(nh)}_{s_1 s_2 s_3}(-i)^{-s_T}\delta^{[-\alpha]}(W_1\cdot W_2)\delta^{[-\beta]}(W_2\cdot W_3)\delta^{[-\gamma]}(W_3\cdot W_1)
\end{align}
One can check that under a CPT transformation defined in \cite{Bala:2025gmz} the results of sub-appendix \ref{subapp:AllDualtwistor} go over to those in \ref{subapp:AllTwistor}.

\section{$\langle 0|\hat{J}(Z_1)\hat{J}(Z_2)O_1(Z_3)|0\rangle$: A comparison between the Penrose and Witten transforms}\label{app:FixingWightman}
The aim of this appendix is to show that two distinct twistor space expressions can yield the same position space expression after a Penrose transform but yield different expressions in spinor helicity when Witten transformed. 

For concreteness, we focus on the correlator $\langle 0|\hat{J}(Z_1)\hat{J}(Z_2) O_1(Z_3)|0\rangle$. Our ansatz for this correlator in twistor space obtained after setting $s_1=s_2=1$ in \eqref{JsO1O1ansatz} is given by,
\begin{align}\label{JJO1twistoransatz}
    \langle 0|\hat{J}^{+}(Z_1)\hat{J}^{+}(Z_2)O_1(Z_3)|0\rangle=-\alpha\delta^{[2]}(Z_1\cdot Z_2)\delta(Z_2\cdot Z_3)\delta(Z_3\cdot Z_1)+\beta \delta^3(Z_1,Z_2,Z_3;-2,0,0).
\end{align}
\subsection{Witten transform of $\langle 0|\hat{J}(Z_1)\hat{J}(Z_2) O_1(Z_3)|0\rangle$}
Let us first perform a half-Fourier transform to obtain the spinor helicity variables result. We obtain,
\begin{align}\label{JJO1twistoransatzinSH}
    \langle 0|\hat{J}^{+}(p_1)\hat{J}^{+}(p_2)O_1(p_3)|0\rangle=\bigg(\alpha\frac{\langle \Bar{1}\Bar{2}\rangle^2}{(p_1+p_2+p_3)^2}+\beta\frac{\langle \Bar{1}\Bar{2}\rangle^2}{(p_1+p_2-p_3)^2}\bigg)\delta^3(p_1^\mu+p_2^\mu+p_3^\mu).
\end{align}
The value $\alpha=\beta$ reproduces the $(++)$ helicity component of the momentum space Wightman function \cite{Bala:2025gmz}\footnote{The conversion to spinor helicity variables from momentum space can be performed following section $2$ of \cite{Bala:2025gmz}.},
\begin{align}\label{JJO1correctWightman}
\notag\langle0| J(z_1,p_1)J(z_2,p_2)O_1(p_3)|0\rangle=&z_1\cdot z_2\Big(\frac{p_1p_2p_3}{E(E-2p_1)(E-2p_2)(E-2p_3)}\Big)\\+z_1\cdot p_2\;z_2\cdot z_1\Big(&\frac{p_1p_2(4(p_3^2-(p_1^2+p_2^2))+E(E-2p_1)(E-2p_2)(E-2p_3))}{p_3E^2(E-2p_1)^2(E-2p_2)^2(E-2p_3)^2}\Big).
\end{align}
The reason is that only for $\alpha=\beta$ is the condition that the Wightman function should vanish when we set the momentum magnitudes $p_1$ or $p_2$ to zero is satisfied\footnote{To see why this property is satisfied by Wightman functions, please refer to \cite{Bautista:2019qxj,Bala:2025gmz}.}. 

Its expression in the $(++)$ helicity given by,
\begin{align}\label{correctWightmanJJO1pp}
    \langle 0|\hat{J}^{+}(p_1)\hat{J}^{+}(p_2)O_1(p_3)|0\rangle=\bigg(\frac{\langle \Bar{1}\Bar{2}\rangle^2}{(p_1+p_2+p_3)^2}+\frac{\langle \Bar{1}\Bar{2}\rangle^2}{(p_1+p_2-p_3)^2}\bigg)\delta^3(p_1^\mu+p_2^\mu+p_3^\mu).
\end{align}
Thus we see that the expression in twistor space that reproduces the correct momentum space Wightman function is,
\begin{align}\label{JJO1twistorresult}
    \langle 0|\hat{J}_s^{+}(Z_1)\hat{J}_s^{+}(Z_2)O_1(Z_3)|0\rangle=-\delta^{[2]}(Z_1\cdot Z_2)\delta(Z_2\cdot Z_3)\delta(Z_3\cdot Z_1)+ \delta^3(Z_1,Z_2,Z_3;-2,0,0).
\end{align}

At this point let us note that in contrast to \eqref{JJO1twistoransatz} where we have two terms with $\alpha=\beta$ required to produce the correct spinor helicity variables Wightman function, in \cite{Baumann:2024ttn}, the authors instead do not consider the $\delta^3$ solution and instead take,
\begin{align}\label{JJOBaumann}
    \langle 0|\hat{J}^{+}(Z_1)\hat{J}^{+}(Z_2)O_1(Z_3)|0\rangle_{\cite{Baumann:2024ttn}}=-\alpha \delta^{[2]}(Z_1\cdot Z_2)\delta(Z_2\cdot Z_3)\delta(Z_3\cdot Z_1).
\end{align}
Converting \eqref{JJOBaumann} to spinor helicity variables results in,
\begin{align}
    \langle 0|\hat{J}^{+}(p_1)\hat{J}^{+}(p_2)O_1(p_3)|0\rangle_{\cite{Baumann:2024ttn}}=\bigg(\alpha\frac{\langle \Bar{1}\Bar{2}\rangle^2}{(p_1+p_2+p_3)^2}\bigg)\delta^3(p_1^\mu+p_2^\mu+p_3^\mu),
\end{align}
which is not the Wightman function in the $(++)$ helicity as it is missing the second term in \eqref{correctWightmanJJO1pp} which is required to satisfy all properties of the Wightman function, please refer to the discussion below \eqref{JJO1correctWightman}. However, the authors of \cite{Baumann:2024ttn} show that starting with just \eqref{JJOBaumann} and converting to position space using a Penrose transform \eqref{PenroseTransform} yields the correct expression! The reconciliation between these facts as we shall prove now is that inside a Penrose transform, the two terms in our solution \eqref{JJO1twistorresult} are equivalent and thus give rise to the same position space result that one would obtain by keeping just the first term like in \cite{Baumann:2024ttn}.

\subsection{Penrose transform of $\langle 0|\hat{J}(Z_1) \hat{J}(Z_2) O_1(Z_3)|0\rangle$}
Let us start with our expression with the coefficients $\alpha$ and $\beta$ intact \eqref{JJO1twistoransatz} for keeping track of the two terms. The Penrose transform \eqref{PenroseTransform} after contracting with arbitrary polraization spinors for this correlator reads,
\begin{align}\label{JJO1Penrose1a}
    &\langle 0|J(x_1,\zeta_1)J(x_2,\zeta_2)O_1(x_3)|0\rangle\notag\\&=\int \prod_{i=1}^{3}\langle \lambda_i d\lambda_i\rangle~(\lambda_1\cdot \zeta_1)^2(\lambda_2\cdot \zeta_2)^2\bigg(-\alpha \delta^{[2]}(Z_1\cdot Z_2)\delta(Z_2\cdot Z_3)\delta(Z_3\cdot Z_1)+\beta \delta^3(Z_1,Z_2,Z_3;-2,0,0)\bigg)\bigg|_{X}.
\end{align}
 Let us thus split \eqref{JJO1Penrose1a} into a sum of these two terms,
\begin{align}
    \langle 0|J(x_1,\zeta_1)J(x_2,\zeta_2)O_1(x_3)|0\rangle=\alpha\langle 0|J(x_1,\zeta_1)J(x_2,\zeta_2)O_1(x_3)|0\rangle_{\alpha}+\beta \langle 0|J(x_1,\zeta_1)J(x_2,\zeta_2)O_1(x_3)|0\rangle_{\beta},
\end{align}
where,
\begin{align}\label{JJO1alphacoeff}
    &\langle 0|J(x_1,\zeta_1)J(x_2,\zeta_2)O_1(x_3)|0\rangle_{\alpha}=-\int \prod_{i=1}^{3}\langle \lambda_i d\lambda_i\rangle~(\lambda_1\cdot \zeta_1)^2(\lambda_2\cdot \zeta_2)^2\bigg( \delta^{[2]}(Z_1\cdot Z_2)\delta(Z_2\cdot Z_3)\delta(Z_3\cdot Z_1)\bigg)\bigg|_{X},
\end{align}
and
\begin{align}\label{JJO1betacoeff}
    &\langle 0|J(x_1,\zeta_1)J(x_2,\zeta_2)O_1(x_3)|0\rangle_{\beta}=\int \prod_{i=1}^{3}\langle \lambda_i d\lambda_i\rangle~(\lambda_1\cdot \zeta_1)^2(\lambda_2\cdot \zeta_2)^2\bigg( \delta^3(Z_1,Z_2,Z_3;-2,0,0)\bigg)\bigg|_{X}.
\end{align}
We want to show that the coefficient of $\beta$ \eqref{JJO1betacoeff} is precisely equal to the coefficient of $\alpha$ \eqref{JJO1alphacoeff} i.e.,
\begin{align}\label{proofrequired}
    \langle 0|J(x_1,\zeta_1)J(x_2,\zeta_2)O_1(x_3)|0\rangle_{\beta}=\langle 0|J(x_1,\zeta_1)J(x_2,\zeta_2)O_1(x_3)|0\rangle_{\alpha}.
\end{align}
Lets start with the coefficient of $\beta$ viz \eqref{JJO1betacoeff}. First, of all we have,
\begin{align}\label{JJO1betanextstep}
    &\langle 0|J(x_1,\zeta_1)J(x_2,\zeta_2)O_1(x_3)|0\rangle_{\beta}=\int \prod_{i=1}^{3}\langle \lambda_i d\lambda_i\rangle~(\lambda_1\cdot \zeta_1)^2(\lambda_2\cdot \zeta_2)^2\bigg( \delta^3(Z_1,Z_2,Z_3;-2,0,0)\bigg)\bigg|_{X}.\notag\\
    &=\int \prod_{i=1}^{3}\langle \lambda_i d\lambda_i\rangle~(\lambda_1\cdot \zeta_1)^2(\lambda_2\cdot \zeta_2)^2 \prod_{i=1}^{3}\int d^2\Bar{\mu}_i \delta^2(\Bar{\mu}_i^a-x_i^{ab}\lambda_{ib}) \delta^3(Z_1,Z_2,Z_3;-2,0,0)\notag\\&=\int \prod_{i=1}^{3}\langle \lambda_i d\lambda_i\rangle~(\lambda_1\cdot \zeta_1)^2(\lambda_2\cdot \zeta_2)^2 \prod_{i=1}^{3}\int d^2\Bar{\mu}_i \delta^2(\Bar{\mu}_i^a-x_i^{ab}\lambda_{ib})\notag\\&~~~~~~~~~~~~~~~~~~~~~~\qquad\times\bigg(\prod_{i=1}^{3}\int \frac{d^2\Bar{\lambda}_i}{(2\pi)^2}e^{i\Bar{\lambda}_i\cdot \Bar{\mu}_i}\langle \Bar{1}\Bar{2}\rangle^2\bigg(\frac{1}{(p_1+p_2-p_3)^2}\bigg)\delta^3(p_1^\mu+p_2^\mu+p_3^\mu).
\end{align}
To go from the first step to the second one, we introduced delta functions that impose the incidence relations \eqref{IncidenceRelation} and from that line to the next, we expressed the twistor space correlator \eqref{JJO1twistoransatz} as a inverse half-Fourier transform of its spinor helicity counterpart viz \eqref{JJO1twistoransatzinSH} with $p_i=-\frac{1}{2}\lambda_i\cdot \Bar{\lambda}_i$.

Inside the integrals in  \eqref{JJO1betanextstep} we are free to re-label $\lambda_3\leftrightarrow \Bar{\lambda}_3$. This flips the magnitude $p_3=-\frac{1}{2}\lambda_3\cdot \Bar{\lambda}_3\to -p_3$ but leaves the momentum conserving delta function invariant. Thus we obtain,
\begin{align}
      &\langle 0|J(x_1,\zeta_1)J(x_2,\zeta_2)O_1(x_3)|0\rangle_{\beta}=\int \prod_{i=1}^{2}\langle \lambda_i d\lambda_i\rangle~(\lambda_i\cdot \zeta_i)^2\prod_{i=1}^{2}\int d^2\Bar{\mu}_i \delta^2(\Bar{\mu}_i^a-x_i^{ab}\lambda_{ib})\prod_{i=1}^{2}\frac{d^2\Bar{\lambda}_i}{(2\pi)^2}e^{i\Bar{\lambda}_i\cdot \Bar{\mu}_i}\langle \Bar{1}\Bar{2}\rangle^2\notag\\&\times \int \langle \Bar{\lambda}_3d\Bar{\lambda}_3\rangle\int d^2\Bar{\mu}_3\delta^2(\Bar{\mu}_3^a-x^{ab}\Bar{\lambda}_{3b})\times\int \frac{d^2\lambda_3}{(2\pi)^2}e^{i\lambda_3\cdot \Bar{\mu}_3}\frac{1}{(p_1+p_2+p_3)^2}\delta^3(p_1^\mu+p_2^\mu+p_3^\mu).
\end{align}
Re-labelling $\Bar{\mu}^a=-\mu^a$ yields,
\begin{align}\label{JJO1betanextstepnext}
      &\langle 0|J(x_1,\zeta_1)J(x_2,\zeta_2)O_1(x_3)|0\rangle_{\beta}=\int \prod_{i=1}^{2}\langle \lambda_i d\lambda_i\rangle~(\lambda_i\cdot \zeta_i)^2\prod_{i=1}^{2}\int d^2\Bar{\mu}_i \delta^2(\Bar{\mu}_i^a-x_i^{ab}\lambda_{ib})\prod_{i=1}^{2}\frac{d^2\Bar{\lambda}_i}{(2\pi)^2}e^{i\Bar{\lambda}_i\cdot \Bar{\mu}_i}\langle \Bar{1}\Bar{2}\rangle^2\notag\\&\times \int \langle \Bar{\lambda}_3d\Bar{\lambda}_3\rangle\int d^2\mu_3\delta^2(-\mu_3^a-x^{ab}\Bar{\lambda}_{3b})\times\int \frac{d^2\lambda_3}{(2\pi)^2}e^{-i\lambda_3\cdot \mu_3}\frac{1}{(p_1+p_2+p_3)^2}\delta^3(p_1^\mu+p_2^\mu+p_3^\mu)\bigg)\notag\\
      &=\prod_{i=1}^{2}\int \langle \lambda_i d\lambda_i\rangle(\lambda_i\cdot \zeta_i)^2\int  d^2\Bar{\mu}_i \delta^{2}(\Bar{\mu}_i^a-x_i^{ab}\lambda_{ib}) \int \langle \Bar{\lambda}_3d\Bar{\lambda}_3\rangle\int d^2 \mu_3 \delta^2(-\mu_3^a-x^{ab}\Bar{\lambda}_{3b})\notag\\
      &\times \int \frac{d^2\Bar{\lambda}_1 d^2\Bar{\lambda}_2 d^2 \lambda_3}{(2\pi)^6}e^{i\big(\Bar{\lambda}_1\cdot \Bar{\mu}_1+\Bar{\lambda}_2\cdot \Bar{\mu}_2-\lambda_3\cdot \mu_3\big)}\frac{\langle \Bar{1}\Bar{2}\rangle^{2}}{(p_1+p_2+p_3)^2}\delta^3(p_1^\mu+p_2^\mu+p_3^\mu)\notag\\
      &=\prod_{i=1}^{2}\int \langle \lambda_i d\lambda_i\rangle(\lambda_i\cdot \zeta_i)^2\int  d^2\Bar{\mu}_i \delta^{2}(\Bar{\mu}_i^a-x_i^{ab}\lambda_{ib}) \int \langle \Bar{\lambda}_3d\Bar{\lambda}_3\rangle\int d^2 \mu_3 \delta^2(-\mu_3^a-x^{ab}\Bar{\lambda}_{3b})\notag\\
      &\times \bigg(- \delta^{[2]}(\lambda_1\cdot \Bar{\mu}_2-\lambda_2\cdot \Bar{\mu}_1)\delta(\lambda_2\cdot \mu_3-\Bar{\lambda}_3\cdot \Bar{\mu}_2)\delta(\Bar{\lambda}_3\cdot \Bar{\mu}_1-\lambda_1\cdot \mu_3)\bigg),
\end{align}
where to go from the penultimate step of \eqref{JJO1betanextstep} to the last line of the same equation, we explicitly performed the integrals over $\Bar{\lambda}_1,\Bar{\lambda}_2$ and $\lambda_3$. Now we just re-label back $\Bar{\lambda}_3\leftrightarrow \lambda_3$ and $\Bar{\mu}_3\leftrightarrow \mu_3$ and identify $Z_i\cdot Z_j=\lambda_i\cdot \Bar{\mu}_j-\lambda_j\cdot \Bar{\mu}_i$. This yields,
\begin{align}\label{JJO1alphaequalsbeta}
    &\langle 0|J(x_1,\zeta_1)J(x_2,\zeta_2)O_1(x_3)|0\rangle_{\beta}=-\prod_{i=1}^{3}\langle \lambda_i d\lambda_i\rangle\prod_{i=1}^{2}(\lambda_i\cdot \zeta_i)^2 \delta^{[2]}(Z_1\cdot Z_2)\delta(Z_2\cdot Z_3)\delta(Z_3\cdot Z_1)|_X\notag\\
    &~~~~~~~~~~~~~~~~~~~~~~~~~~~~~~~~~~~~~~~~=\langle 0|J(x_1,\zeta_1)J(x_2,\zeta_2)O_1(x_3)|0\rangle_{\alpha}.
\end{align}
where in the last line we identified the coefficient of $\alpha$ viz \eqref{JJO1alphacoeff}. This is precisely what we sought out to prove \eqref{proofrequired}. Therefore, inside the Penrose transform \eqref{JJO1Penrose1a} we are free to use the following identity:
\begin{align}\label{JJO1Penrose1b}
    &\langle 0|J(x_1,\zeta_1)J(x_2,\zeta_2)O_1(x_3)|0\rangle\notag\\&=\int \prod_{i=1}^{3}\langle \lambda_i d\lambda_i\rangle~(\lambda_1\cdot \zeta_1)^2(\lambda_2\cdot \zeta_2)^2\bigg(-\alpha \delta^{[2]}(Z_1\cdot Z_2)\delta(Z_2\cdot Z_3)\delta(Z_3\cdot Z_1)+\beta \delta^3(Z_1,Z_2,Z_3;-2,0,0)\bigg)\bigg|_{X}\notag\\
    &=-(\alpha+\beta)\int \prod_{i=1}^{3}\langle \lambda_i d\lambda_i\rangle~(\lambda_1\cdot \zeta_1)^2(\lambda_2\cdot \zeta_2)^2\bigg(\delta^{[2]}(Z_1\cdot Z_2)\delta(Z_2\cdot Z_3)\delta(Z_3\cdot Z_1)\bigg)\bigg|_{X}.
\end{align}

To summarize, the two different solutions in \eqref{JJO1twistoransatz} both give rise to the same answer inside a Penrose transform \eqref{JJO1alphaequalsbeta} and thus either of them can be used to obtain the position space correlator. However, in order to obtain the correct momentum space Wightman function \eqref{correctWightmanJJO1pp} we require both terms in \eqref{JJO1alphaequalsbeta} with the coefficients satisfying $\alpha=\beta$. This is summarized in figure \ref{fig:twistor-transform-flow}.

\section{Details of epsilon transform in Twistor space}\label{app:EPTdetails}
In this appendix, we derive the epsilon transformation in the twistor space using the epsilon transformation in the position space \eqref{EPTspinornotation}. We do so for the spin-1 case for illustrative purposes, since a generalization to higher spin is simple. Our starting point is the Penrose transform \eqref{PenroseTransform} for spin-1:
\begin{align}
    J_{ab}(x) = \int \langle\lambda d\lambda\rangle \lambda_{(a} \lambda_{b)} \int d^2\bar{\mu}~\delta^2(\bar{\mu}-x^{ab}\lambda_b)\hat{J}_s^{+}(\lambda,\bar{\mu}) 
\end{align}
Performing \eqref{EPT} on both sides of the above equation, we get,
\begin{align}
    (\epsilon\cdot J)_{ab}(x)&= \int \langle\lambda d\lambda\rangle \lambda^c \lambda_{\left(a\right.}\int d^2\bar{\mu}\bigg(-i\int \frac{d^3y}{|y-x|^2}\frac{\partial}{\partial y^{\left. b \right)c}}\bigg) \delta^2(\bar{\mu}^a-y^{ab}\lambda_b) \hat{J}_s^{+}(\lambda,\bar{\mu}) \notag \\
    &= \int \langle\lambda d\lambda\rangle \lambda^c \lambda_{\left(a\right.}\int d^2\bar{\mu}\bigg(-i\int \frac{d^3y}{|y-x|^2}\frac{\partial}{\partial y^{\left. b \right)c}}\bigg) \int d^2\lambda_1 e^{i \lambda_{1a}(\bar{\mu}^a-y^{ab}\lambda_b)} \hat{J}_s^{+}(\lambda,\bar{\mu})\notag \\
    &= \int \langle\lambda d\lambda\rangle \lambda^c \lambda_{(\left(a\right.}\int d^2\bar{\mu}(-i)\int \frac{d^3y}{|y-x|^2} \int d^2\lambda_1 (-i\lambda_{\left. b \right)}\lambda_{1c})  e^{i \lambda_{1a}(\bar{\mu}^a-y^{ab}\lambda_b)} \hat{J}_s^{+}(\lambda,\bar{\mu})\notag \\
    &= \int \langle\lambda d\lambda\rangle \lambda^c \lambda_{\left(a\right.}\lambda_{\left. b \right)}\int d^2\bar{\mu}(-i)\int \frac{d^3y}{|y-x|^2} \int d^2\lambda_1 \big(-\frac{\partial}{\partial\bar{\mu}^{c}}e^{i \lambda_{1a}(\bar{\mu}^a-y^{ab}\lambda_b)}\big)   \hat{J}_s^{+}(\lambda,\bar{\mu})\notag \\
    &= \int \langle\lambda d\lambda\rangle \lambda_{\left(a\right.}\lambda_{\left. b \right)}\int d^2\bar{\mu}(-i)\int \frac{d^3y}{|y-x|^2} \int d^2\lambda_1 e^{i \lambda_{1a}(\bar{\mu}^a-y^{ab}\lambda_b)}   \lambda^c\frac{\partial}{\partial\bar{\mu}^{c}}\hat{J}_s^{+}(\lambda,\bar{\mu})\notag \\
    &= \int \langle\lambda d\lambda\rangle \lambda_{\left(a\right.}\lambda_{\left. b \right)}\int d^2\bar{\mu}(-i)\int \frac{d^3 \bar{y}}{\bar{y}^2} \delta^2(\bar{\mu}^a-x^{ad}\lambda_d-\bar{y}^{ad}\lambda_d)   \lambda^c\frac{\partial}{\partial\bar{\mu}^{c}}\hat{J}_s^{+}(\lambda,\bar{\mu}) \notag \\
    (\epsilon.J)_{ab}(x)&=\int \langle\lambda d\lambda\rangle \lambda_{\left(a\right.}\lambda_{\left. b \right)} \int d^2 \bar{\mu}~\delta^2(\bar{\mu}^a-x^{ad}\lambda_d) (-i)\int \frac{d^3\bar{y}}{\bar{y}^2} Z^AI_{AB} \frac{\partial}{\partial Z_B} \hat{J}_s^{+}(\lambda,\bar{\mu}')\bigg|_{\bar{\mu}^a\to\bar{\mu}^a + \bar{y}^{ad}\lambda_d}     
\end{align}    
Therefore, the epsilon transform in twistor space is given by
\begin{align}
    (\epsilon\cdot \hat{J}_s)^{+}(\lambda,\bar{\mu})= (-i)\int \frac{d^3 y}{y^2} Z^AI_{AB} \frac{\partial}{\partial Z_B} \hat{J}_s^{+}(\lambda,\bar{\mu})\bigg|_{\bar{\mu}^a\to\bar{\mu}^a + y^{ad}\lambda_d}.
\end{align}

\section{The non-local special conformal generator acting on $\hat{O}_{\Delta}$ in twistor space}\label{app:TwistorGenerators}
Here, we present the derivation of the dilatation and SCT operators \eqref{DODelta} and \eqref{KODelta} acting on arbitrary $\Delta$ scalars in twistor space. Recall from subsection \ref{subsec:genscalar} that the rescaled momentum space scalar operator is given by,
\begin{align}\label{ODeltarescaledop}
    \hat{O}_{\Delta}(\lambda,\Bar{\lambda})=|\frac{\lambda\cdot \Bar{\lambda}}{2}|O_{\Delta}(\lambda,\Bar{\lambda}).
\end{align}
The action of the dilatation and SCT operators on the unrescaled scalars $O_{\Delta}$ are given by\footnote{For generic spinning operators the action of the SCT involves an additional term $K_{ab}\supset -2i(\sigma^\mu)_{ab}\mathcal{M}_{\mu\nu}\frac{\partial}{\partial p_\nu}$}\label{footnoteODeltas},
\begin{align}
    &[D,O_{\Delta}]=\frac{i}{2}\big(\lambda^a\frac{\partial}{\partial\lambda^a}+\Bar{\lambda}^a\frac{\partial}{\partial\Bar{\lambda}^a}+2(3-\Delta)\big)O_{\Delta}\notag\\
    &[K_{ab},O_{\Delta}]=\bigg(2\frac{\partial}{\partial\lambda^{(a}\partial\Bar{\lambda}^{b)}}+\frac{(\Delta-2)}{p}\big(\Bar{\lambda}_{(a}\frac{\partial}{\partial\Bar{\lambda}^{b)}}-\lambda^{(a}\frac{\partial}{\partial\lambda^{b)}}\big)\bigg)O_{\Delta}.
\end{align}
These equations can be used to derive the action of these operator on the rescaled scalar operators \eqref{ODeltarescaledop} and then via the Witten transform \eqref{ODeltaWitten}, one obtains the results \eqref{DODelta} and \eqref{KODelta}. It is also easy to check that the conformal algebra is satisfied with the SCT generator \eqref{KODelta} and dilatation generator \eqref{DODelta} with translations and Lorentz transformations given in \eqref{TABcomponents}.

\section{Conformal invariance of two point Wightman functions involving $\hat{O}_\Delta$}\label{app:ODeltaConf}
Our formula for the $O_{\Delta}$ two point function in twistor space is given in \eqref{ODelta2point}. Let us now prove that it is conformally invariant, using the SCT generator \eqref{KODelta}. This will serve to illustrate how the non-local terms in the definition are put to practice. For convinience we present the correlator here again,
\begin{align}\label{ODelta2pointagain}
    \langle 0|\hat{O}_{\Delta}(Z_1)\hat{O}_{\Delta}(Z_2)|0\rangle=\frac{(\langle Z_1 I Z_2\rangle)^{2})^{(\Delta-1)}}{(Z_1\cdot Z_2)^{2\Delta}}.
\end{align}
The SCT operator \eqref{KODelta} acting on the Wightman function can be divided into three terms: local, non-local of order $1$ and non-local of order $2$ denoted by superscript below:
\begin{align}
    K_{ab}=2i\big(K_{ ab}^{\text{loc}}+K_{ ab}^{\text{non-loc,1}}+K_{ab}^{\text{non-loc,2}}\big),
\end{align}
where,
\begin{align}\label{Klocact}
    &K_{ab}^{\text{loc}}=\sum_{i=1}^{2}\Bar{\mu}_{i(a}\frac{\partial}{\partial\lambda_i^{b)}},~K_{ ab}^{\text{non-loc,1}}=(\Delta-1)\sum_{i=1}^{2}(\lambda_i\cdot\frac{\partial}{\partial\Bar{\mu}_i})^{-1}\big(\lambda_{i(a}\frac{\partial}{\partial\lambda_i^{b)}}-\Bar{\mu}_{i(a}\frac{\partial}{\partial\Bar{\mu}_i^{b)}}\big)\notag\\
    &K_{ab}^{\text{non-loc,2}}=-2(\Delta-1)\sum_{i=1}^{2}(\lambda_{i}\cdot\frac{\partial}{\partial\Bar{\mu}_i})^{-2}\lambda_{i(a}\frac{\partial}{\partial\Bar{\mu}_i^{b)}}.
\end{align}
The action of $K_{ab}^{\text{loc}}$ on \eqref{ODelta2pointagain} yields,
\begin{align}
    K_{ab}^{\text{loc}}\langle 0|\hat{O}_{\Delta}(Z_1)\hat{O}_{\Delta}(Z_2)|0\rangle=\frac{2(\Delta-1)(\lambda_{1(a}\Bar{\mu}_{2b)}-\lambda_{2(a}\Bar{\mu}_{1b)})}{|\langle 12\rangle|}\langle 0|\hat{O}_{\Delta}(Z_1)\hat{O}_{\Delta}(Z_2)|0\rangle.
\end{align}
The first order non-local term results in,
\begin{align}\label{firstordernonlocKact}
     &K_{ab}^{\text{non-loc,1}}\langle 0|\hat{O}_{\Delta}(Z_1)\hat{O}_{\Delta}(Z_2)|0\rangle\notag\\&=(\Delta-1)(\lambda_1\cdot\frac{\partial}{\partial\Bar{\mu}_1})^{-1}\bigg(-\frac{2(\Delta-1)\lambda_{1(a}\lambda_{2b)}}{|\langle 1 2\rangle|}+\frac{2\Delta(\lambda_{1(a}\Bar{\mu}_{2b)}-\lambda_{2(a}\Bar{\mu}_{1b)}}{Z_1\cdot Z_2}\bigg)\langle 0|\hat{O}_{\Delta}(Z_1)\hat{O}_{\Delta}(Z_2)|0\rangle\notag\\
     &+(\Delta-1)(\lambda_2\cdot\frac{\partial}{\partial\Bar{\mu}_2})^{-1}\bigg(\frac{2(\Delta-1)\lambda_{1(a}\lambda_{2b)}}{|\langle 1 2\rangle|}+\frac{2\Delta(\lambda_{1(a}\Bar{\mu}_{2b)}-\lambda_{2(a}\Bar{\mu}_{1b)}}{Z_1\cdot Z_2}\bigg)\langle 0|\hat{O}_{\Delta}(Z_1)\hat{O}_{\Delta}(Z_2)|0\rangle\notag\\
     &=-4\Delta(\Delta-1)(\langle 1 2\rangle^2)^{\Delta-1}(\lambda_{1(a}\Bar{\mu}_{2b)}-\lambda_{2(a}\Bar{\mu}_{1b)})\int_{0}^{\infty}\frac{ds}{(\lambda_1\cdot\Bar{\mu}_2-\lambda_2\cdot\Bar{\mu}_1+s\langle 1 2\rangle)^{2\Delta+1}}\notag\\
     &=-\frac{2(\Delta-1)(\lambda_{1(a}\Bar{\mu}_{2b)}-\lambda_{2(a}\Bar{\mu}_{1b)})}{|\langle 12\rangle|}\langle 0|\hat{O}_{\Delta}(Z_1)\hat{O}_{\Delta}(Z_2)|0\rangle,
\end{align}
where we use the definition of the inverse derivative \eqref{inversederivative}. Using \eqref{inversederivative} it is also very easy to see that the action of the second order non-local operator cancel out between the two operators yielding,
\begin{align}
    K_{ab}^{\text{non-loc,2}}\langle 0|\hat{O}_{\Delta}(Z_1)\hat{O}_{\Delta}(Z_2)|0\rangle=0
\end{align}
Putting this together with \eqref{Klocact} and \eqref{firstordernonlocKact} proves the conformal invariance viz,
\begin{align}
    K_{ab}\langle 0|\hat{O}_{\Delta}(Z_1)\hat{O}_{\Delta}(Z_2)|0\rangle=0
\end{align}
A similar exercise which also involves non trivial contributions from the second order non-local part of $K$ shows that $\langle 0|\hat{O}_2(Z_1)\hat{O}_1(Z_1)|0\rangle$ \eqref{O2O1Twistor} is conformally invariant as well. 
\section{Some explicit Two point Penrose transforms}\label{app:ExplicitPenrose}
In this appendix, we explicitly evaluate some two point Penrose transforms which also serve as additional applications of the identities of appendix \ref{app:ProjectiveVSnonproj}.
\subsection{$\langle O_1 O_1\rangle$}
The Penrose transform \eqref{PenroseTransform} for the $\langle O_1 O_1\rangle$ twistor space correlator \eqref{O1O1twopoint} reads,
\begin{align}
    \langle 0|O_1(x_1)O_1(x_2)|0\rangle=\int \langle \lambda_1 d\lambda_1\rangle\langle\lambda_2 d\lambda_2\rangle\frac{1}{(\lambda_{1a}\lambda_{2b}(x_1-x_2)^{ab})^{2}}.
\end{align}
We can locally parametetrize $\lambda_{ia}=(\xi_i,1)$. This converts the above integral into,
\begin{align}
     \langle 0|O_1(x_1)O_1(x_2)|0\rangle=\int_{-\infty}^{\infty}d\xi_1 \int_{-\infty}^{\infty} d\xi_2 \frac{1}{(\xi_1 \xi_2 (x_1-x_2)^{11}+(\xi_1+\xi_2)(x_1-x_2)^{12}+(x_1-x_2)^{22})^2}.
\end{align}
Evaluating these integrals by taking the limits to be $-\Lambda$ and $+\Lambda$ and evaluating these integrals using say Mathematica and taking the limit $\Lambda\to \infty$ in the end results in the correct conformally invariant result,
\begin{align}\label{O1O1PenrosetoPositionSpace}
     \langle 0|O_1(x_1)O_1(x_2)|0\rangle=\frac{1}{|x_1-x_2|^2}.
\end{align}
\subsection{$\langle O_1 O_2\rangle$}
The two point function is given by \eqref{O2O1Twistor}. Its Penrose transform \eqref{O1PenroseTrans}, \eqref{O2PenroseTrans} using \eqref{claim1step1} can be written as,
\begin{align}
    \frac{1}{\text{Vol}(GL(1,\mathbb{R}))^2}\int d^2\lambda_1 d^2\lambda_2 |\lambda_1\cdot \lambda_2|\delta^{[2]}(\lambda_{1a}\lambda_{2b}(x_2-x_1)^{ab}).
\end{align}
Let us define a bi-spinor (vector) $v_{ab}=\lambda_{1(a}\lambda_{2b)}$. Using the formula \eqref{integralclaim2} we can re-write the above expression as,
\begin{align}
    &\frac{4}{\text{Vol}(GL(1,\mathbb{R}))}\int d^3 v \delta^{[2]}(v_{ab}(x_2-x_1)^{ab})=-\frac{4}{\text{Vol}(GL(1,\mathbb{R}))}\int d^3 v \int dc c^2 e^{-i c v_{ab}(x_2-x_1)^{ab}}\notag\\
    &=-\frac{4}{\text{Vol}(GL(1,\mathbb{R}))}\int dc~c^2~\delta^3(-2 c(x_2-x_1))=\delta^3(x_1-x_2)\frac{-1}{2\text{Vol}(GL(1,\mathbb{R}))}\int \frac{dc}{|c|}=-\frac{1}{2}\delta^3(x_1-x_2).
\end{align}
Thus, we see that the volume factor is canceled using the definition \eqref{volumeGL1R} and leaves us with the correct conformally invariant contact term \cite{Nakayama:2019mpz}.
\subsection{$\langle J_s J_s\rangle_{\text{odd}}$}
The aim of this subsection is to explictly evaluate the Penrose transform \eqref{PenroseTransform} for parity odd two point functions. First consider an integer ($s\in \mathbb{Z}_{>0}$) spin parity odd two point function\eqref{twopointodd}. Its Penrose transform after contracting the free indices with auxiliary polarization spinors $\zeta$ is as follows,
    \begin{align}
        \langle J_{s} J_{s}\rangle & =\int \langle\lambda_1 d\lambda_1\rangle \langle\lambda_2 d\lambda_2\rangle (\zeta_1\cdot\lambda_1)^{2s}(\zeta_2\cdot\lambda_2)^{2s} \text{Sign}(\langle12\rangle) (-i)^{2s+1} \delta^{[2s+1]}(Z_1\cdot Z_2)\big|_X \notag \\
        & \int \langle\lambda_1 d\lambda_1\rangle \langle\lambda_2 d\lambda_2\rangle \frac{((\zeta_1\cdot\lambda_1) (\zeta_2\cdot\lambda_2)+(\zeta_1\cdot\lambda_2) (\zeta_2\cdot\lambda_1)+ \zeta_1\cdot\zeta_2 \langle12\rangle)^{2s}}{2^{2s}} \text{Sign}(\langle12\rangle) \delta^{[2s+1]}(\lambda_{1a}\lambda_{2b}x_{12}^{ab}),
    \end{align}
    where $x_{12}^{ab}=x_2^{ab}-x_1^{ab}$. Using \eqref{integralclaim2} in the above equation, we can convert a projective four-dimensional integral into a three-dimensional volume integral. Using the notation $\lambda_{1(a}\lambda_{2b)}=\frac{\lambda_{1a}\lambda_{2b}+\lambda_{1b}\lambda_{2a}}{2}=V_{12ab}$, $\langle\lambda\rho,x_{12}\rangle \equiv \lambda_a\rho_b x_{12}^{ab}$ and $V=|V^{\mu}_{12}|= \sqrt{-2V_{12ab}V_{12}^{ab}}=|\langle12\rangle|$ we get,
\begin{align}
    &=\frac{4 Vol(GL(1,\mathbb{R}))}{Vol(GL(1,\mathbb{R}))^2} \int \frac{d^3V_{12}}{|\langle12\rangle|}\bigg(\langle\zeta_1\zeta_2,V_{12}\rangle+ \frac{\zeta_1\cdot\zeta_2 \langle12\rangle}{2}\bigg)^{2s} \text{Sign}(\langle12\rangle) (-i)^{2s+1} \delta^{[2s+1]}(V_{12}\cdot x_{12}) \notag \\
    & =\frac{4}{Vol(GL(1,\mathbb{R}))} \int \frac{d^3V_{12}}{\langle12\rangle}\langle\zeta_1\zeta_2,V_{12}\rangle^{2s}\bigg(1+ \frac{\zeta_1\cdot\zeta_2 \langle12\rangle}{2\langle\zeta_1\zeta_2,V_{12}\rangle}\bigg)^{2s} (-i)^{2s+1} \delta^{[2s+1]}(V_{12}\cdot x_{12}) \notag \\
    & =\frac{4}{Vol(GL(1,\mathbb{R}))} \int \frac{d^3V_{12}}{\langle12\rangle}\langle\zeta_1\zeta_2,V_{12}\rangle^{2s}\bigg(\sum_{r=0}^{2s} \binom{2s}{r} \bigg( \frac{\zeta_1\cdot\zeta_2 \langle12\rangle}{2\langle\zeta_1\zeta_2,V_{12}\rangle}\bigg)^{2s-r} \bigg)(-i)^{2s+1} \delta^{[2s+1]}(V_{12}\cdot x_{12})
\end{align}
We can observe that the above integral will be non-zero only for the even powers of $\langle12\rangle$. Thus, we define $r= (2k+1)|_{k\in\mathbb{Z}_{>0}}$ and proceed, 

\begin{align}
   & =\frac{4}{Vol(GL(1,\mathbb{R}))} \int \frac{d^3V_{12}}{\langle12\rangle}\langle\zeta_1\zeta_2,V_{12}\rangle^{2s}\bigg(\sum_{k=0}^{s-1} \binom{2s}{2k+1} \bigg( \frac{\zeta_1\cdot\zeta_2 \langle12\rangle}{2\langle\zeta_1\zeta_2,V_{12}\rangle}\bigg)^{2(s-k)-1} \bigg)(-i)^{2s+1} \delta^{[2s+1]}(V_{12}\cdot x_{12}) \notag \\
   & =\frac{4}{Vol(GL(1,\mathbb{R}))} \sum_{k=0}^{s-1}\binom{2s}{2k+1}\bigg( \frac{\zeta_1\cdot\zeta_2}{2}\bigg)^{2(s-k)-1} \int d^3V_{12}\langle\zeta_1\zeta_2,V_{12}\rangle^{2k+1} \langle12\rangle^{2(s-k)-2}(-i)^{2s+1} \notag \\ & \hspace{12cm} \delta^{[2s+1]}(V_{12}\cdot x_{12})\notag \\
   & =\frac{4}{Vol(GL(1,\mathbb{R}))} \sum_{k=0}^{s-1}\binom{2s}{2k+1}\bigg( \frac{\zeta_1\cdot\zeta_2}{2}\bigg)^{2(s-k)-1} \int d^3V_{12}\langle\zeta_1\zeta_2,V_{12}\rangle^{2k+1} V_{12}^{2(s-k)-2}  \int \frac{dc_{12}}{2\pi} c_{12}^{2s+1} e^{i c_{12} V_{12}\cdot x_{12} }\notag \\
   & =\frac{4}{Vol(GL(1,\mathbb{R}))} \sum_{k=0}^{s-1}\binom{2s}{2k+1}\bigg( \frac{\zeta_1\cdot\zeta_2}{2}\bigg)^{2(s-k)-1} \int d^3V_{12} \int \frac{dc_{12}}{2\pi} c_{12}^{2s+1} \bigg(\frac{-i}{c_{12}}\bigg)^{2k+1}\bigg(\frac{-i}{c_{12}}\bigg)^{2(s-k-1)} \notag \\ &\hspace{9cm}\bigg(\zeta_{1}^{a}\zeta_{2}^{b}\frac{\partial}{\partial x_{12}^{ab}}\bigg)^{2k+1}\bigg(\frac{\partial^2}{\partial x_{12}\cdot\partial x_{12}}\bigg)^{s-k-1}e^{i c_{12} V_{12}\cdot x_{12} }\notag \\
    & =\frac{4}{Vol(GL(1,\mathbb{R}))} \sum_{k=0}^{s-1}\binom{2s}{2k+1}\bigg( \frac{\zeta_1\cdot\zeta_2}{2}\bigg)^{2(s-k)-1} (-i)^{-2s+1} \int  \frac{dc_{12}}{2\pi} c_{12}^{2} \bigg(\zeta_{1}^{a}\zeta_{2}^{b}\frac{\partial}{\partial x_{12}^{ab}}\bigg)^{2k+1} \notag \\ &\hspace{9cm}\bigg(\frac{\partial^2}{\partial x_{12}\cdot\partial x_{12}}\bigg)^{s-k-1} \delta^3(c_{12} x_{12}^{\mu}) 
\end{align}
\begin{align}
   & =\frac{4(-i)^{-2s+1}}{Vol(GL(1,\mathbb{R}))}\bigg(  \int\frac{dc_{12}}{2\pi} \frac{1}{|c_{12}|}\bigg) \sum_{k=0}^{s-1}\binom{2s}{2k+1}\bigg(\frac{\zeta_1\cdot\zeta_2}{2}\bigg)^{2(s-k)-1} \bigg(\zeta_{1}^{a}\zeta_{2}^{b}\frac{\partial}{\partial x_{12}^{ab}}\bigg)^{2k+1}\bigg(\frac{\partial^2}{\partial x_{12}\cdot\partial x_{12}}\bigg)^{s-k-1} \notag \\ &\hspace{15cm} \delta^3(x_{12}^{\mu}) \notag \\
   & =\frac{2(-i)^{-2s+1}}{\pi} \sum_{k=0}^{s-1}\binom{2s}{2k+1}\bigg(\frac{\zeta_1\cdot\zeta_2}{2}\bigg)^{2(s-k)-1}\bigg(\zeta_{1}^{a}\zeta_{2}^{b}\frac{\partial}{\partial x_{12}^{ab}}\bigg)^{2k+1}\bigg(\frac{\partial^2}{\partial x_{12}\cdot\partial x_{12}}\bigg)^{s-k-1} \delta^3(x_{12}^{\mu})
\end{align}
The odd number of $\zeta_{1}\cdot\zeta_{2}$ in the above expression shows the presence of an odd no. of $\epsilon$ symbols and hence indicates the parity odd nature of the correlator.

Similarly, one can perform the similar steps as above to obtain the parity odd two point function for half-integer spins:
\begin{align}
    \langle J_s J_s\rangle = \frac{2(-i)^{2s-1}}{\pi} \sum_{k=0}^{s-\frac{1}{2}}\binom{2s}{2k}\bigg(\frac{\zeta_1\cdot\zeta_2}{2}\bigg)^{2s-2k}\bigg(\zeta_{1}^{a}\zeta_{2}^{b}\frac{\partial}{\partial x_{12}^{ab}}\bigg)^{2k}\bigg(\frac{\partial^2}{\partial x_{12}\cdot\partial x_{12}}\bigg)^{2s-2k-1} \delta^3(x_{12}^{\mu})
\end{align}
Here also, we can observe the odd number of $\zeta_1\cdot\zeta_2$ products indicating the oddness of the expression under parity.

\subsection{$\langle 0|O_{\Delta}O_{\Delta}|0\rangle$ two point function}
In this subappendix, we shall evaluate the Penrose transform \eqref{ODeltaPenrose} for the two point function of arbitrary scalars \eqref{ODelta2point}. We have,
\begin{align}\label{ODeltaODeltaPenrosedetail1}
    \langle 0|O_{\Delta}(x_1)O_{\Delta}(x_2)|0\rangle &=\int \langle \lambda_1d\lambda_1\rangle \langle\lambda_2d\lambda_2\rangle\bigg(\bigg(\frac{\langle Z_1 IZ_2\rangle}{Z_1\cdot Z_2}\bigg)^{2(\Delta-1)}\frac{1}{(Z_1\cdot Z_2)^2}
    \bigg)\bigg|_{X} \notag \\
    &= \int\langle\lambda_1d\lambda_1\rangle\langle\lambda_2d\lambda_2\rangle \bigg(\frac{\langle 12\rangle}{\lambda_{1a}\lambda_{2b}(x_1-x_2)^{ab}}\bigg)^{2(\Delta-1)} \frac{1}{(\lambda_{1a}\lambda_{2b}(x_1-x_2)^{ab})^2} \notag \\
    &= \int\langle\lambda_1d\lambda_1\rangle\langle\lambda_2d\lambda_2\rangle \frac{1}{(-1)^{2\Delta-2}(2\Delta-2)!}\bigg(\frac{\partial^2}{\partial(x_{12})^{cd}\partial(x_{12})_{cd}}\bigg)^{(\Delta-1)} \frac{1}{(\lambda_{1a}\lambda_{2b}(x_{12})^{ab})^2} \notag \\
    &= \frac{1}{(-1)^{2\Delta-2}(2\Delta-2)!}\bigg(\frac{\partial^2}{\partial(x_{12})^{cd}\partial(x_{12})_{cd}}\bigg)^{(\Delta-1)}\int\langle\lambda_1d\lambda_1\rangle\langle\lambda_2d\lambda_2\rangle \frac{1}{(\lambda_{1a}\lambda_{2b}(x_{12})^{ab})^2}.
\end{align}
Therefore, we see that,
\begin{align}\label{ODeltaODeltaPenrosedetail2}
   \langle 0|O_{\Delta}(x_1)O_{\Delta}(x_2)|0\rangle & = \frac{1}{(-1)^{2\Delta-2}(2\Delta-2)!}\bigg(\frac{\partial^2}{\partial(x_{12})^{cd}\partial(x_{12})_{cd}}\bigg)^{(\Delta-1)} \langle 0|O_1(x_1)O_1(x_2)|0\rangle \notag \\ &= \frac{1}{(-1)^{2\Delta-2}(2\Delta-2)!} \frac{1}{|x_{12}|^{2\Delta}},
\end{align}
where we used the result \eqref{O1O1PenrosetoPositionSpace}. This is the traditional position space two point function thus validating the result.

\bibliographystyle{JHEP}
\bibliography{biblio}
\end{document}